\documentclass[nonacm, sigconf]{acmart}

\renewcommand\footnotetextcopyrightpermission[1]{}
\setcopyright{none}

\usepackage{booktabs}
\usepackage{graphicx}
\usepackage{algorithm}
\usepackage[noend]{algpseudocode}
\usepackage{multirow}
\usepackage{graphicx}
\usepackage{amsmath}
\usepackage{float}

\usepackage[inline]{enumitem}

\usepackage{hyperref}
\usepackage{url}

\usepackage{xspace}
\usepackage{subfig}
\usepackage{xcolor}
\usepackage{tabularx,colortbl}
\usepackage[inline]{enumitem}

\usepackage{hyperref}
\hypersetup{
  colorlinks=true,      % false: boxed links; true: colored links
  linkcolor=blue,       % color of internal links
  citecolor=magenta,    % color of links to bibliography
  filecolor=cyan,       % color of file links
  urlcolor=red          % color of external links
}

\usepackage[compact,small]{titlesec}
\usepackage[font={small}, skip=1pt]{caption}

\usepackage{algorithm, algorithmicx,algpseudocode}
\algrenewcommand\algorithmicindent{0.5em}%

\newenvironment{denseitemize}{
\begin{itemize}[topsep=2pt, partopsep=0pt, leftmargin=1em]
  \setlength{\itemsep}{2pt}
  \setlength{\parskip}{0pt}
  \setlength{\parsep}{0pt}
}{\end{itemize}}

\newenvironment{denseenum}{
\begin{enumerate}[topsep=2pt, partopsep=0pt, leftmargin=1em]
  \setlength{\itemsep}{2pt}
  \setlength{\parskip}{0pt}
  \setlength{\parsep}{0pt}
}{\end{enumerate}}

\def\eg{{e.g.\xspace}}

\def\etc{etc.\xspace}

\def\spotframework{Memtrade\xspace}
\def\manager{manager\xspace}
\def\daemon{producer store\xspace}
\def\harvester{harvester\xspace}
\def\producer{producer\xspace}
\def\Producer{Producer\xspace}
\def\broker{broker\xspace}
\def\client{consumer\xspace}
\def\Client{Consumer\xspace}
\def\tswap{Silo\xspace}

\begin{document}
\sloppy

% \title{\bf {\spotframework}: Disaggregated-Memory Marketplace for Public Clouds}
\title{ {\spotframework}: A Disaggregated-Memory Marketplace for Public Clouds}

\author{Hasan Al Maruf$^*$, Yuhong Zhong$^{\dagger}$, Hongyi Wang$^{\dagger}$, Mosharaf Chowdhury$^*$,  Asaf Cidon$^{\dagger}$, Carl Waldspurger$^{\ddagger}$ }

\affiliation{\institution{University of Michigan$^*$ \hspace{0.4em} Columbia University$^\dagger$
\hspace{0.4em} Carl Waldspurger Consulting$^\ddagger$}
}
%\author{\rm Paper \#41, \pageref{lastpage} pages}

\begin{abstract}
We present {\em \spotframework}, the first memory disaggregation system for public clouds. Public clouds introduce a set of unique challenges for resource disaggregation across different tenants, including security, isolation and pricing. \spotframework allows producer virtual machines (VMs) to lease both their unallocated memory and allocated-but-idle application memory to remote consumer VMs for a limited period of time.
%Memory disaggregation is realized without requiring
% in user space and can be deployed in both virtualized and containerized public clouds.
\spotframework does not require
any modifications to host-level system software or support from the cloud provider.
It harvests producer memory using an application-aware control loop to form a distributed transient remote memory pool with minimal performance impact; it employs a broker to match producers with consumers while satisfying performance constraints; and it exposes the matched memory to consumers as a secure KV cache.
Our evaluation using real-world cluster traces shows that \spotframework provides significant performance benefit for consumers (improving average read latency up to 2.8$\times$)
% a memory-craving
while preserving confidentiality and integrity, with little impact on producer applications (degrading performance by less than 2.1\%). 
\end{abstract}

\settopmatter{printfolios=true}
\maketitle
\thispagestyle{empty}

\section{Introduction}

Cloud resources are increasingly being offered in an elastic and disaggregated manner.
Examples include serverless computing~\cite{awslambda,azure-functions} and disaggregated storage~\cite{ebs,reflex,snowset,flash-disaggre,alibaba-disagg} that scale rapidly and adapt to highly dynamic workloads~\cite{muller2020serverless, pocket, granular, selecta, aurora-serverless}. 

Memory, however, is still largely provisioned statically, especially in public cloud environments.
In public clouds, a user launching a new VM typically selects from a set of static, pre-configured instance types, each with a fixed number of cores and a fixed amount of DRAM~\cite{ec2pricing,gcppricing,azurepricing}.
Although some platforms allow users to customize the amount of virtual CPU and DRAM~\cite{gcpcustom}, the amount remains static throughout the lifetime of the instance.
Even in serverless frameworks, which offer elasticity and auto-scaling, a function has a static limit on its allocation of CPU and memory~\cite{awslambdalimits}.

At the same time, long-running applications deployed on both public and private clouds are commonly highly over-provisioned relative to their typical memory usage.
For example, cluster-wide memory utilization in Google, Alibaba, and Facebook datacenters %running a variety of workloads 
hovers around 40\%--60\% \cite{google-trace-analysis, LegoOS, infiniswap, autopilot}.
Large-scale analytics service providers that run on public clouds, such as Snowflake, fare even worse -- on average 70\%--80\% of their memory remains unutilized~\cite{snowset}.
Moreover, in many real-world deployments, workloads rarely use all of their allocated memory all of the time.
Often, an application allocates a large amount of memory but accesses it infrequently
% during its remaining life-cycle
(\S\ref{subsec:feasibility}).
%Often, large allocations are initially touched by an application, but are subsequently unused or accessed infrequently
% during its remaining life-cycle
%(\S\ref{subsec:feasibility}).
For example, in Google's datacenters, up to 61\% of allocated memory remains idle~\cite{googledisagg}.
Since DRAM is a significant driver of infrastructure cost and power consumption~\cite{MyNVM,memorycost,bandana}, excessive underutilization leads to high capital and operating expenditures, as well as wasted energy (and carbon emissions).
Although recent memory-disaggregation systems address this problem by satisfying an application's excess memory demand from an underutilized server~\cite{infiniswap, googledisagg, remote-regions, leap,AIFM,farmemory-throughput},
existing frameworks are designed for private datacenters.

In this paper, we harvest both unallocated and allocated-but-idle application memory to enable memory disaggregation in public clouds.  
We propose a new memory consumption model that allows over-provisioned and/or idle applications ({\em producers}) to offer excess idle memory to memory-intensive applications ({\em consumers}) that are willing to pay for additional memory for a limited period of time at an attractive price, via a trusted third-party ({\em broker}).
Participation is voluntary, and either party can leave at any time.
Practical realization of this vision must address following challenges:
%Specifically, we must address the following challenges to practically realize this vision:
%how to harvest memory from producers, how to ensure performance benefit with security to consumers, and how to optimally match producers to consumers in the market ensuring economic benefit to both ends. 

%Applying them in public cloud scenarios faces many roadblocks:

\begin{denseenum}
  \item \emph{Immediately Deployable.} Our goal is that \spotframework is immediately deployable on existing public clouds.
  Prior frameworks depend on host kernel or hypervisor modifications~\cite{infiniswap, LegoOS, googledisagg, remote-regions, leap,farmemory-throughput}. In a public cloud setting this would require the operator to manage the memory disaggregation service, since a tenant cannot modify host-level system software. %, which means these frameworks cannot be deployed today.
  In addition, prior work assumes the latest networking hardware and protocols (\eg, RDMA)~\cite{infiniswap, LegoOS, googledisagg, remote-regions, leap,AIFM};
  %In public clouds, a tenant cannot modify host-level system software, and may be unwilling to make guest kernel modifications.
  availability of these features in public clouds is limited, restricting adoption.
  % of ultra-fast protocols such as RDMA in public clouds would restrict their adoption.

	\item \emph{Efficient Harvesting.}
        Memory harvesting needs to be lightweight, transparent and easily deployable without impacting performance.
        Most prior work includes only a VM's {\em unallocated} memory in the remote memory pool.
          Leveraging idle application-level memory -- allocated to an application but later unused or accessed infrequently -- significantly enhances disaggregated memory capacity. This is especially challenging in public clouds, where a third-party provider has limited visibility of tenant workloads, and workloads may shift at any time.
          Existing {\em cold page detection}-based~\cite{googledisagg, thermostat} proactive page reclamation techniques need significant CPU and memory resources, along with host kernel or hypervisor modifications.
% to the workload run by each tenant.
%          can enhance the capacity of disaggregated memory significantly.

  %\item %\emph{Memory allocation granularity.} 
    % \emph{Concerns in remote memory consumption.}
    % [carl] this point is still a bit unclear to me -- exactly what are we trying to say?
    %\emph{Granularity and availability.} Private datacenters can enforce participation in memory disaggregation, enabling
    %servers to share remote memory in large chunks~\cite{infiniswap,remote-regions} with availability guarantees to reduce fragmentation and memory-management overheads.
    %In contrast, in a disaggregated public cloud, tenants may disappear at any time, so large remote allocations
    %may cause performance issues.
    % Unlike private datacenters, a large chunk of remote memory allocation may cause performance issues here.
  % [carl] separated out security concerns, since previous item was too long and muddled
  \item \emph{Performant and Secure Consumption.} 
    %Security assumptions for private datacenters no longer hold. 
    %Prior disaggregation frameworks assume applications all belong to the same organiation.
    To ensure producer-side performance, \spotframework must return a producer's harvested memory seamlessly when needed. 
    Memory offered to consumers may also disappear due to a sudden burst in the producer's own demand for memory, or if a producer disappears unexpectedly.
    \spotframework needs to manage this unavailability to provide a high-performance memory interface.
    % makes remote memory transient in nature -- it can be reclaimed from the consumer without any notice.
    Furthermore, in public clouds, an application's memory may reside in a remote VM that belongs to a different organization, which may expose or corrupt the memory at any time.
    Existing frameworks lack data confidentiality and integrity, and provide poor client accountability for CPU bypass operations~\cite{secured-rdma}, restricting their adoption in public clouds. 
    %In addition, memory offered to consumers may disappear due to a sudden burst in the producer's own demand for memory, or if a producer leaves the system unexpectedly.
    %    This uncertain availability makes remote memory transient in nature -- it can be reclaimed from the consumer without any notice.

  \item \emph{Incentivization and Resource Matching.} 
  Unlike prior work, which assumes cooperative applications, in a public cloud setting we need to create a \emph{market} where producers and consumers have monetary incentives to participate. Producers must be compensated for leasing memory, and the price must be attractive to consumers compared to alternatives (\eg, existing in-memory caching services or spot instances).
  %Existing systems are cooperative, and do not consider incentives in non-cooperative environments.
    %Without attractive incentives, public cloud tenants may not participate in memory disaggregation. 
    %The system also needs to prevent abuse from consumers and producers.
    In addition, producers have varied availability and bursty workload demands, while consumers may have their own preferences regarding remote memory availability, fragmentation, network overhead, and application-level performance, all which must be considered when matching producers to consumers.
\end{denseenum}

We design and develop {\em \spotframework}, an immediately-deployable realization of memory disaggregation in public clouds that addresses these challenges without any host kernel or hypervisor modifications.  %~\carl{``User-space'' isn't quite
%  accurate since the harvester uses a loadable kernel module, as noted later in Sec~\ref{sec:harvester}.}
% the above mentioned issues.  
\spotframework employs a {\em harvester} in each producer VM to monitor its application-level performance and adaptively control resource consumption by dynamically setting the application's Linux control group (cgroup) limits.
% It, therefore, pragmatically harvests
The harvester uses an adaptive control loop that decides when to harvest from and when to return memory to the producer.

To prevent performance degradation in memory-sensitive applications, we design a novel in-memory swap space, {\em \tswap}, which serves as a temporary victim cache for harvested pages. In the case of a sudden loss of performance, the harvester proactively prefetches previously-harvested pages back into memory. 
The combination of these mechanisms allows \spotframework to harvest idle pages with low impact to producer workload performance and offer them to consumers.
%Given the transient nature of harvested memory, 
Consumers of \spotframework can access the harvested memory through a key-value (KV) cache or a swap interface, with cryptographic protections for the confidentiality and integrity of data stored in the untrusted producer VM.

To maximize cluster-wide utilization, \spotframework employs a {\em broker} -- a central coordinator that manages the disaggregated-memory market and matches producers and consumers based on their supply and demand, and helps facilitate their direct communication.
The broker sets the price per unit of remote memory and is incentivized by receiving a cut of monetary transactions.
%\mosharaf{How to set price?}
Although we focus primarily on virtualized public clouds, {\spotframework} can be deployed in other settings, such as private datacenters and containerized clouds.
We plan to open-source \spotframework. 

Overall, we make the following research contributions: 

\begin{denseitemize}

  \item We introduce {\spotframework}, the first end-to-end system that enables memory disaggregation in public clouds (\S\ref{sec:framework}). \spotframework is easily-deployable and does not require any support from the cloud provider.
%  \spotframework provides consumers with a familiar key-value cache interface to access transient remote memory.
    
  \item We design a system to identify and harvest idle memory with minimal overhead and negligible performance impact (\S\ref{sec:harvester}), which uses {\tswap} -- a novel in-memory victim cache for swapped-out pages to reduce performance loss during harvesting an application's idle memory. 
  
  \item We design a broker that arbitrates between consumers and producers, enables their direct communication and implements a placement policy based on consumer preferences, fragmentation, and producer availability (\S\ref{sec:broker}).
  
  %\item We implement consumer interfaces for {\spotframework}, which protect the confidentiality and integrity of data stored in remote memory on untrusted servers (\S\ref{sec:consumer}).
   %\hasan{Should we remove the mention of swap interface from contribution?}
   \item \spotframework improves consumer average latency by up to 2.8$\times$,
     % at an average cost of 3\% degradation in the latency of producer applications.
     while impacting producer latency by less than 2.1\% (\S\ref{sec:eval}), and significantly increases memory utilization (up to 97.9\%). 
      %Memory utilization in public clouds can increase by up to 97.9\%.
     We are the first to evaluate pricing strategies for resource disaggregation in public clouds (\S\ref{subsec:eval-pricing}).
\end{denseitemize}

% \begin{figure}[!t]
% 	\centering
% 	\includegraphics[scale=0.6]{figure/avg-mem.pdf}
% 	\caption{Peak cluster-wide memory utilization in datacenters (over one month for Google \cite{google-trace-analysis} and Facebook \cite{pacman}; 12 hours for Alibaba \cite{LegoOS}).\carl{Please fix y-axis label formatting.}}
% 	\label{fig:avg-mem}
% \end{figure}	

\section{Background and Motivation}
\label{sec:motivation}

\subsection{Memory Disaggregation}
Memory disaggregation exposes capacity available in remote hosts as a pool of memory shared among many machines. 
It is often implemented logically by leveraging unallocated memory in remote machines via well-known abstractions, such as files~\cite{remote-regions}, remote memory paging \cite{infiniswap, app-performance-disagg-dc, swapping-infiniband,farmemory-throughput}, distributed OS virtual memory management \cite{LegoOS} and the C++ Standard Library data structures~\cite{AIFM}.
Existing % memory disaggregation
frameworks require specialized kernels, hypervisors or hardware that might not be available in public clouds.
Prior works focus on private-cloud use cases~\cite{googledisagg,infiniswap,remote-regions,farmemory-throughput,AIFM} and do not consider the transient nature of public-cloud remote memory, nor the isolation and security challenges when consumers and producers belong to different organizations.

\begin{figure}[h]
	\centering
	\subfloat[][\textbf{Google Cluster}]{
		\label{fig:google}
		\includegraphics[width=0.33\columnwidth]{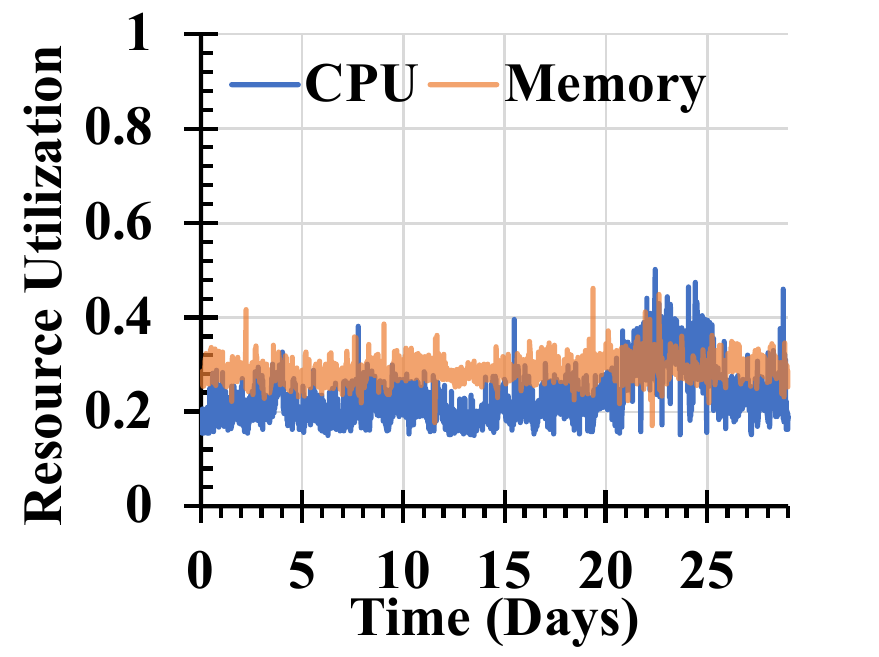}%
	}
%  \hspace{1cm}
	\subfloat[][\textbf{Alibaba Cluster}]{
		\label{fig:alibaba}
		\includegraphics[width=0.33\columnwidth]{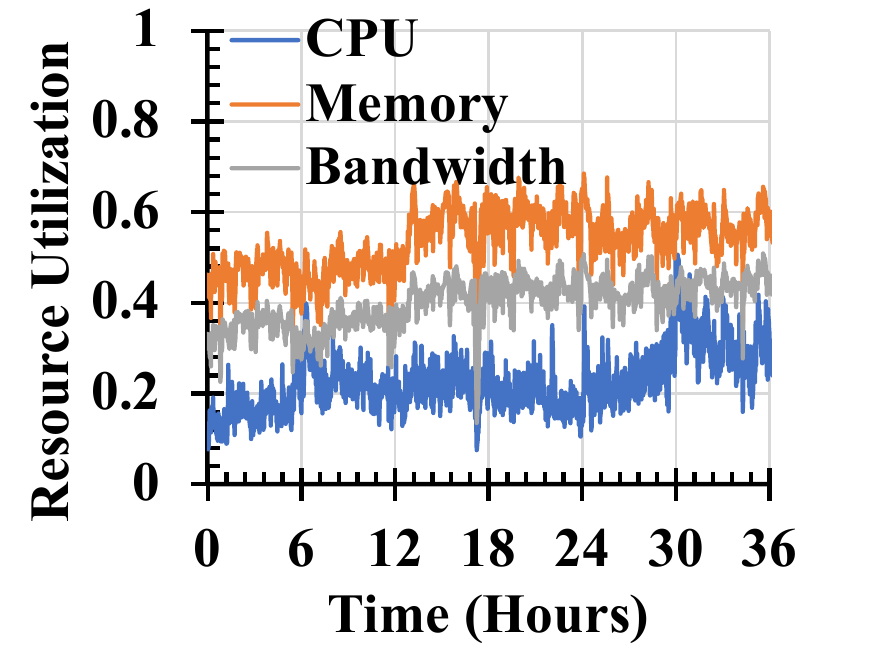}%
	}
	\subfloat[][\textbf{Snowflake Cloud}]{
		\label{fig:snowflake}
		\includegraphics[width=0.33\columnwidth]{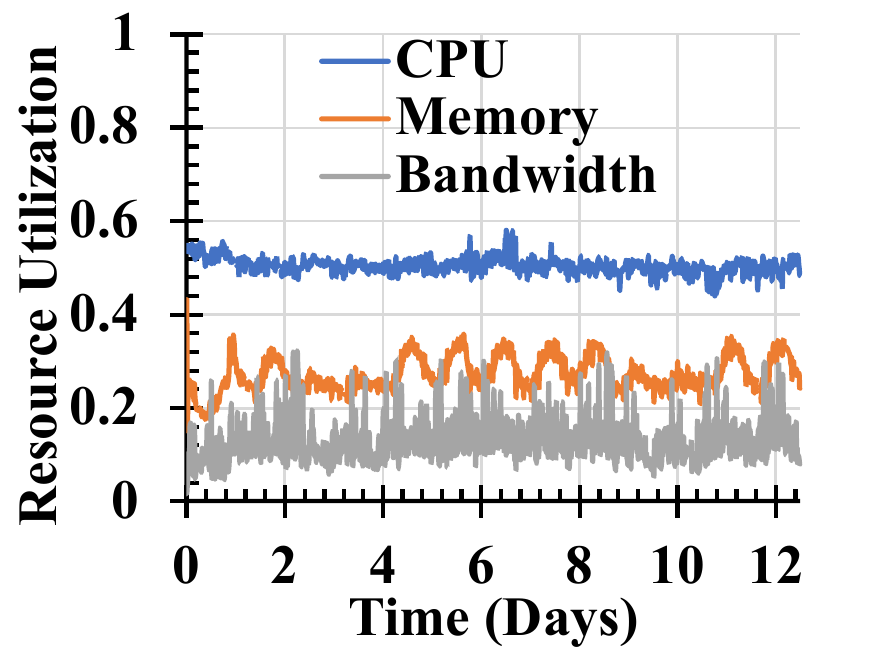}%
	}
	\caption{Cluster resources remain significantly unallocated in (a) Google (bandwidth not reported), (b) Alibaba, and (c) Snowflake.}
  \label{fig:cluster-utilization}
\end{figure}

\subsection{Resource Underutilization in Cloud Computing}
\label{subsec:feasibility}

\paragraph{Underutilized Resources.}
Due to over-provisioning, a significant portion of resources remains idle in private and public clouds that run a diverse mix of workloads. 
To demonstrate this, we analyze production traces of Google~\cite{google-cluster-trace}, Alibaba~\cite{alibaba-cluster-trace},  and Snowflake~\cite{snowset} clusters for periods of 29 days, 36 hours, and 14 days, respectively (Figure~\ref{fig:cluster-utilization}).
In Google's cluster, averaging over one-hour windows, memory usage never exceeds 50\% of cluster capacity. 
In Alibaba's cluster, at least 30\% of the total memory capacity always remains unused.
Even worse, in Snowflake's cluster, which runs on public clouds, 80\% of memory is unutilized on average.

However, the existence of idle memory is not sufficient for providing 
remote memory access; in the absence of dedicated hardware such as RDMA, it also requires additional CPU and network-bandwidth both at the consumer and the producer.
Fortunately, the production traces also show that a significant
portion of these resources are underutilized.  
Approximately 50--85\% of overall cluster CPU capacity remains unused in all of these traces; Alibaba and Snowflake traces, which include bandwidth usage, show that 50--75\% of network capacity remains idle.

\begin{figure}[!t]
	\centering
	\subfloat[][\textbf{Google Cluster Memory}]{
	\label{fig:google-free-duration}
		\includegraphics[width=0.5\columnwidth]{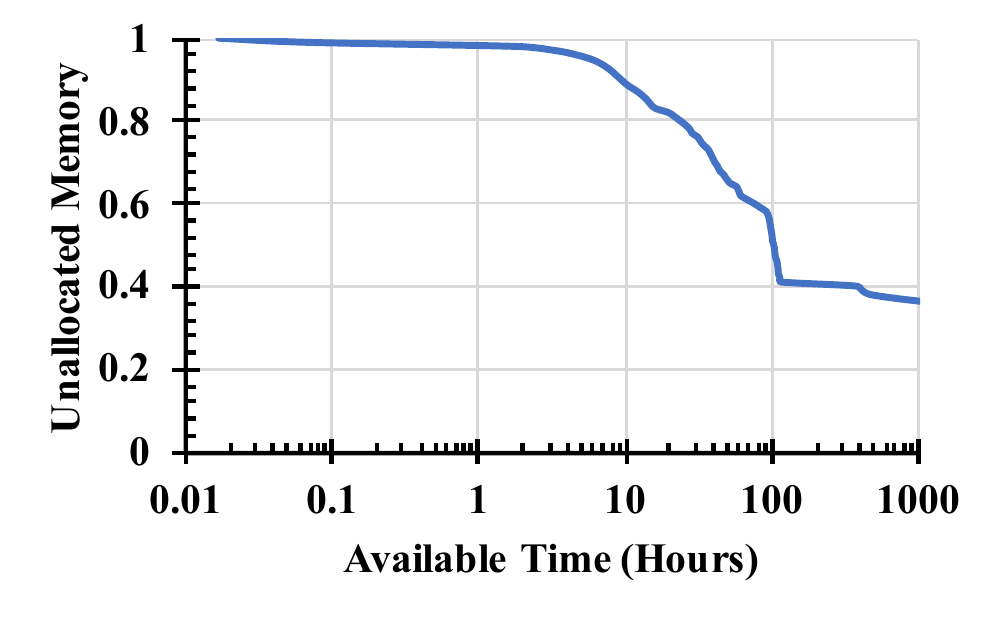}
	}
	\subfloat[][\textbf{Idle Application Memory}]{
	\label{fig:google-cold-duration}
		\includegraphics[width=0.5\columnwidth]{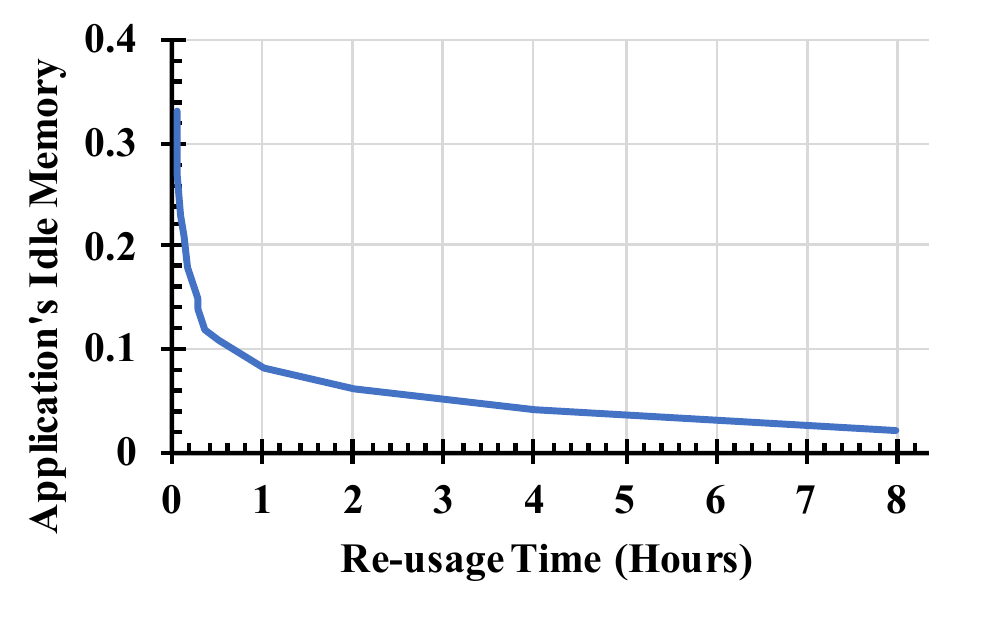}
	}
	\caption{(a) Unallocated memory remains available for long periods,
          but (b) idle memory pages are reused quickly. 
          %\asaf{Figure a here should say "Unallocated memory", not unutilized}
          }
\end{figure}

\paragraph{Availability of Unallocated and Idle Memory.}
Another important consideration
% Another question determining the feasibility of spot memory
is whether unutilized memory remains available for sufficiently long periods of time to enable other applications to access it productively. 
%If unused memory remains idle only for short periods of time, it would be difficult to consume efficiently.
Figure~\ref{fig:google-free-duration} shows that 99\% of the unallocated memory in the Google cluster remains available for at least an hour.
Beyond unallocated memory, which constitutes 40\% of the memory in the Google traces, a significant pool of memory is allocated to applications but remains idle~\cite{farmemory-throughput}. 
Figure~\ref{fig:google-cold-duration} shows that an additional 8\% of total memory is application memory that is not touched for an hour or more.
In public clouds, where many tenants are over-provisioned,
the proportion of application idle memory may be much higher~\cite{snowset}.

% \paragraph{Use Case of Transient Remote Memory.}
\paragraph{Uses for Transient Remote Memory.}
Transient remote memory seems attractive for numerous uses in many environments.
KV caches are widely used in cloud applications~\cite{scalingmemcached,memshare,who-using-redis,facebookworkload,twitter-cache,cachelib},
and many service providers offer popular in-memory cache-as-a-service systems~\cite{elasticache,memcachier,redislabs,azure-cache}.
Similarly, transient memory can be used for filesystem-as-a-service~\cite{deltafs} in serverless computing.
Application developers routinely deploy remote caches in front of persistent storage systems, and in-memory caches are a key driver of memory consumption in clouds~\cite{scalingmemcached,facebookworkload}.
%Alternatively, recent memory disaggregation frameworks ~\cite{remote-regions, infiniswap, app-performance-disagg-dc, LegoOS, googledisagg, leap} allow an unmodified application to access remote memory transparently via demand paging to remote swap space.

% \subsection{Challenges in Disaggregated Public Cloud}
\subsection{Disaggregation Challenges in Public Clouds}
\label{subsec:challenge}

%\subsubsection{Harvesting Spot Memory}
%\label{subsubsec:harvester-challenges}
% \paragraph{Harvesting User-space Memory.}
\paragraph{Harvesting Application Memory.}

\begin{figure}[!t]
	\centering
	\subfloat[][\textbf{Zipfian on Redis}]{
		\label{fig:curve-redis-single}
		\includegraphics[width=0.485\columnwidth]{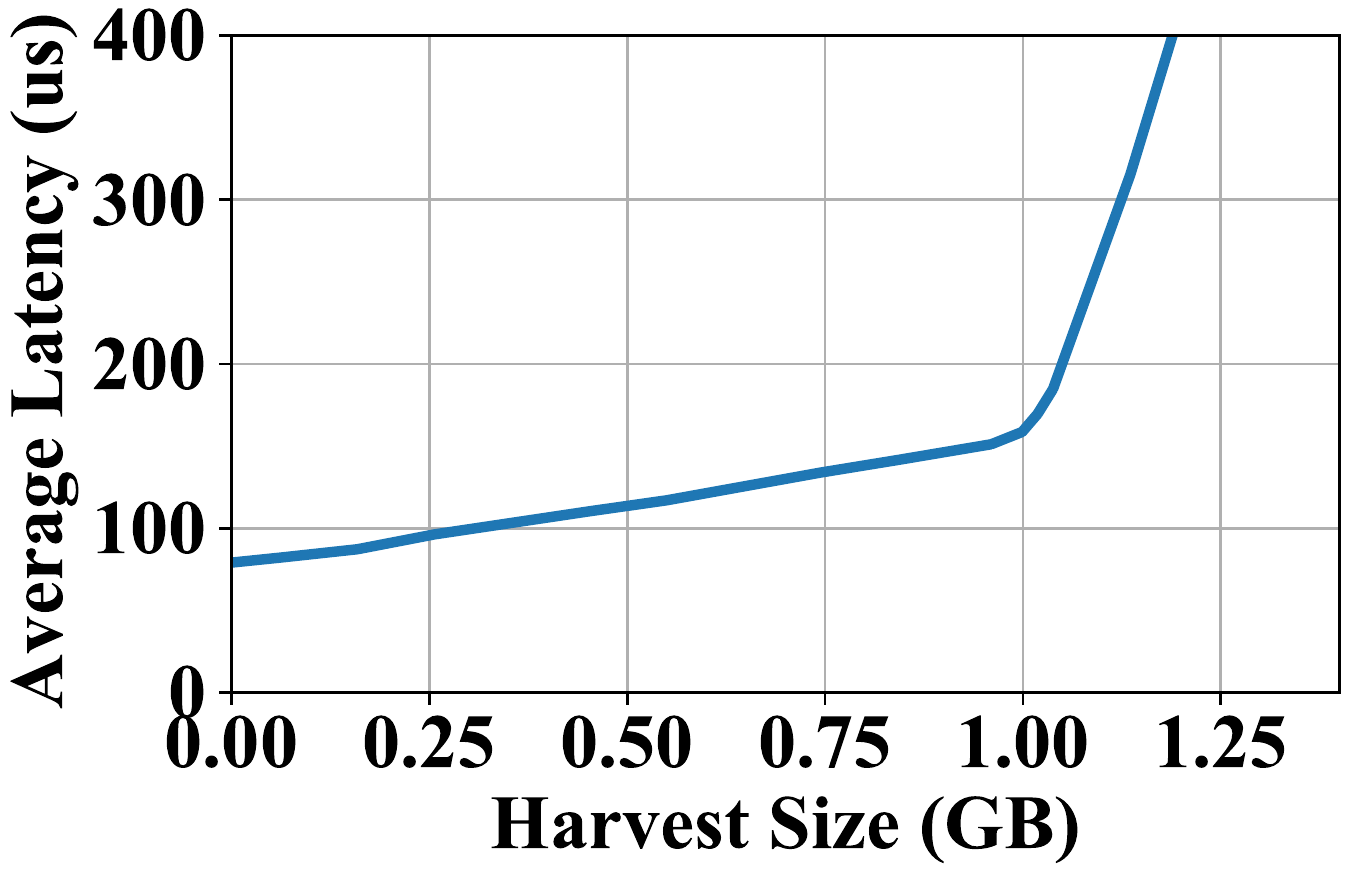}%
	}
%  \hspace{1cm}
	\subfloat[][\textbf{XGBoost}]{
		\label{fig:curve-xgboost-single}
		\includegraphics[width=0.485\columnwidth]{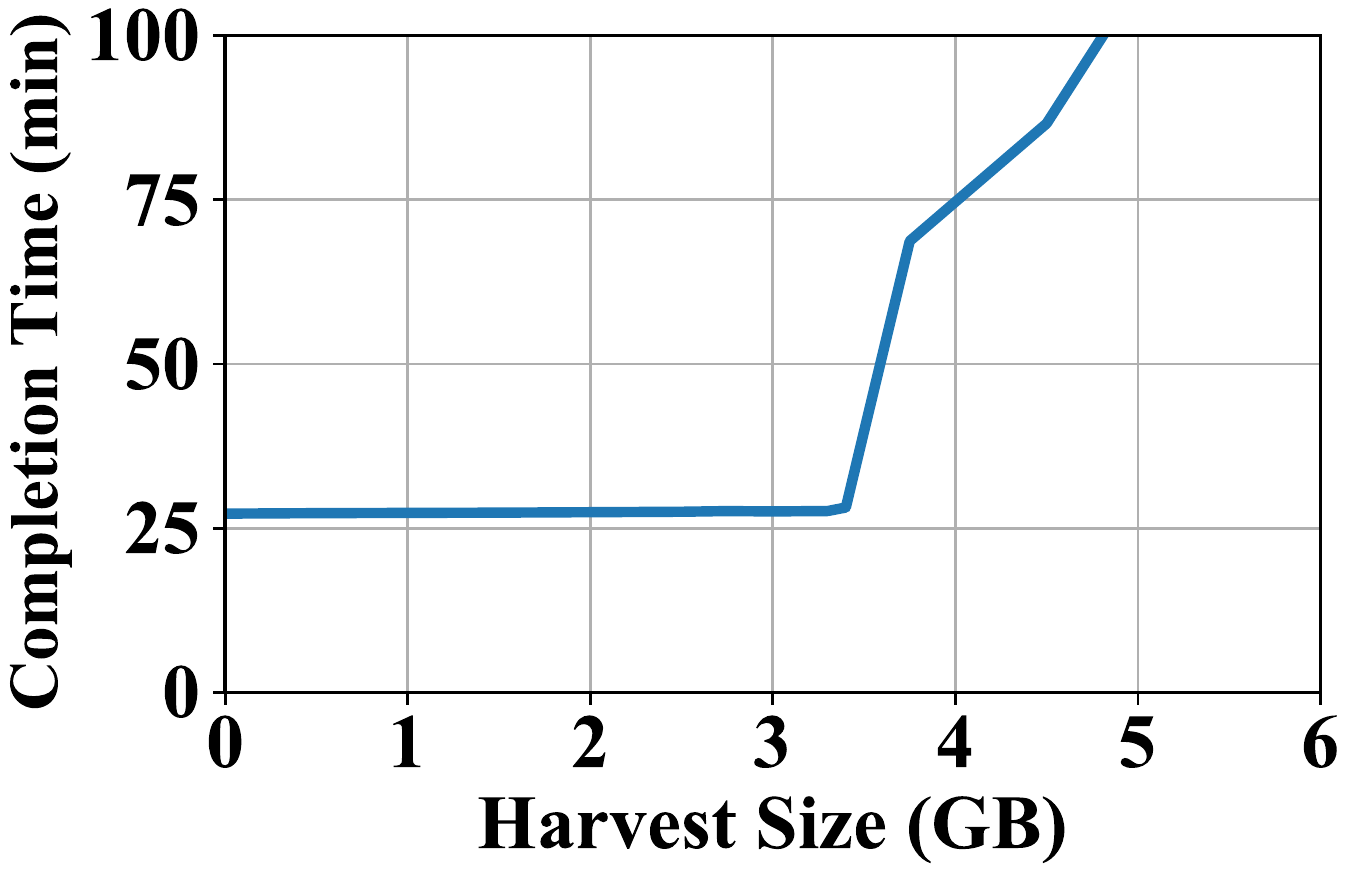}%
	}
	\caption{Performance drop while harvesting different amounts of memory for (a) Zipfian trace on Redis%with 1~GB of DRAM
, and (b) XGBoost training.
}
  \label{fig:performance-curve-mc-tf}
\end{figure}

Beyond unallocated memory, a large amount of potentially-idle memory is allocated to user VMs. %, especially in public clouds.
In many cases, harvesting such idle memory has only minimal performance impacts.
However, harvesting too aggressively can easily result in severe performance degradation, or even crash applications.
%Determining a safe amount of memory to harvest is challenging,
% due to the
%and is exacerbated by the
%cost of fine-grained profiling and the lack of visibility inside producer VMs.
%\paragraph{Impact on Application-level Performance}
Figure~\ref{fig:performance-curve-mc-tf} shows the performance degradation while harvesting memory from two applications.
% For both of the highlighted applications,
We can harvest
% several hundred MBs or few GBs
a substantial amount
of memory from each without much performance loss. However,
performance can quickly fall off a cliff,
% due to % unexpected
and dynamic application load changes %, such as diurnal or weekly load patterns~\cite{googledisagg, autopilot}, which preclude inferring a static harvesting size for a given workload. 
necessitate adaptive harvesting decisions in real-time.

%\paragraph{Overhead of Real-time Detection}
To reclaim an application's idle memory, existing solutions use kernel modifications~\cite{googledisagg, thermostat} to determine the age of pages mapped by the application.
%A well-explored technique is to periodically scan the \textit{accessed bit} present in page table entries to infer if a physical page is accessed in a given time period.
%If a page's age goes beyond a threshold, then it is marked as a cold page and, therefore, reclaimed.
Such an approach is difficult to deploy in public clouds, where each user controls its own kernel distribution. %One major pitfall of this approach in the public cloud is this demands the modification of hypervisor and even the host kernel, which may not always be feasible.
Moreover, continuous page tracking can consume significant CPU and memory and require elevated permissions from the host kernel~\cite{ipt}.

%\subsubsection{Placement of Consumer Requests}
%\label{subsubsec:place-spot}
%\paragraph{Consumption of Transient Remote Memory.}
%Each producer and consumer have their own respective supply and demand characteristics.
%Given the remote I/O cost, % cost of reading and writing over the network,
%too much churn in memory availability may deteriorate a consumer's performance. 
%On the other hand, multiple consumers may contend for CPU and bandwidth resources when accessing remote memory from the same producer.
%Moreover, the producers may demand for varied incentive for leasing spot memory based on their availability and performance guarantee.
%Matching consumers with available producers considering the costs and benefits to both parties is challenging.
%
%\paragraph{Uncertainty in Re-usage.}
%\subsubsection{Consumption of Spot Memory}
%\label{subsubsec:consume-spot}
%
\paragraph{Transience of Remote Memory.}
Each producer and consumer have their own respective supply and demand characteristics.
At the same time, producers may disappear at any time,
and the amount of unallocated and idle memory that can be harvested safely from an application varies over time.
Given the remote I/O cost, % cost of reading and writing over the network,
too much churn in memory availability may deteriorate a consumer's performance. 
%These effects combine to cause significant uncertainty; remote memory in a public cloud is inherently transient in nature.
% Asaf: commenting these two lines out (repetitive with previous paragraph)
%Frequent eviction of consumer state due to bursts in producer memory access can nullify the benefit of disaggregated memory.
%Consumers also have their own performance requirements while accessing remote memory, and the broker needs to respect these constraints when assigning memory.

%\paragraph{Fair Sharing and Eviction}
%When a single producer's harvested memory is assigned to multiple consumers, there will be contention for other producer resources, such as CPU and network bandwidth.  
%A natural question that arises is how to ensure fair share of these resources. 
%There can be contention for the producer's memory as well. 
%For example, when a producer needs to quickly reclaim its memory, it needs to decide which consumer's memory it should evict.

\paragraph{Security.}
%In contrast to deployments in private data centers, 
VMs that do not belong to the same organization in the public cloud are completely untrusted.  
Since consumer data residing in remote memory may be read or corrupted % at any time
due to accidents or malicious behavior, its confidentiality and integrity must be protected.
%As a result, we need to rely on cryptographic techniques to protect its confidentiality and integrity.
Producer applications must also be protected from malicious remote memory operations, and the impact of overly-aggressive or malicious consumers on producers must be limited.

%Since the contract between producers and consumers allows remote memory to be reclaimed without notice\mosharaf{It was not clear in the intro. I think we need to give a precise model in the intro given that section 2 is not spot memory definition any more.} -- even in the absence of any misbehavior -- we do not attempt to ensure availability for consumer data residing in remote memory.  
%However, to help prevent remote memory operations from harming the availability of producer applications from which memory has been harvested, it is important to limit the impact of overly-aggressive or malicious consumers on producers.
%\mosharaf{Too much talking. Just highlight the points.}

\section{\spotframework: Overview}
\label{sec:framework}
\begin{figure}[!t]
	\centering
	\includegraphics[scale=0.6]{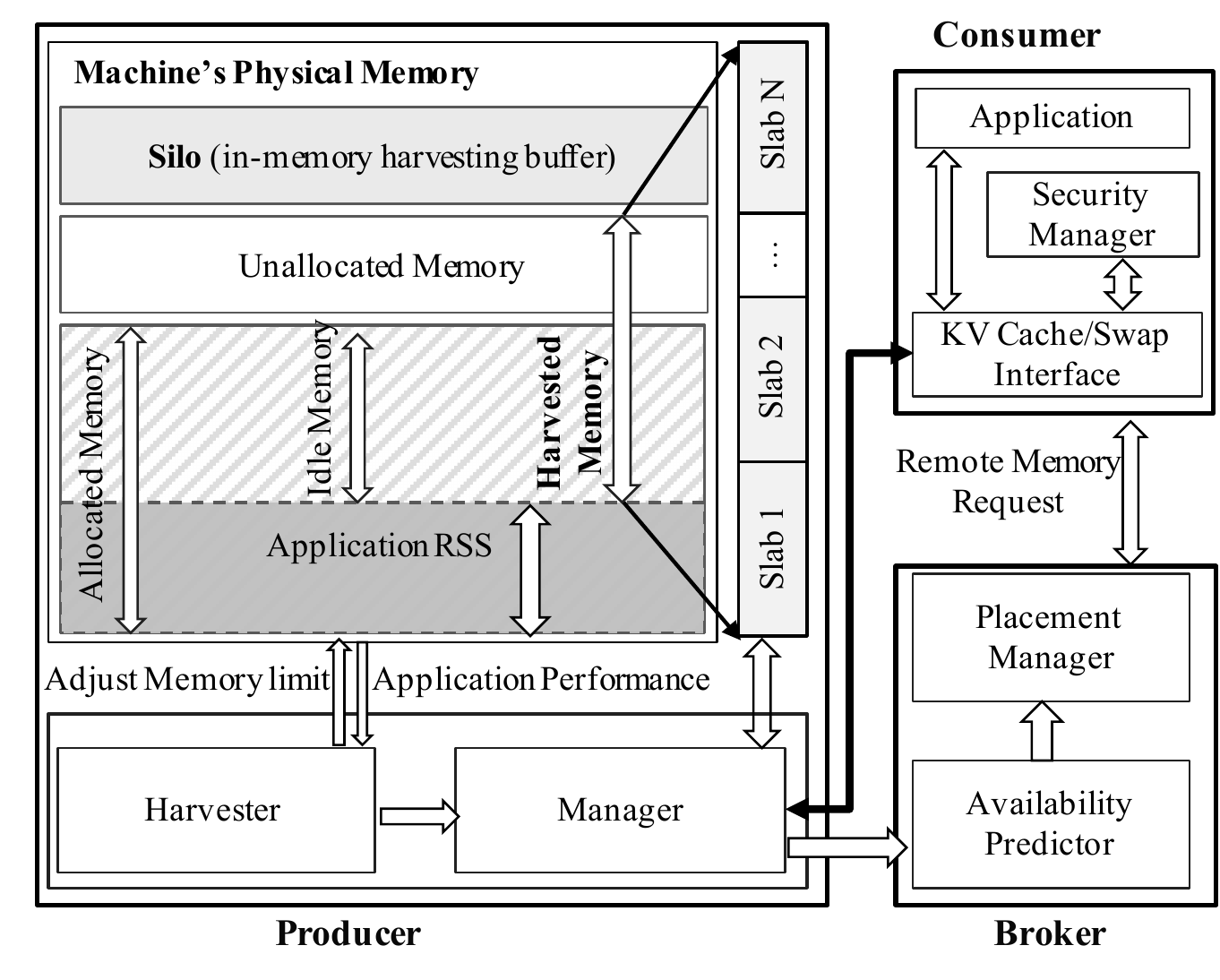}
	\caption{%In {\spotframework}, a \producer monitors application-level performance to make harvesting decisions and exposes the harvested memory through a key-value cache interface. 	The \broker arbitrates between producers and consumers.	The consumer interface ensures the confidentiality and integrity of data placed in remote memory.
	{\spotframework} architecture overview.}
	\label{fig:architecture}
\end{figure}

{\em \spotframework} is a system that realizes memory disaggregation in public clouds.
It consists of three core components (Figure~\ref{fig:architecture}):
{\bf (i) \producer{}s}, which expose their harvested idle memory to the disaggregated-memory market (\S\ref{sec:harvester});
{\bf (ii) the \broker}, which pairs producers with consumers while optimizing cluster-wide objectives, such as maximizing resource utilization (\S\ref{sec:broker}); and 
{\bf (iii) \client{}s}, which request remote-memory allocations based on their demand and desired performance characteristics (\S\ref{sec:consumer}).
This section provides an overview of these components and their interactions; more details appear in subsequent sections.

\paragraph{Producers.}
A \producer employs a collection of
%user-level~\carl{not quite; Silo requires a kernel module} 
processes to harvest idle memory within a VM, making it available to the disaggregated-memory market.
A producer voluntarily participates in the market by first registering with the \broker.
%\mosharaf{This is something that can be in the appendix. What does this API look like?}~\carl{Seems fine as-is to me; not clear it's worth specifying a more complete API in an appendix.}
Next, the \producer monitors its resource usage and application-level performance metrics, periodically notifying the \broker about its resource availability.
The \producer harvests memory slowly until it detects a possible performance degradation, causing it to back off and enter recovery mode. 
During recovery, memory is returned to the producer application proactively until its original performance is restored.
Once the \producer determines that it is safe to resume harvesting, it transitions back to harvesting mode.
% All these producer functionalities are implemented without guest kernel or hypervisor modifications.

When the \broker matches a consumer's remote memory request to the producer, it is notified with the consumer's connection credentials and the amount of requested memory.
%The \producer runs a key-value cache similar to Redis~\cite{redis} or Memcached~\cite{memcached} which is dedicated to that consumer, providing a {\tt GET} / {\tt PUT} / {\tt  DELETE} interface to consume the assigned memory.
The \producer then exposes harvested memory through fixed-sized slabs dedicated to that consumer. %, that can  be accessed either through a KV interface (\eg, {\tt GET} / {\tt PUT} / {\tt  DELETE} interface) or remote swap.
A \producer may stop participating at any time by deregistering with the \broker.
%However, one can replace the consumption interface by a transient file system or swap-device to handle different aspects of disaggregation. 

\paragraph{Broker.}
%The producers and consumers communicate to the \broker with their supply and demand information, respectively. 
The \broker arbitrates between producers and consumers, matching supply and demand for harvested remote memory while considering consumer preferences and constraints. 
While \spotframework supports untrusted producers and consumers from diverse tenants, its logically-centralized \broker component should be run by a trusted third party -- such as a caching-as-a-service provider~\cite{memcachier,redislabs} or the public cloud operator.
The \broker facilitates the direct connection between the consumer and producer using
virtual private cloud interconnection mechanisms~\cite{transit-gateway,vnet-peering}.
The \broker decides on the per-unit remote memory price for a given lease time in the disaggregated system, based
in part on monitoring the current price of spot instances offered in the same public cloud.
Appropriate pricing provides incentives for both producers and consumers to participate in the market;
the \broker receives a cut of the monetary transactions it brokers as commission.

To improve the availability of transient remote memory, % during a matching decision,
the \broker relies on historical resource usage information for producers to predict their future availability.  
It additionally considers producer reputations, based on the frequency of any prior broken leases, in order to reduce
occurrences of unexpected remote memory revocations.
Finally, it assigns producers to consumers in a manner that maximizes the overall cluster-wide resource utilization.

\paragraph{Consumers.}
A consumer voluntarily participates in the disaggregated-memory market by registering its connection credentials with the \broker.
%Once approved by the {\broker}, the consumer can submit a remote memory request by invoking the function {\tt remote\_memory\_request(memory size, lease time, constraints)}.~\carl{As the only code-like function reference, this seems oddly out of place.  Suggest just replacing with text.}
Once approved by the {\broker}, the consumer can submit a remote memory request by specifying its required remote memory, lease time, and preferences.
After matching the request with one or more producers, the {\broker} sends a message to the \client with connection credentials for the assigned producer(s).

The {\client} then communicates directly with assigned producers through a simple KV cache {\tt GET} / {\tt PUT} / {\tt  DELETE} interface
% or a remote-paging mechanism
to access remote memory.
We also implement a transparent remote-paging interface for the {\client}. However, since memory will be occasionally evicted by producers, and a swap interface assumes data is stored persistently, we do not focus on this interface. Conveniently, applications using caches assume that data is not persistent, and may be evicted asynchronously.
%o
The confidentiality and integrity of {\client} data stored in producer memory is ensured cryptographically in a
transparent manner (\S\ref{subsec:security-design}).

\section{Producer}
\label{sec:harvester}
The \producer consists of two key components:
% The core functionalities of the \producer can be divided into two pieces: 
% {\bf (i) \harvester},
the {\em \harvester}, which employs a control loop to harvest application memory, and
% {\bf (ii) \manager}, that
the {\em \manager}, which 
% transforms the harvested memory into
exposes harvested memory to {\client}s as remote memory.
The \producer does not require modifying host-level software,
facilitating deployment in existing public clouds.
Our current \producer implementation only supports Linux VMs.
The \harvester coordinates with a loadable
kernel module within the VM to make harvesting decisions, without recompiling the guest kernel.

\begin{algorithm}[!t]
	%\algsetup{linenosize=\tiny}
	\small
   \caption{Harvester Pseudocode}
   \label{alg:controller}
   \begin{algorithmic}[1]
     \Procedure{ DoHarvest}{} %\Comment{}
        \State Decrease cgroup memory limit by ChunkSize %\Comment{default ChunkSize 64MB}
        %\Comment{adaptively decrease}
        \State $sleep(CoolingPeriod)$  \Comment{wait for performance impact}
     \EndProcedure
    
     %\Statex
     
     \Procedure{DoRecovery}{} %\Comment{}
        \While{{RecoveryPeriod} not elapsed}
        \State Disable cgroup memory limit
        \EndWhile
     \EndProcedure
     
     %\Statex
          
     \Procedure{RunHarvester}{} %\Comment{}
        \For{each performance monitor epoch}
            \If{no page-in}
                \State Add performance data point to baseline estimator 
                \State Generate baseline performance distribution
            \EndIf
            \State Generate recent performance distribution

            \If{ performance drop detected } 
            \State DoRecovery()
            \Else
            \State DoHarvest()
            \EndIf
            \If{ severe performance drop detected }
            \State Prefetch from disk
            \EndIf
        \EndFor
     \EndProcedure
     \end{algorithmic}
 \end{algorithm}

The \harvester runs producer applications within a Linux control group (cgroup)~\cite{cgroup} to monitor and limit the VM's consumption of resources, including DRAM and CPU; network bandwidth is managed by a custom traffic shaper (\S\ref{subsec:manager}).
Based on application performance, the \harvester decides whether to harvest more memory or release already-harvested memory. 
% and, therefore, adaptively adjusts the cgroup memory limit to swap out idle application pages. 
%
Besides unallocated memory which is immediately available for {\client}s, the \harvester can increase the free memory within the VM by reducing the resident set size (RSS) of the application.
In {\em harvesting mode}, the cgroup limit is decreased incrementally to reclaim memory in relatively small chunks; the default \texttt{ChunkSize} is 64~MB.
If a performance drop is detected, 
the \harvester stops harvesting and enters {\em recovery mode},
% Inspired by TCP congestion control~\cite{rfc5681}, the \harvester applies an additive-decrease, multiplicative-increase (ADMI) control loop to determine the \emph{recovery period} needed to improve the application-level performance.\footnote{TCP congestion control uses AIMD while we use ADMI. 
% This is due to the fact that TCP is searching for the maximal bandwidth, while we are probing for the minimal recovery period.}
% 
% During the recovery mode, it disables the cgroup memory limit, which allows the application to fully recover.
disabling the cgroup memory limit and allowing the application to fully recover.

Because directly reclaiming memory from an application address space can result in performance cliffs if hot pages are swapped to disk, we introduce
{\em \tswap}, a novel in-memory region that serves as a temporary buffer, or victim cache, holding harvested pages
before they are made available as remote memory.  
\tswap allows the \harvester to return recently-reclaimed pages to applications efficiently.
In addition, when it detects a significant performance drop due to unexpected load, \tswap proactively prefetches swapped-out pages from disk, which helps mitigate performance cliffs.
Algorithm~\ref{alg:controller} presents a high-level sketch of the {\harvester}'s behavior.

The \manager exposes the harvested memory via a key-value cache {\tt GET} / {\tt PUT} / {\tt DELETE} interface,
by simply running a Redis server for each consumer.  
A key challenge for the \manager is handling the scenario where the \harvester needs to evict memory.
We leverage the existing Redis LRU cache-eviction policy, % that prioritizes eviction of LRU items,
which helps reduce the impact on consumers when the producer suddenly needs more memory.

% \begin{figure}[!t]
% 	\centering
% 	\includegraphics[width=\columnwidth]{figure/state-diagram}
% 	\caption{The \producer iteratively decreases cgroup memory limit unless it faces performance drop or workload bursts. During recovery state it iteratively increases cgroup limit. Recovery period is determined by an adaptive control loop.\mosharaf{The states should be separate from the actions in the state. Right now its more confusing than helpful.}
% 	}
% 	\label{fig:state-diagram}
% \end{figure}	

%\mosharaf{We need a clear overview first and then the rest of the section should address the details we omitted in the overview. Right now, one has to read the entire section to understand even the high level.}

\subsection{Adaptive Harvesting of Remote Memory}

\paragraph{Monitoring Application Performance.} 

The \harvester provides an interface for applications to periodically report their performance,
with a metric such as latency or throughput.
Without loss of generality, our description uses a performance
metric where higher values are better.
Many applications already expose performance metrics that
the \harvester can leverage.  
% For applications with standard interface for monitoring its performance metrics, the \harvester can simply leverage it.
Otherwise, the \harvester uses the swapped-in page count (promotion rate) as a proxy for performance~\cite{googledisagg}.

\begin{figure}[!t]
  \centering
  \subfloat[][\textbf{\tswap first swaps to RAM.}]{
  \includegraphics[width=0.75\columnwidth]{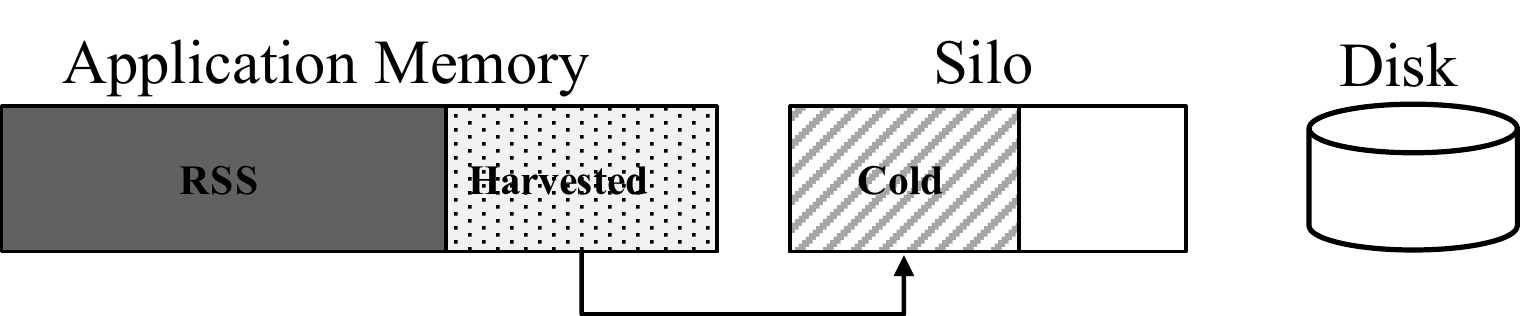}
  \label{fig:harvester-swap}
  }\\[-0.4ex]
 %\hspace{1em}
   \subfloat[][\textbf{Keeps hot pages in RAM, evicts cold pages to disk.}]{
  \includegraphics[width=0.75\columnwidth]{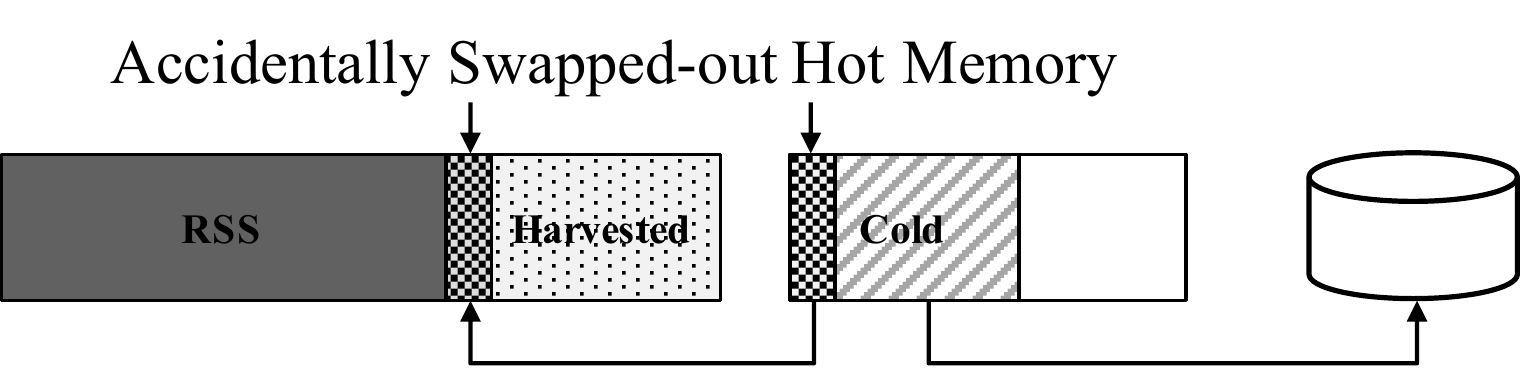}
   \label{fig:harvester-cooling}
  }\\[-0.4ex]
 %\hspace{1em}
   \subfloat[][\textbf{Proactive prefetching during bursts.}]{
  \includegraphics[width=0.75\columnwidth]{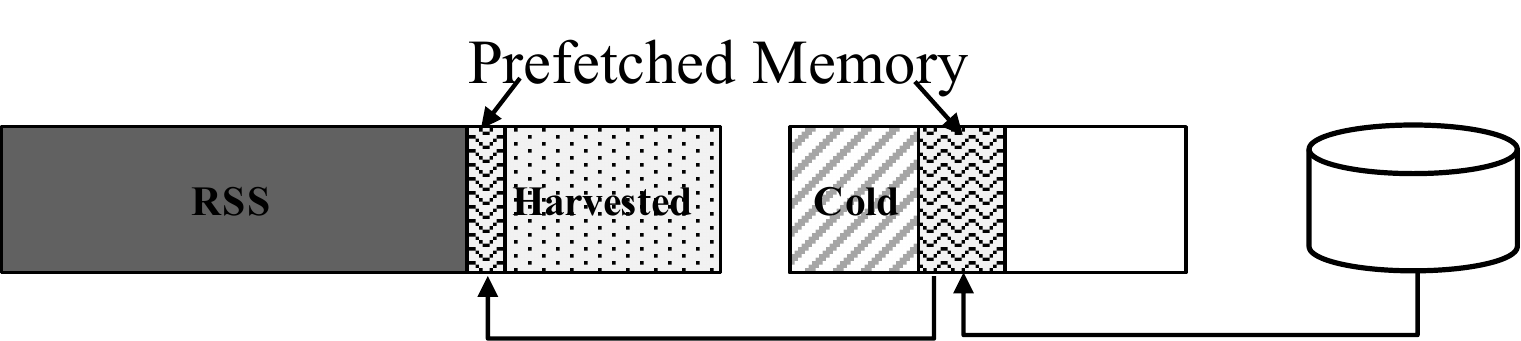}
   \label{fig:harvester-prefetch}
  }
   \caption{Memory harvesting and recovery using \tswap.}
  \label{fig:harvester}
\end{figure}

\paragraph{Estimating the Baseline.} To determine whether the memory limit should be decreased or increased, the \harvester compares the current application performance metric to baseline values observed \emph{without memory harvesting}. 
Of course, measuring performance without memory harvesting is difficult while the \producer is actively reclaiming memory.  
To estimate the baseline performance without harvesting, we use statistics for swap-in events.
% While monitoring application performance,
When there are no swap-in events, the application has enough memory to run its workload.
Therefore, the \harvester includes the performance metric collected during these times
% of that particular moment
as a baseline.  An efficient AVL-tree data structure is used to track
these points, which are discarded after an expiration time.  Our current implementation
adds a new data point every second, which expires after a 6-hour \texttt{WindowSize}. We found this yielded good
performance estimates; shorter data-collection intervals or longer expiration times could
further improve estimates, at the cost of higher resource consumption~(\S\ref{subsec:eval-harvester}).%\mosharaf{Sensitivity analysis?} 
% However, to accommodate workload shifts, the expiration times cannot be too long. \hasan{any insight on the range of a good expiration time?}

% It inserts the candidate into an AVL tree data structure and generates the distribution of the baseline performance.
% The AVL tree allows the \harvester to provide an online estimation of the baseline performance even if the producer wants to consider a large number of performance data points for more accurate estimation.

% Each entry in the AVL tree has an associated expiration time, and data points are discarded after they expire. 
% In the current implementation of \spotframework, the default time gap between monitoring the performance metric is 1 second, and it uses 1 hour as the baseline data point expiration time. 
% Using smaller time gaps and larger expiration time will increase the number of data points and result in a better performa% nce estimator at the cost of higher CPU consumption.

%\paragraph{Estimating Baseline Performance.}

% \paragraph{Performance Drop Detection.}
\paragraph{Detecting Performance Drops.}

To decide if it can safely reduce the cgroup memory limit, the \harvester checks whether performance
has degraded more than expected from its estimated baseline performance.
Similar to baseline estimation, 
the \harvester maintains another AVL tree to track % the set of
application performance values over the same period.

After each performance-monitoring epoch, it calculates the 99th percentile (p99)
% XXX [carl] -- footnote seems unnecessary to me
% \footnote{We define p99 as the performance measurement that is {\em worse} than 99\% of the measurements within the distribution.}
of the recent performance distribution.
The \harvester assumes performance has dropped if the recent p99 is worse than baseline p99
by \texttt{P99Threshold} (by default, 1\%), and it
stops reducing the cgroup size, entering a recovery state.
It then releases harvested memory adaptively to minimize the performance drop. 
Different percentiles or additional criteria can be used to detect performance drops.

\paragraph{Effective Harvesting with \tswap.}
The \harvester reclaims memory until a performance drop is detected.
However, some workloads are extremely sensitive, and losing even a small amount of hot memory can result in severe performance degradation.
Also, because the \harvester adjusts the application memory size via a cgroup, it relies on the Linux kernel's Page Frame Reclamation Algorithm (PFRA) to make decisions.
Unfortunately, PFRA is not perfect and sometimes reclaims hot pages, even with an appropriate memory limit.

\begin{figure}[!t]
	\centering
	\subfloat[][\textbf{Zipfian on Redis}]{
		\label{fig:curve-redis}
		\includegraphics[width=0.49\columnwidth]{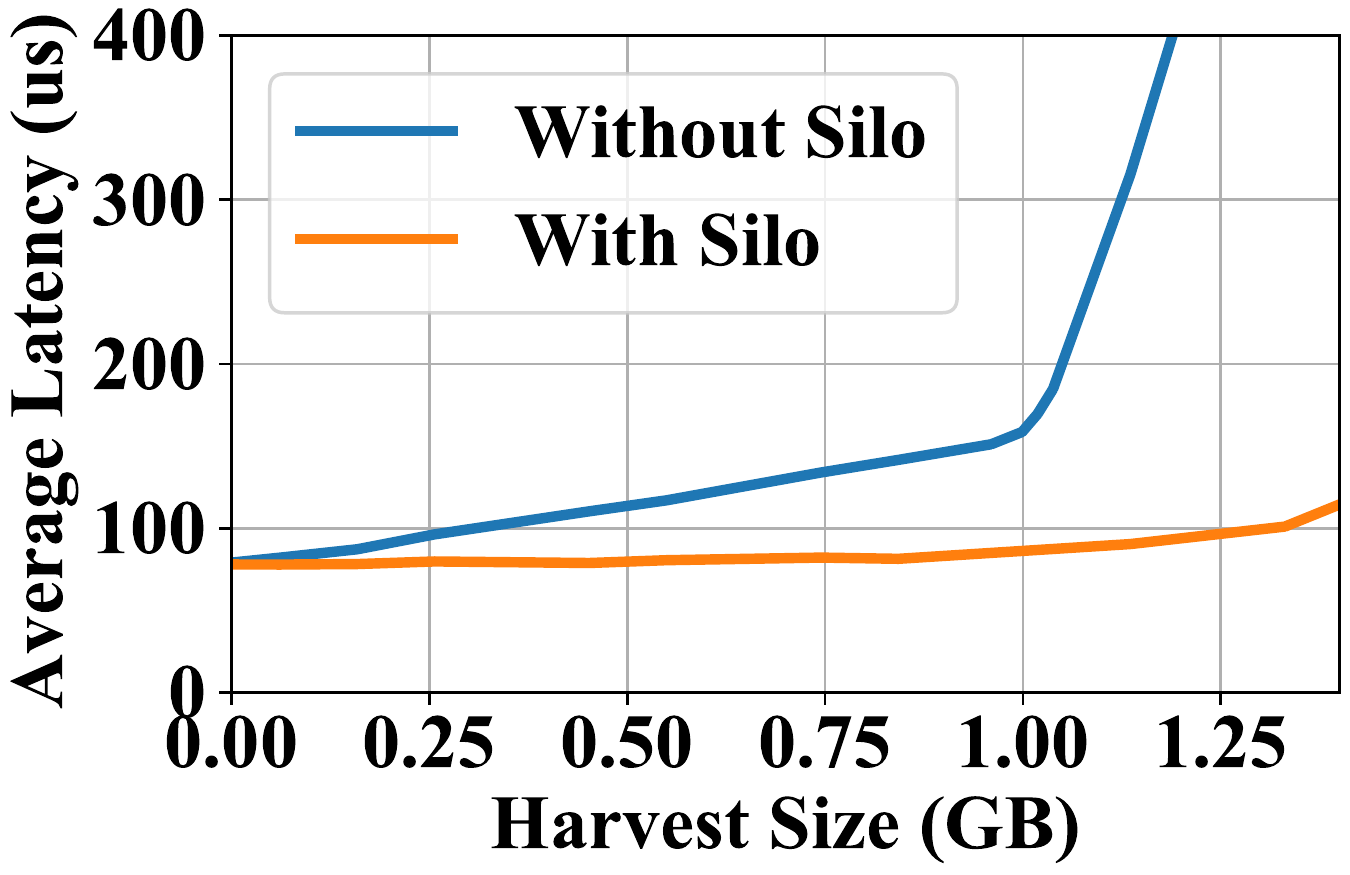}%
	}
%  \hspace{1cm}
	\subfloat[][\textbf{XGBoost}]{
		\label{fig:curve-xgboost}
		\includegraphics[width=0.49\columnwidth]{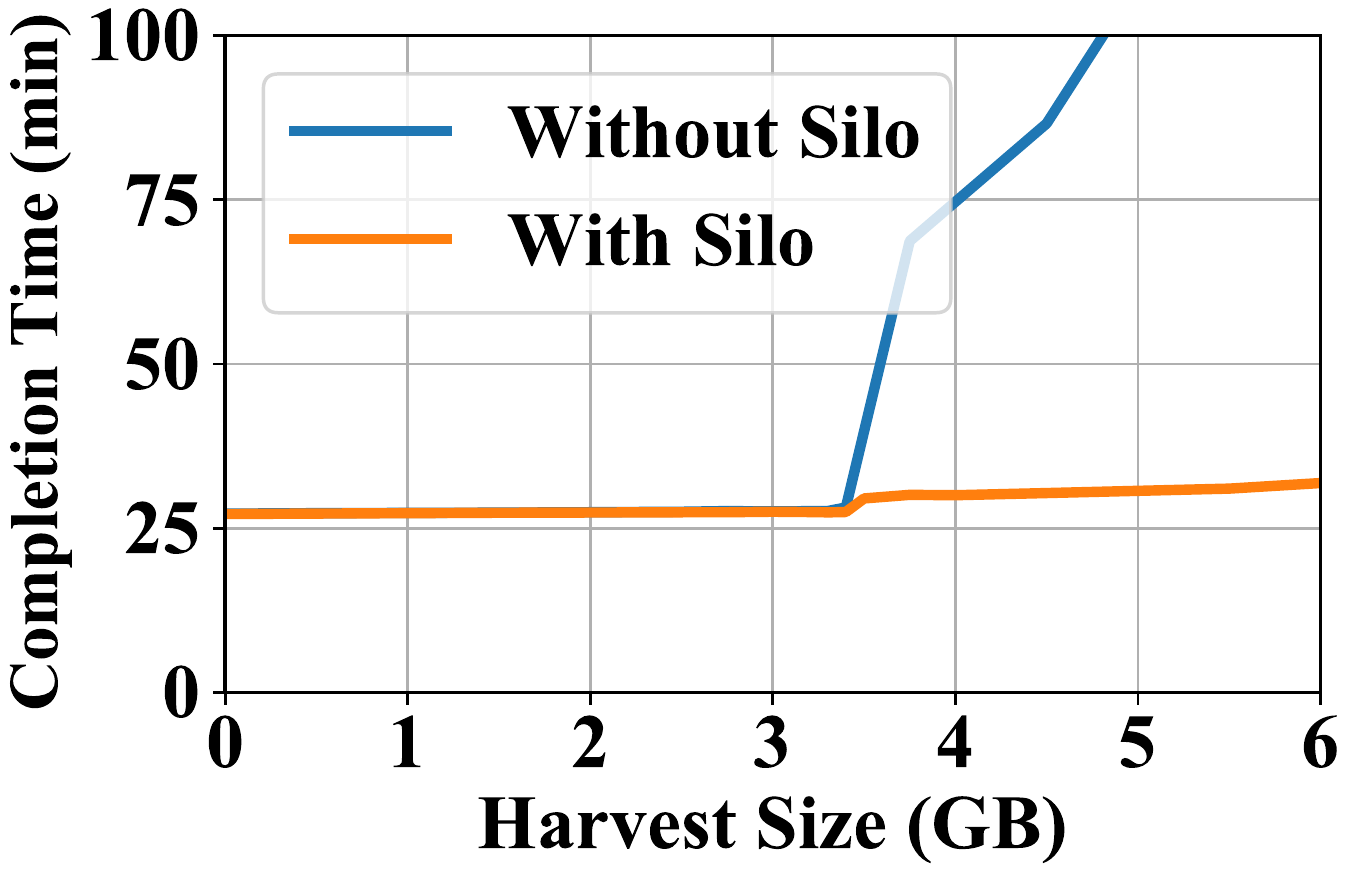}%
	}
	\caption{Performance degradation caused by harvesting different amounts of memory with and without \tswap for (a) Zipfian trace running on Redis%with 1~GB of DRAM
, and (b) XGBoost training.% with 2~GB of DRAM.\mosharaf{Are these X GB DRAM/container or X GB working set? What's the working set size?}
}
  \label{fig:performance-curve}
\end{figure}

To address these problems, we design {\em \tswap}, a novel in-memory area for temporarily storing swapped-out pages. 
We implement \tswap as a loadable kernel module that is a backend for the Linux frontswap interface~\cite{frontswap}.  
The guest kernel swaps pages to \tswap instead of disk, thus reducing the cost of swapping (Figure~\ref{fig:harvester-swap}).
If a page in \tswap is not accessed for a configurable \texttt{CoolingPeriod} (by default, 5 minutes), it is evicted to disk.
Otherwise, an access causes it to be efficiently mapped back into the application's address space (Figure~\ref{fig:harvester-cooling}). 
In effect, \tswap is an in-memory victim cache, preventing hot pages from being swapped to disk.
Figure~\ref{fig:performance-curve} shows that \tswap can prevent performance cliffs, allowing the \harvester to avoid significant performance degradation.

%\tswap evicts cold application pages to disk only after a \texttt{CoolingPeriod}.
When the \harvester initially causes some pages to be swapped out to \tswap, it cannot determine the performance impact
% of swapping them to disk
until the \texttt{CoolingPeriod} ends.
As a result, it harvests cautiously, monitoring the application's RSS, which is available in the memory stats for its cgroup.
If it enters harvesting mode, triggering the PFRA to move some pages to \tswap, then the \harvester refrains from further decreasing the cgroup limit for at least the \texttt{CoolingPeriod}.
% Once this period has elapsed, and
Afterwards, if performance is still stable, the \harvester may resume decreasing the limit.
This ensures that the \harvester will not explore the harvesting size too aggressively without considering the performance impact caused by any disk I/O.

% \paragraph{Recovery from Performance Drop.}

% When the \harvester detects a large performance drop, it stops harvesting and releases already-harvested memory
% incrementally over a period of time. 
% This recovery period is determined using an ADMI control loop.
% During harvesting, each successive decrease to the cgroup limit without a corresponding performance drop
% implies a lower probability of entering recovery mode.
% means the application's currently available memory is enough to avoid any performance drop; as such, the chance of entering into the recovery mode is low.
% Thus, after each successful harvesting decision, the control loop adaptively decreases the recovery period by \texttt{StepSize} (by default, 30 seconds).
% To conservatively minimize the impact of potential performance losses due to harvesting, the control loop
% On the other hand, producer applications should experience minimal performance loss due to harvesting.
% To this end, being conservative, the control loop
% multiplicatively increases the recovery period by a constant (by default, 2) with the detection of any performance loss.
% \carl{Should mention multiplicative factor used by default.}

\paragraph{Handling Workload Bursts.}
Simply disabling the cgroup memory limit may not prevent performance drops in the face of sudden bursts.
\spotframework addresses this issue by prefetching previously-reclaimed pages, proactively swapping them in from disk.  
If the current performance is worse than all the recorded baseline data points for consecutive epochs, the \harvester instructs \tswap to prefetch \texttt{ChunkSize} of the most recently swapped-out pages (Figure~\ref{fig:harvester-prefetch}).
% Prefetching helps reduce performance degradation through quick, proactive recovery.
Producers with a low tolerance for performance degradation and compressible data could
alternatively use a compressed RAM disk~\cite{zram} instead of a disk-based swap device.
%, as in Figure~\ref{fig:with-zram}.
This would provide more rapid recovery, trading off total harvestable memory.

\subsection{Exposing Remote Memory to Consumers}
\label{subsec:manager}
%\mosharaf{This would be a good place to compress because many of the things we are using are standard.}

The \manager communicates with the \broker to report resource availability, and it exposes a KV interface to the \client.
The entire harvested memory space is logically partitioned into fixed-size slabs; a slab (by default, 64~MB) is the granularity at which memory is leased to consumers.
%\mosharaf{Slab and chunk are similar things but on different sides.Can be a little confusing given that both have the same size. Is there a 1-1 mapping? Probably not.}
Different slabs from the same producer can be mapped to multiple consumers for performance and load balancing.  
Upon receiving an assignment message from the \broker, the \manager can instantly create a lightweight \daemon in the producer VM, dedicated to serving remote memory for that consumer. %\mosharaf{How fast can you create one?}

In \spotframework, the \daemon is implemented by running a Redis~\cite{redis} server within a cgroup in the producer VM, providing a familiar KV cache interface to consumers.
Since an empty Redis server consumes only 3~MB of memory and 
% 0.5\% of CPU,
negligible CPU, for simplicity, the \manager runs a separate \daemon for each \client.
The producer can limit the maximum CPU used by {\daemon}s via cgroup controls.
However, producer-side CPU consumption is typically modest; YCSB on Redis uses 3.1\% of a core on average.

%\begin{figure}[!t]
%	\centering
%	\includegraphics[scale=0.6]{figure/manager}
%	\caption{\asaf{We can probably kill this figure to save space} The \manager runs a \daemon for each \client. A \client can communicate with multiple {\manager}s.}
%	\label{fig:manager}
%\end{figure}	

When the lease period expires, before terminating the Redis server, the \manager checks with the \broker to determine if the consumer wants to extend its lease (at the current market price).
%This means, if the consumer is willing to pay the current market price to renew its lease for more time, the \manager allows it to do so to ensure the cluster-wide efficiency.
Otherwise, the \daemon is terminated and its memory slabs are returned to the remote memory pool.

%~\carl{Should provide a {\em right of first refusal} for consumers who want to extend their current lease, i.e., if the consumer is willing to pay the current market price to renew its lease for more time, it should be allowed to do so, for efficiency. I did the same thing for time slices in Spawn~\cite{spawn}.}

\paragraph{Network Rate Limiter.}
The \manager limits the amount of network bandwidth used by each {\client}. 
We implemented a standard token-bucket algorithm~\cite{keshav1997engineering} to limit {\client} bandwidth.
The \manager periodically adds tokens to each \client bucket, in proportion to its allotted bandwidth specified in the \client request for remote memory.
Before serving a request, the \daemon first queries the \manager to check the {\client}'s available token count; if the I/O size exceeds the number of available tokens, it refuses to execute the request and notifies the \client.

% [carl] comment resolved by mentioning "a custom traffic shaper" in para 2 of sec 4 overview
% \carl{Need to clarify if the token-bucket rate limiter is something we implemented, or  simply using standard cgroup network bandwidth control, as mentioned briefly earlier in this section.   If it's vanilla cgroup behavior, can just add cite and cut most of this paragraph.}~\hasan{We are not using the cgroup network control here. Do you want to have more details on it?}

%~\carl{Does it return an error to the \client?}
%Therefore, the \manager notifies the respective \client and shuts down the connection between the daemon and the \client.
%Here, if the producer have more idle bandwidth to lease, the consumer can extend the token rate by paying more.
%
%~\carl{Would be simpler to implement the consumer rate limiter  using a standard token-bucket algorithm. Should also mention an extension that charges consumers more for higher rates (e.g. per-token).}

%\carl{Should clearly state that we are using Redis, add an appropriate cite, and shorten this description. Should probably also cite Psounis/Prabhakar/Engler CISS '00 paper, which introduced the k-random LRU approximation.}

\paragraph{Eviction.}
The size of a \daemon is determined by the amount of memory leased by each \client.
Once a \daemon is full, it uses the default Redis eviction policy, a probabilistic LRU approximation~\cite{k-lru}.
%The {\daemon}, therefore, removes the corresponding entry from its storage.
%If the {\client} exhibits a power law distribution in the popularity of the spot-memory requests, then LRU provides with better performance benefit.
%Here, approximation LRU reduces the memory and CPU overhead.
%One can tune the precision of the algorithm by modifying  the number of samples to check for every eviction, at the cost of additional CPU usages.  
%If the \client has a random access pattern or the producer is not willing to spend extra CPU, the \daemon employs random eviction policy.
%Here, data is evicted randomly.
%However, when the \client evicts its data mapped to a remote {\daemon}, an explicit {\tt DELETE} request is generated for the respective {\daemon}.
%The {\daemon}, therefore, removes the corresponding entry from its storage.
In case of sudden memory bursts, the \manager must release memory back to the producer rapidly.
In this scenario, the {\harvester} asks the \manager to reclaim an aggregate amount of remote memory allocated to consumers. 
The \manager then generates per-consumer eviction requests proportional to their corresponding {\daemon} sizes and employs the approximate-LRU-based eviction for each {\daemon}. %\mosharaf{How does proportional and approximate LRU work together?}
%An ordinary {\client}-side eviction sends an explicit {\tt DELETE} request to ensure the consumer- and producer-store contents remain synchronized.~\carl{This is an important point to make, but it's not clear here, since the consumer hasn't yet been explained;  consider moving or adding forward reference to Sec~\ref{sec:consumer}.}
%
%\carl{Also need to describe how {\em ordinary} eviction works  somewhere, as we discussed in Slack, i.e. consumer evicts from  consumer Redis, and sends {\tt DELETE} for locally-evicted key to corresponding producer Redis, so that the keys in both stay in sync.}
%
%\paragraph{Defragmentation}
%\carl{Again, much of this is describing vanilla Redis defragmentation  behavior, so should shorten and add an appropriate cite.}
%

\paragraph{Defragmentation.}
The size of KV pairs may be smaller than the OS page size~\cite{scalingmemcached,cliffhanger,fb2020}, which means
%Due to the buddy memory allocation technique, freed up memory may not always return to the OS. 
%Especially, when the allocated data size is smaller than the OS page size, multiple allocation may share the same page. 
that an application-level eviction will not necessarily free up the underlying OS page if other data in the same page has not been evicted.
%
%When a consumer reaches its maximum allocation, further allocation request will trigger the eviction in that consumer's spot-memory space.
%The \daemon tries to reuse the logically freed up memory to allocate the new requests.
Fortunately, Redis supports memory defragmentation, which
the \daemon uses to compact memory.

%Asaf: I am commenting out the following paragraph for now... Not sure it's critical to discuss this.
% Considering the extra CPU usage during online defragmentation, by default, the \daemon disables this feature.  
% If the producer is willing to spend extra CPU cycles, \daemon can enable this feature for certain consumers that are prone to have fragmentation issues.
% However, producer can restrict the CPU usage for online
% defragmantation operations through \texttt{MaxCPUforDefrag} knob.
% Note that lower CPU usage may take longer time to reclaim memory for the producer.
% ~\carl{Should mention how this interacts with CPU cgroup.}

\section{Broker}
\label{sec:broker}

The \spotframework \broker is a trusted third-party that facilitates transactions between producers and consumers.
It can be operated by the cloud provider, or by another company that runs the market as a service, similar to existing
caching-as-a-service providers~\cite{memcachier,redislabs}. 
Producers and consumers participate in the disaggregated market voluntarily, by registering their respective credentials with the \broker.
Each producer periodically sends its resource utilization metrics to the \broker, which uses the resulting 
historical time series to predict future remote-memory availability over requested lease periods.
Consumers request remote memory by sending allocation requests to the \broker with the desired number of slabs and lease time,
along with other preferences such as acceptable latency and bandwidth.
The \broker connects producers and consumers that may reside in separate virtual private clouds (VPCs)
via existing high-bandwidth peering services~\cite{vpc-peering,vnet-peering}.
The \broker maps consumer requests to producer slabs using an assignment algorithm that satisfies consumer preferences, while minimizing producer overhead and ensuring system-wide wellness objectives (\eg, load balancing and utilization).
%as well as fair access to and fair eviction from spot memory.

Our current design runs the \broker on a single node and can handle a market with thousands of
% [carl] I thought "consumer-producer VMs" looked odd, so replaced with "participating VMs"
% {\client}-{\producer}
participating VMs (\S\ref{subsec:eval-broker}).
Since consumers communicate directly with assigned producers until their leases expire,
even if the \broker is temporarily unavailable, the system can still continue to operate normally,
except for the allocation of new remote memory.
For higher availability, the \broker state could be replicated using distributed consensus, \eg, leveraging
Raft~\cite{raft,etcd} or ZooKeeper~\cite{zookeeper}.
% We assume that
The \spotframework operator may also run several \broker instances, each
serving a disjoint set of consumers and producers (\eg, one broker per region or datacenter).

\subsection{Availability Predictor}
\label{subsec:arima}
Remote memory is transient by nature and can be evicted at any time to protect the performance of producer applications.
Hence, allocating remote memory without considering its availability may result in frequent evictions that degrade consumer performance.
Fortunately, application memory usage often follows a predictable long-term pattern, such as exhibiting diurnal fluctuations~\cite{cortez2017resource}. 
The \broker capitalizes on historical time series data for producer memory consumption, predicting the availability of offered remote memory using an Auto Regressive Integrated Moving Average (ARIMA) model~\cite{arima}.  
{\Producer}s with completely unpredictable usage patterns are not suitable for \spotframework.
ARIMA model parameters are tuned daily via a grid search over a hyperparameter space to minimize the mean squared error of the prediction.

%The \broker periodically communicates with the {\manager}s to collect the producer VM's maximum resource usage (\eg, memory, bandwidth and CPU consumption)
%over an hour.
%To do so, the \broker uses an Auto Regressive Integrated Moving Average (ARIMA)~\cite{arima} model with orders $p,q,d$ (\texttt{ARIMA(p,q,d)}) on top of each producer's time series data of memory usage.
% To do so, the \broker uses an Auto Regressive Integrated Moving Average (ARIMA)~\cite{arima} model on the time series data of memory usage of each producer.
% Here, $p$ is the order of Auto Regression (AR) that refers to the number of lags of the time series data to be used as predictors. 
% $q$ is the order of Moving Average (MA), \ie, the number of lagged forecast errors.
% $d$ is the minimum number of differencing needed to make the time series stationary.
% For $d=1$, the model for a time series $X_t$ is as follow:
% \begin{multline*}
% 	\label{eqn:arima}
% 	X_t = \alpha + \beta_1X_{t-1} + \beta_2X_{t-2} + \cdots + \beta_pX_{t-p} \\ + \epsilon_t + \phi_1\epsilon_{t-1} + \phi_2\epsilon_{t-2} + \cdots + \phi_q\epsilon_{t-q}
% \end{multline*}
% 
% \carl{It's not clear to me that this level of detail is needed here, since
%   ARIMA is a standard modeling approach; the cite and a brief explanation
%   seems sufficient.}
% 
% Here, $\beta$ and $\phi$ are the parameters/coefficients of AR and MA terms with order p and q, respectively. 
% $\epsilon_t$ are the errors of the autoregressive models of the respective lags - 
% usually a white noise with intensity $\sigma^{2}$.

\subsection{Remote Memory Allocation}

%The \broker searches for the best producer mappings that satisfy consumer requirements while also minimizing producer overhead.

\paragraph{Constraints and Assumptions.}
While matching a consumer's remote memory request, the \broker tries to achieve the aforementioned goals under the following assumptions:

\begin{denseitemize}
	\item[1.] \textit{Online requests:} Consumers submit remote memory requests in an online manner.
	During a placement decision, new or pending requests may be queued.
	
	\item[2.] \textit{Uncertain availability:} It is unknown exactly how long producer remote memory slabs will remain available.
	
	\item[3.] \textit{Partial allocation:} 
	The \broker may allocate fewer slabs than requested, as long as it satisfies the minimum amount specified by the consumer.
          %\carl{Seems better to let the consumer specify both minimum and desired slabs; please update if still accurate.}
	
%     \item[4.] \textit{Dynamic pricing:} Different producer may have different price schemes for the same amount of spot memory.
%     A producer's price may vary over time based on the amount and span of lease.
%     Once an allocation is made, producer can't change the price for that allocation.
\end{denseitemize} 

\paragraph{Placement Algorithm.}
When the \broker receives an allocation request from a consumer, it checks whether at least one producer is expected to have at least one slab available for the entire lease duration (\S\ref{subsec:arima}),
at a price that would not exceed the consumer budget (\S\ref{subsec:pricing}).
The \broker calculates the placement cost of the requested slabs based on the current state of all potential producers with availability,
as a weighted sum of the following metrics: number of slabs available at a given producer, predicted availability (based on ARIMA modeling), available bandwidth and CPU, network latency between the consumer and producer, and producer reputation (fraction of remote memory not prematurely evicted during past lease periods).
A consumer may optionally specify weights for each of these placement desirability metrics with its request.%~\carl{The term ``cost'' here is potentially confusing given our use of monetary prices.  Consider something like ``placement desirability'' instead?}

The \broker selects the producer with the lowest placement cost, greedily assigning the desired number of slabs.  
If the producer cannot allocate the entire request, the \broker selects the producer with the next-lowest cost and continues iteratively until there are no slabs left to allocate or no available producers.  
When fewer than the requested number are allocated, a request for the remaining slabs is appended to a queue.  
Pending requests are serviced in FIFO order until they are satisfied, or they are discarded after a specified timeout.

\subsection{Remote Memory Pricing}
\label{subsec:pricing}
Remote memory must be offered at a price that is attractive to both producers and consumers, providing incentives to participate in the disaggregated memory market.  
Considering the transient nature of harvested memory, any monetary incentive for leasing otherwise-wasted resources
is beneficial to a producer, provided its own application-level performance is not impacted.
This incentive can help the producer defray the expense of running its VM.
A consumer must weigh the monetary cost of leasing remote memory against the cost of running a static or spot instance with larger memory capacity.

% In this regards, the \broker can continuously monitor the price of different instance types in the same public cloud.
% Specifically, for a given lease, the \broker can set the per-unit remote memory price to a value below that associated with the lowest available instance price at that time. 
% Of course, this simple pricing model can be replaced with a customized pricing model designed to address different economic objectives.
The \broker sets a price for leasing a unit of remote memory (GB/hour) and makes it visible to all {\client}s.%\mosharaf{This is using another size: 1GB, different than slab or chunk sizes; I think we want to say price in GB/hr unit but you don't need to get in 1 GB incremenets.}
Various economic objectives could be optimized (\eg, total trading volume, total revenue of {\producer}s, \etc)
% Since the primary {\spotframework} goal is improving cluster-wide utilization,
% the \broker focuses on maximizing the total trading volume that can be achieved by
% setting a market-clearing price for remote memory.
We assume the \broker optimizes the total revenue of {\producer}s by default, since this strategy maximizes the {\broker}'s cut of the revenue.

From a {\client}'s perspective, an alternative to \spotframework is running a separate spot instance and
consuming its memory remotely~\cite{spot-cache}.  
Thus, to be economically viable to {\client}s, the price of remote memory
in \spotframework should never exceed the corresponding spot instance price.
%This provides a useful upper bound on the reasonable price that can be charged for remote memory; higher prices would present an arbitrage opportunity, making it profitable to rent spot instances and resell their memory capacity as remote memory.%\footnote{This is analogous to breaking up a company when the sum of its parts is more valuable than the whole; instance CPU  time could be resold as well.}
%
%Of course, in real-world scenarios exploring every possible prices to optimize the pricing objective is prohibitive, since it introduces too much variance to the market price.
For simplicity, the \broker initially sets the price for each unit of remote memory to
one quarter of the current market price for a spot instance, normalized by its size.
%\mosharaf{Why one quarter?}
This initial lower price makes remote memory attractive to {\client}s.
Afterwards, the price is adjusted to approximate the maximal total \producer revenue by searching for a better price locally.
%
% Whenever the total supply of memory harvested from {\producer}s differs from the aggregate {\client} demand by
% more than a specified threshold (by default, 10\%), the \broker adjusts the current market price
% by a predefined step size (by default, 0.001 cent/GB$\cdot$hour) until the market reaches approximate equilibrium. %\mosharaf{Probably give forward pointer to where we evaluate this or what we find.}
In each iteration, the \broker considers the current market price $p$, $p + \Delta p$, and $p - \Delta p$ as the candidates for the price in the next iteration, where $\Delta p$ is the step size (by default, 0.002 cent/GB$\cdot$hour).
Then the \broker chooses the one that generates the maximal total revenue for {\producer}s.
Our pricing model yields good performance in real-world traces (\S\ref{subsec:eval-pricing}).
% Of course, this price-adjustment mechanism can be replaced with alternative pricing models designed to
% achieve different economic objectives.
Of course, alternative price-adjustment mechanisms can be designed to achieve different economic objectives.
%What would be the best way to figure out the optimal price and budget for computational resources in a competitive spot market is a separate research question and beyond the scope of this work. 
%In \S~\ref{sec:price-model}, we discuss possible pricing models for spot memory that can benefit both consumers and producers.

\section{Consumer}
\label{sec:consumer}

% The \client enables remote memory consumption.
A \client uses remote memory.
It first sends its remote memory demand to the \broker, based on the current market price given by the \broker and its expected performance benefit.
After receiving an assignment from the \broker, the \client communicates with {\producer}s directly during the lease period.
To ensure the confidentiality and integrity of its remote data, the \client employs standard cryptographic methods
during this communication. 
Rate-limiting techniques are used to protect {\producer}s from misbehaving or malicious consumers.

The \client can use either a KV cache or a swap interface, which we built on top of
Redis~\cite{redis} and Infiniswap~\cite{infiniswap} clients, respectively. By default, \spotframework uses the key-value interface, because in contrast
to swapping to disk, applications
using a KV cache naturally assume cached data can disappear. We have also found
the KV interface performs better than the swap interface (\S\ref{subsec:eval-end2end}), due
to the added overhead of going through the block layer when swapping.  
For the sake of brevity, we focus our description on the KV interface.

\subsection{Confidentiality and Integrity}
\label{subsec:security-design}
This section explains how the \client ensures data confidentiality and integrity during its KV operations.
The subscripts $C$ and $P$ are used to denote consumer-visible and producer-visible data, respectively.

\paragraph{PUT Operations.} 
To perform a {\tt{}PUT}, the consumer prepares a KV pair ($K_C$, $V_C$) to be stored at a remote producer.
%\footnote{During swapping, $K_C$ indicates {\tt{}swap\_index} and $V_C$ indicates page data.} 
First, the value $V_C$ is encrypted using the consumer's secret key and a fresh, randomly-generated initialization vector ($IV$).  
The $IV$ is prepended to the resulting ciphertext, yielding the value $V_P$ to be stored at the producer. 
Next, a secure hash $H$ is generated for $V_P$, to verify its integrity and defend against accidental or malicious corruption by the producer.

To avoid exposing the lookup key $K_C$, the consumer substitutes a different key $K_P$.  
Since $K_P$ need only be unique, it can be generated efficiently by simply incrementing a counter for each new key stored at a producer.
The \daemon storing the KV pair can be identified using an index $P_i$ into a small table containing producer information.

The consumer stores the metadata tuple $M_C$ = ($K_P$, $H$, $P_i$) locally, associating it with $K_C$. 
While many implementations are possible, this can be accomplished conveniently by adding ($K_C$, $M_C$) to a local KV store, where an entry serves as a proxy for obtaining the corresponding original value.
Significantly, this approach also enables range queries, as all original keys are local.
% [carl] moved footnote into main text, since locally have plenty of space on page
% \footnote{This approach also enables range queries, as all original keys are local.}

\paragraph{GET Operations.}
To perform a {\tt{}GET}, the consumer first performs a local lookup using $K_C$ to retrieve its associated metadata $M_C$, and sends a request to the producer using substitute key $K_P$.  
The consumer verifies that the value $V_P$ returned by the producer has the correct hash $H$; if verification fails, the corrupted value is discarded.  
The value $V_P$ is then decrypted using $IV$ with the consumer's encryption key, yielding $V_C$.

\paragraph{DELETE Operations.}
To perform a {\client}-side eviction, the {\client} first removes the metadata tuple $M_C$ from its local store. 
It then sends an explicit {\tt DELETE} request to the respective {\daemon} so that the {\client} and {\daemon} contents remain synchronized.

\paragraph{Metadata Overhead.}
In our current prototype, each consumer uses a single secret key to encrypt all values.  
Encryption uses AES-128 in CBC mode, and hashing uses SHA-256, both standard constructions. 
By default, the integrity hash is truncated to 128 bits to save space.  
A 64-bit counter is employed to generate compact producer lookup keys. 
%Since the prototype uses a separate local key-value store for each producer, no per-entry producer index is needed.~\carl{How do we avoid needing to query {\em each} local store to find $K_C$, e.g. by sharding based on $K_C$? Should we explain this here?}
The resulting space overhead for the metadata $M_C$ corresponding to a single KV pair ($K_C$, $V_C$) is 24 bytes; the $IV$ consumes an additional 16 bytes at the producer.
%~\carl{As we've discussed,  the experimental evaluation should clearly account for the per-KV-pair memory overhead,  the per-Redis-instance overhead, Redis memory fragmentation, etc.}

For applications where consumer data is not sensitive, value encryption and key substitution are unnecessary.
Such an integrity-only mode requires only the integrity hash, reducing the metadata overhead to 16 bytes.

%\asaf{We could consider removing the availability section if we need to save space}
%
%\subsection{Availability}
%
%Another security concern is protecting the producer from malicious consumers that may try to mount Distributed Denial of Service (DDoS) attacks.  
%The major public clouds provide layer 3 and 4 DDoS protection for free~\cite{aws-shield,azure-ddos,google-ddos}. 
%For application-layer protection, the \producer applies a rate limit on the
%amount of bandwidth and number of requests processed for each consumer
%(\S\ref{subsec:manager}).  
%Once a \client exceeds its limit, the \manager will start dropping its requests.

\subsection{Purchasing Strategy}
A \client must determine a cost-effective amount of memory to lease
to meet its application-level performance goals.
In general, it may be difficult to estimate the monetary value of additional memory.
However, when its application is a cache, lightweight sampling-based techniques~\cite{SHARDS,kinetic,cachesims} 
%such as
%SHARDS~\cite{SHARDS}, AET~\cite{kinetic}, or miniature
%simulation~\cite{cachesims} 
can estimate miss ratio curves (MRCs) accurately, yielding the expected performance benefit from a larger cache size.

The \client estimates the value of additional cache space using the current {\em price-per-hit} from the known cost of running its VM, and its observed hit rate.
The expected increase in hits is computed from its MRC, and valued based on the per-hit price.
When remote memory is more valuable to the \client than its cost at the current market price, it should be leased, yielding an economic \client surplus.

\section{Evaluation}
\label{sec:eval}
We evaluate {\spotframework} on %a \todo{10-machine}\mosharaf{N-VM} 
a CloudLab~\cite{cloudlab} cluster using both synthetic and real-world cluster traces.\footnote{\spotframework can be readily deployed on any major cloud provider. We run our evaluation in CloudLab since it is free.}
Our evaluation addresses the following questions:
\begin{denseitemize}
	\item How effectively can memory be harvested? (\S\ref{subsec:eval-harvester})
	
	\item How well does the \broker assign remote memory? (\S\ref{subsec:eval-broker})
	
  	\item What are {\spotframework}'s end-to-end benefits? (\S\ref{subsec:eval-end2end})  
  	
        \item How does pricing % strategy
              affect utility and utilization? (\S\ref{subsec:eval-pricing})
\end{denseitemize}

% \paragraph{Methodology.}
\paragraph{Experimental Setup.}
Unless otherwise specified, we configure \spotframework as follows. 
The \producer averages application-level latency over each second as its performance metric.
We generate both the baseline performance distribution and the recent performance distribution from data points over the previous 6 hours (\texttt{WindowSize}). 
If the recent p99 drops below the baseline p99 by more than 1\% (\texttt{P99Threshold}), it is considered a performance drop.
Harvesting uses a 64~MB {\tt ChunkSize} and a 5-minute \texttt{CoolingPeriod}.
If a severe performance drop occurs for 3 consecutive epochs, \tswap prefetches % 64~MB
{\tt ChunkSize} from disk.

Each physical server is configured with 192~GB DRAM, % 2$\times$
two Intel Xeon Silver 4114 processors with 20 cores (40 hyperthreads), and a 10Gb NIC. We use Intel DC S3520 SSDs and 7200 RPM SAS HDDs.
We run the Xen hypervisor (v4.9.2) with Ubuntu 18.04 (kernel v4.15) as the guest OS.
%~\carl{  Should also list Xen version, plus SSD and HDD details.}

\begin{table}[!t]
\centering
\small
\begin{tabular}{l|r|r|r|r}
& \vtop{\hbox{\strut Total}\hbox{\strut Harvested}} & \vtop{\hbox{\strut Idle}\hbox{\strut Harvested}} & \vtop{\hbox{\strut Workload}\hbox{\strut Harvested}} & \vtop{\hbox{\strut Perf}\hbox{\strut Loss}}\\
\hline
\hline
Redis & ~~3.8 GB & 23.7\% & 17.4\% & 0.0\%\\
\hline 
memcached & ~~8.0 GB & 51.4\% & 14.6\% & 1.1\%\\
\hline
MySQL & ~~4.2 GB & 21.7\% & ~~7.0\% & 1.6\%\\
\hline
XGBoost & 18.3 GB & 15.4\% & 17.8\% & 0.3\%\\
\hline
Storm & ~~3.8 GB & ~~1.1\% & ~~1.4\% & 0.0\%\\
\hline
CloudSuite & ~~3.6 GB & ~~2.5\% & 15.3\% & 0.0\%\\
\hline
\end{tabular}
\caption{Total memory harvested (idle and unallocated), the percentage of memory harvested that was idle, the percentage of application-allocated memory that was harvested, and the performance loss of different workloads.}
\label{table:harvester}
\end{table}

\paragraph{Workloads.}
Consumers run YCSB~\cite{cooper2010benchmarking} on Redis~\cite{redis}.
Producers run the following applications and workloads:
\begin{denseitemize}
	\item \textbf{Redis} running a Zipfian workload using a Zipfian constant of 0.7 with 95\% reads and 5\% updates.
	\item \textbf{memcached} and \textbf{MySQL} running MemCachier~\cite{memcachier,dynacache} for 36 hours and 40 hours, respectively.
        We use 70 million {\tt SET} operations to populate the memcached server, followed by 677 million queries for memcached and 135 million queries for MySQL.	
	\item \textbf{XGBoost}~\cite{xgboost} training an image classification model on images of cats and dogs \cite{cats-vs-dogs} using CPU, with 500 steps.
	\item \textbf{Storm}~\cite{storm} running the Yahoo streaming workload~\cite{yahoo-streaming-benchmark} for 1.5 hours.
	\item \textbf{CloudSuite}~\cite{cloudsuite} executing a web-serving benchmark with memcached as the cache and MySQL as the database, with 1000 users and 200 threads.

\end{denseitemize}

\begin{figure}[!t]
	\centering
	% \subfloat[][Zipfian on Redis]{
	% 	\label{fig:eval-memcachier-rocksdb}
	% 	\includegraphics[width=0.33\textwidth]{figure/eval-redis-mc.pdf}%
	% }
	\subfloat[][\textbf{MemCachier on memcached}]{
		\label{fig:eval-memcachier}
		\includegraphics[width=0.5\columnwidth]{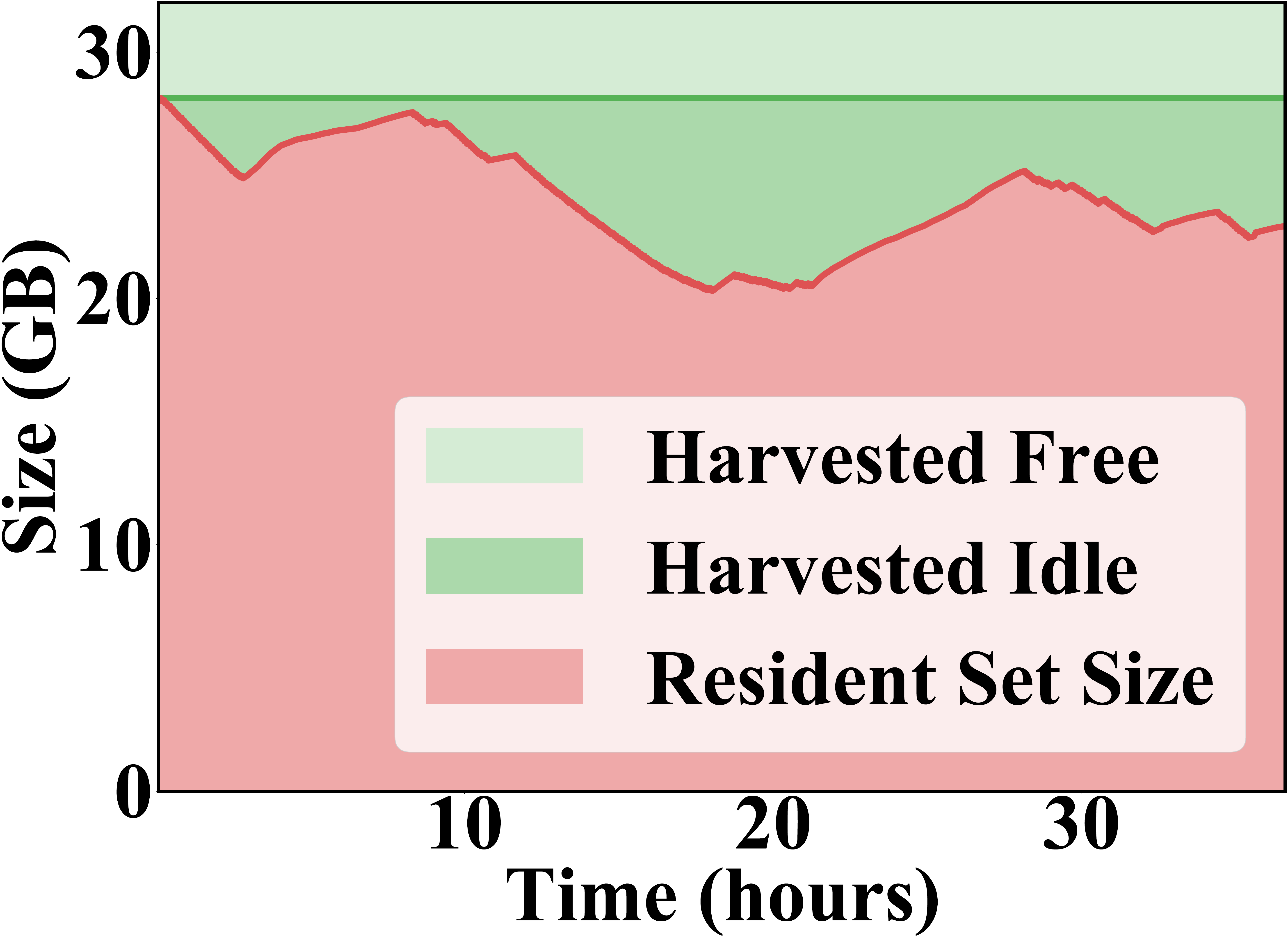}%
	}
	% \subfloat[][MemCachier on MySQL]{
	% 	\label{fig:eval-snowset}
	% 	\includegraphics[width=0.33\textwidth]{figure/eval-mysql-mc.pdf}%
	% }\\[-1em]
	\subfloat[][\textbf{Training classifier on XGBoost}]{
		\label{fig:eval-tf}
		\includegraphics[width=0.5\columnwidth]{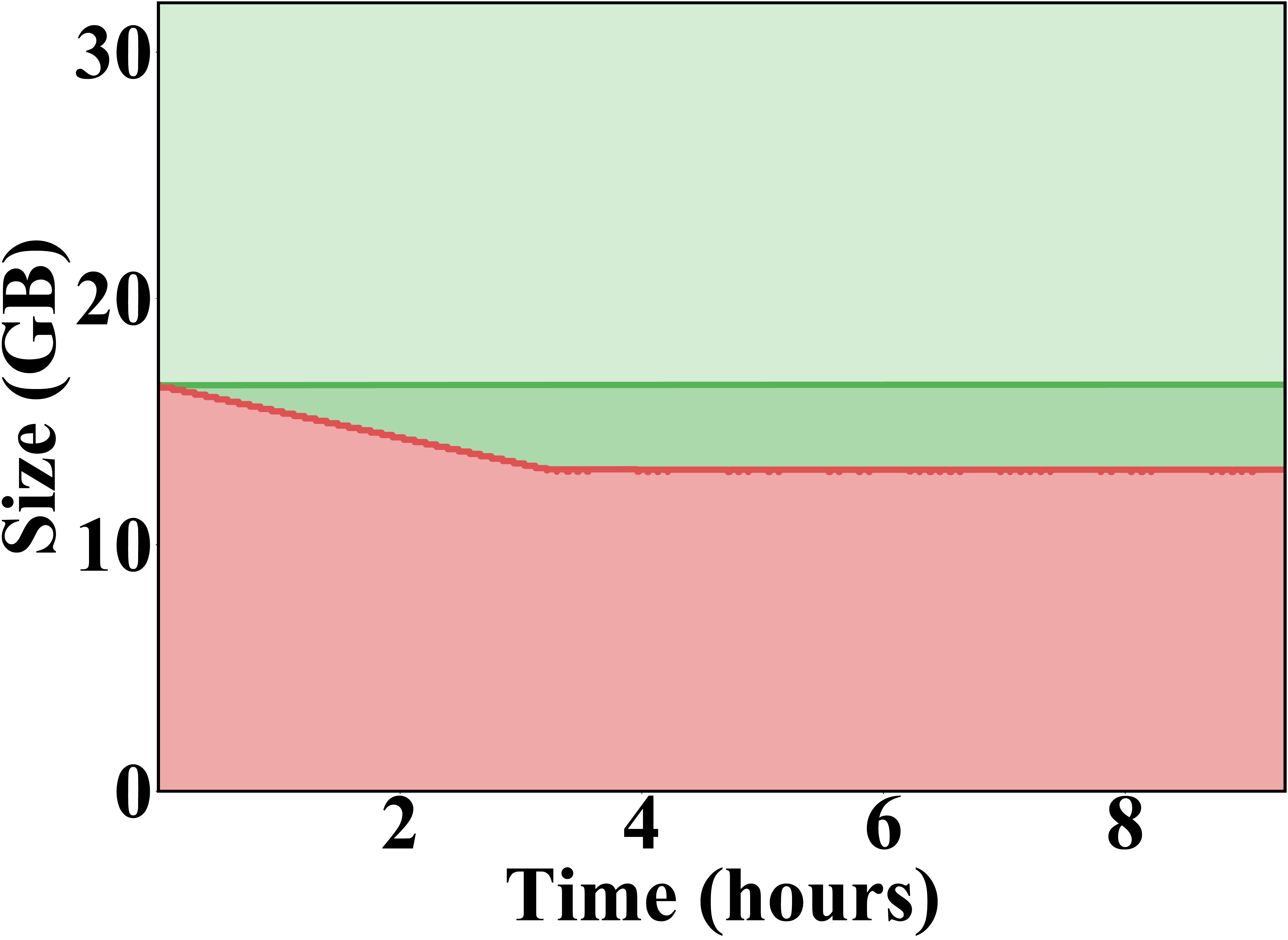}%
	}
	% \subfloat[][Yahoo Streaming on Apache Storm]{
	% 	\label{fig:eval-storm}
	% 	\includegraphics[width=0.33\textwidth]{figure/eval-storm-mc.pdf}%
	% }
	% \subfloat[][CloudSuite Web Serving]{
	% 	\label{fig:eval-cloudsuite}
	% 	\includegraphics[width=0.33\textwidth]{figure/eval-cloudsuite-mc.pdf}%
	% }
	\caption{VM memory composition over time. Additional results are provided in the appendix as Figure~\ref{fig:apdx-harvester-over-time} for the interested.}  %Unallocated represents the part of memory not allocated to the application; harvested means the portion of application's memory which has been swapped to disk; \tswap denotes the part of memory used by \tswap to buffer reclaimed pages; RSS consists of application's anonymous pages, mapped files, and page cache that are collected from the cgroup's stats file. }
	\label{fig:harvester-over-time}
\end{figure}

\paragraph{VM Rightsizing.}
To detemine the VM size for each workload, we find the AWS instance type~\cite{ec2-instance-type} with the minimal number of cores and memory that can fit the workload without affecting its baseline performance. 
We use configurations of M5n.Large (2 vCPU, 8~GB RAM) for Redis, M5n.2xLarge (8 vCPU, 32~GB RAM) for memcached and XGBoost, C6g.2xLarge (8 vCPU, 16~GB RAM) for MySQL, C6g.xLarge (4 vCPU, 8~GB RAM) for Storm, C6g.Large (2 vCPU, 4~GB RAM) for CloudSuite, and T2.xLarge (4 vCPU, 16~GB RAM) for \client YCSB.

\subsection{Harvester}
\label{subsec:eval-harvester}

\paragraph{Effectiveness.}
%Table~\ref{table:harvester} summarizes the amount of memory the \producer can reclaim and the resulting performance degradation.
To observe the effectiveness of the {\harvester}, we run the workloads with their respective producer configurations.
For Redis, memcached, and MySQL we use average latency to measure performance.
Since XGBoost, Storm, and Cloudsuite do not provide any real-time performance metric, we use the promotion rate (number of swapped-in pages) as a proxy for performance.

We find that \spotframework can harvest significant amounts of memory, even from right-sized VMs (Table~\ref{table:harvester}). 
Here, a notable portion of the total harvested memory is extracted from the application's idle memory (on average, 1.1--51.4\% across the entire workload) at a lower performance  degradation cost of 0--1.6\%. 
%The third column shows the percentage of idle memory that is harvested by \spotframework, averaged across the entire workload. Performance degradation is relatively low -- less than 5\% for applications. 
Also, a whole-machine, all-core analysis shows that the producer-side CPU and memory overheads due to the harvester were always less than 1\%. 

Figure~\ref{fig:harvester-over-time} plots memory allocation over time for two representative workloads,
%In the figure, unallocated represents the part of memory not allocated to the application; harvested means the portion of application's memory which has been swapped to disk; \tswap denotes the part of memory used by \tswap to buffer reclaimed pages; RSS consists of application's anonymous pages, mapped files, and page cache that are collected from the cgroup's stats file.
and shows that for workloads such as MemCachier with varying access patterns (Figure~\ref{fig:eval-memcachier}),
the \harvester dynamically adjusts the amount of harvested memory.
For most workloads, the percentage of idle harvested memory is higher at the end of the run. Therefore, we expect that if we ran our workloads longer, the average percentage of idle harvested memory would only increase.

\begin{figure}[!t]
	\centering
%	\subfloat[][Swap w/o prefetch]{
%		\label{fig:with-disk}
%		\includegraphics[width=0.33\columnwidth]{figure/with-disk.pdf}%
%	}
%	\subfloat[][Swap w/ prefetch]{
%		\label{fig:with-prefetch}
%		\includegraphics[width=0.33\columnwidth]{figure/with-prefetch.pdf}%
%	}
%	\subfloat[][Compressed zram]{
%		\label{fig:with-zram}
%		\includegraphics[width=0.33\columnwidth]{figure/with-zram.pdf}%
%	}
	\subfloat[][\textbf{SSD Swap Device}]{
		\label{fig:bursty-zipfian-hdd}
		\includegraphics[width=0.5\columnwidth]{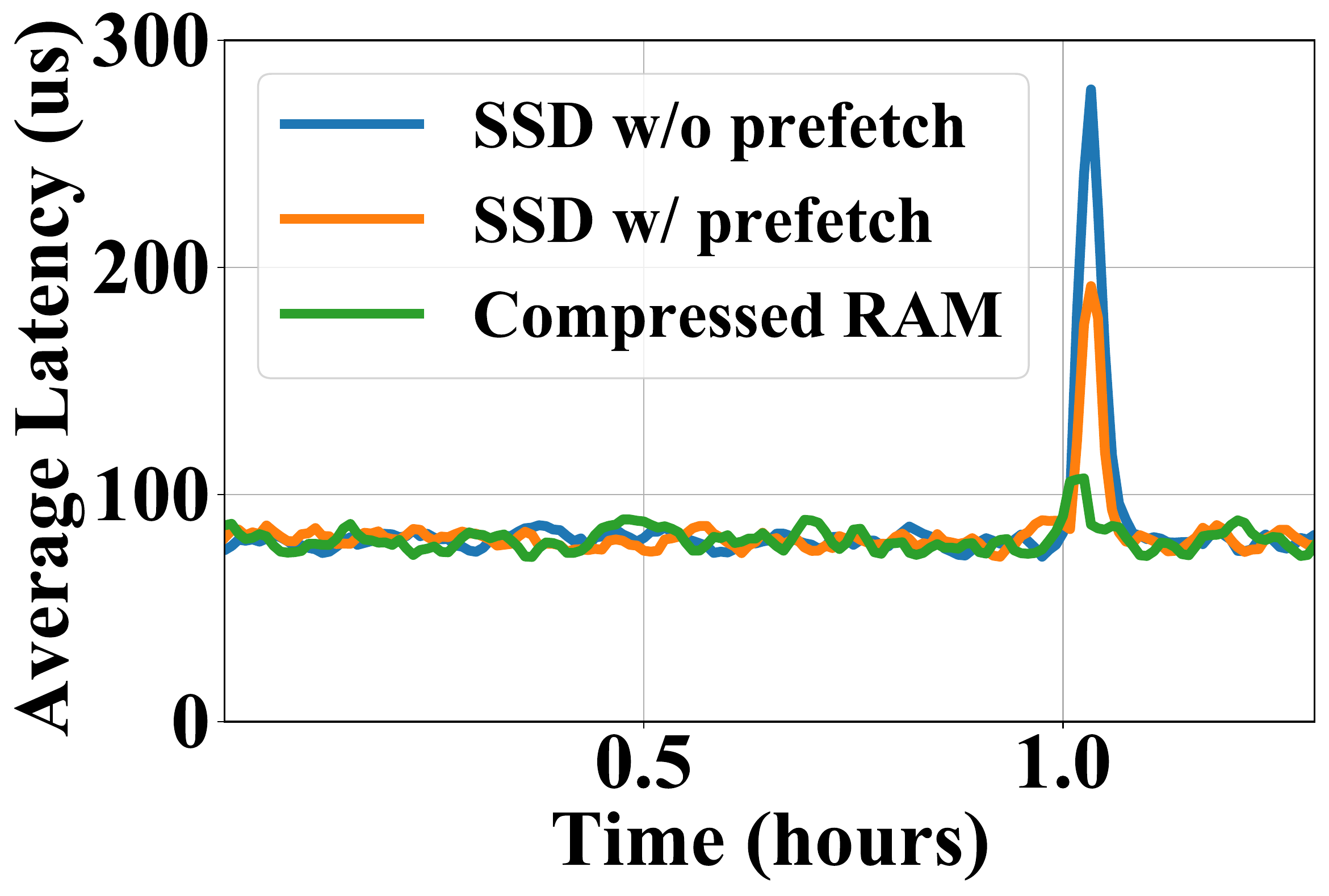}
	}
	\subfloat[][\textbf{HDD Swap Device}]{
		\label{fig:bursty-zipfian-ssd}
		\includegraphics[width=0.5\columnwidth]{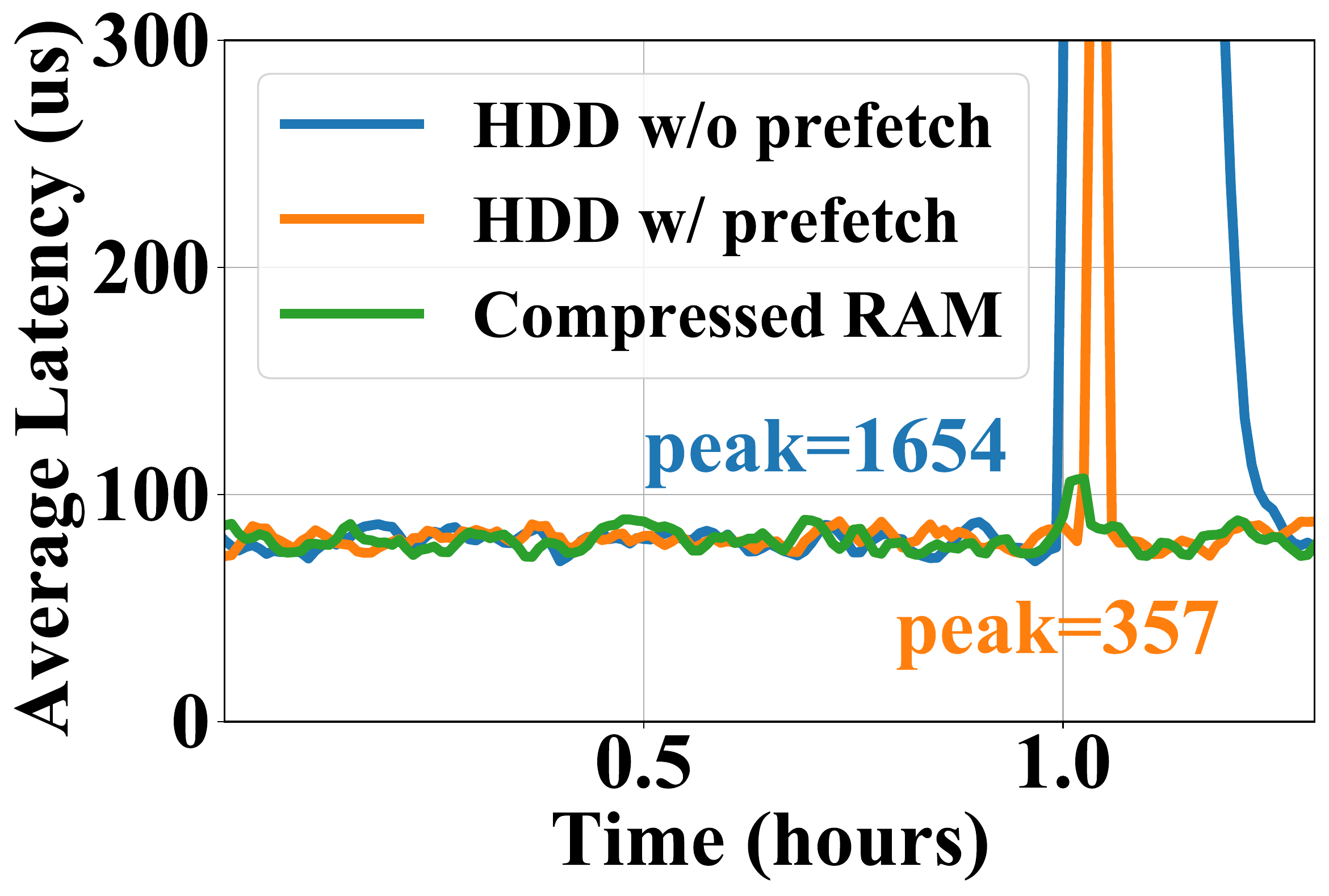}
	}
	% \caption{The control group's memory limit, resident set size and average latency over time of a workload on Redis which shifts from a zipfian distribution to a uniform distribution at the end.}
	%\caption{Memory limit, resident set size, and average latency for a Redis workload distribution shift from Zipfian to uniform after 1 hour.
	\caption{Prefetching enables faster recovery and better latency during workload bursts.
          Using memory compression enables even faster recovery,
          trading off the total amount of harvestable memory.
          }
	\label{fig:bursty-zipfian}
\end{figure}

\begin{figure}[!t]
	\centering
	\subfloat[][\textbf{\tswap \texttt{CoolingPeriod}}]{
		\label{fig:cooling-period}
		\includegraphics[width=0.5\columnwidth]{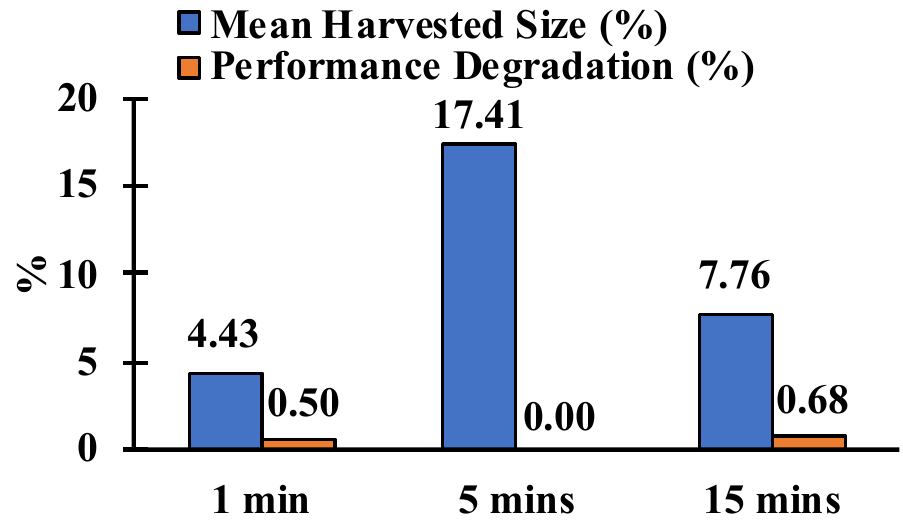}%
	}
	\subfloat[][\textbf{Harvesting \texttt{ChunkSize}}]{
		\label{fig:chunk-size}
		\includegraphics[width=0.5\columnwidth]{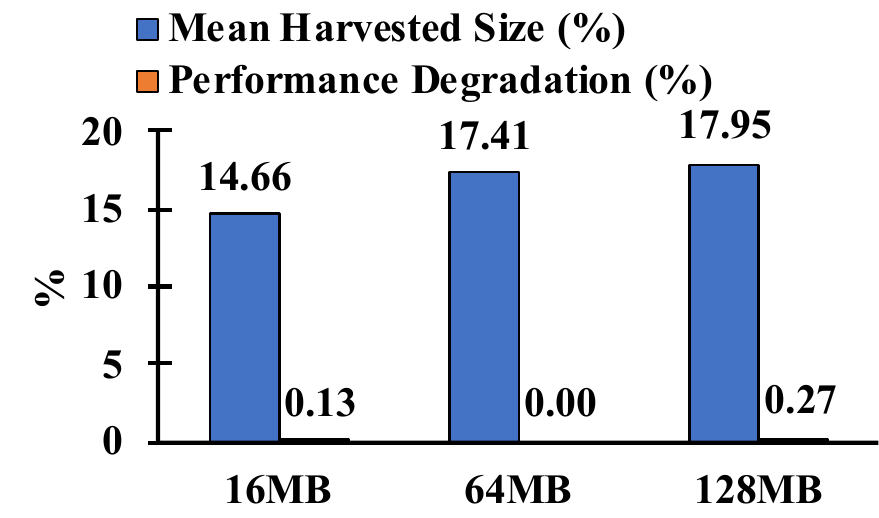}%
	}\\
	\subfloat[][\textbf{\texttt{P99Threshold}}]{
		\label{fig:p99-threshold}
		\includegraphics[width=0.5\columnwidth]{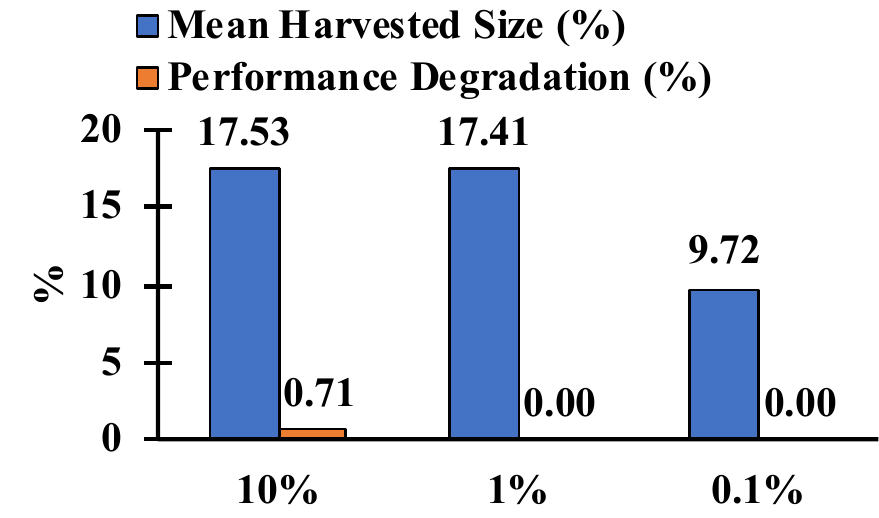}%
	}
	\subfloat[][\textbf{Perf-monitoring \texttt{WindowSize}}]{
		\label{fig:window-size}
		\includegraphics[width=0.5\columnwidth]{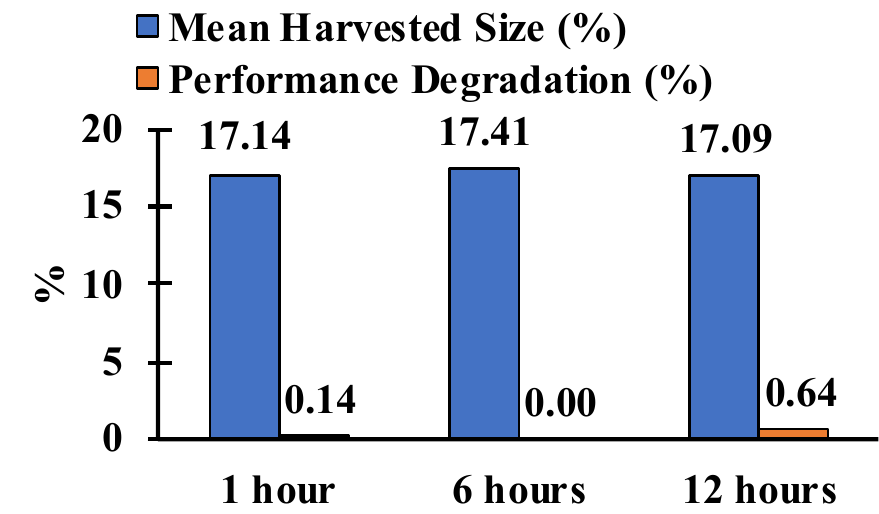}%
	}
	\caption{\textbf{Sensitivity analysis for the \harvester.} Single-machine experiments using Redis with YCSB Zipfian constant 0.7. %(a) Varying \tswap \texttt{CoolingPeriod} (1 min, 5 mins, 30 mins). (b) Varying harvesting \texttt{ChunkSize} (16MB, 64MB, 256MB). (c) Varying \texttt{P99Threshold} (10\%, 1\%, 0.1\%). (d) Varying performance-monitoring \texttt{WindowSize} (1, 6, 12 hours). 
	}
  \label{fig:harvester-sensitivity}
\end{figure}

\paragraph{Impact of Burst-Mitigation Techniques.}

To observe the effectiveness of the \harvester during workload bursts, we run YCSB on Redis using a Zipfian distribution (with constant 0.7).
To create a workload burst, we abruptly shift it to a uniform distribution after one hour of the run. 
Figure~\ref{fig:bursty-zipfian} shows the average latency using different burst-mitigation approaches.
% Without any prefetching, the performance of Redis plummets and takes a long time to recover.
When enabled, \tswap prefetches harvested pages back into memory, which helps the application reduce its recovery time by $22.5\%$ and $94.4\%$ over SSD and HDD, respectively.
A compressed RAM disk exhibits minimal performance degradation period during the workload burst ($68.7\%$ and $10.9\%$ less recovery time over prefetching from SSD and HDD, respectively), at the cost of less harvested memory.

\paragraph{Sensitivity Analysis.}
We run YCSB over Redis with a Zipfian constant of 0.7 to understand the effect of each parameter on harvesting and producer performance.
Each parameter is evaluated in an isolated experiment.
Figure~\ref{fig:harvester-sensitivity} reports the experimental sensitivity results,
using average performance to quantify degradation.

The \tswap \texttt{CoolingPeriod} controls the aggressiveness of harvesting (Figure~\ref{fig:cooling-period}). 
Setting it too high leads to less harvesting, while setting it too low causes performance drops that eventually also leads to less harvested memory.

Both the harvesting \texttt{ChunkSize} (Figure~\ref{fig:chunk-size}) and the \texttt{P99Threshold} (Figure~\ref{fig:p99-threshold}) affect harvesting aggressiveness in a less pronounced way. The amount of harvested memory increases with more aggressive parameters, while the performance impact always remains below 1\%. The performance-monitoring \texttt{WindowSize} does not significantly change either the harvested memory or performance (Figure~\ref{fig:window-size}).

%\begin{figure}[!t]
%	\centering
%	\includegraphics[scale=0.6]{analysis/security-overhead}
%	\caption{Overhead of {\spotframework} for different security modes.}
%	\label{fig:security-overhead}
%\end{figure}

\begin{figure}[!t]
	\centering
%	\subfloat[][\textbf{Cluster-wide Requests}]{
%	\label{fig:simulation-distribution}
%		\includegraphics[width=0.5\columnwidth]{analysis/simulation-memory-distribution}
%	}
	\subfloat[][\textbf{Placement}]{
	\label{fig:simulation-placement}
		\includegraphics[width=0.5\columnwidth]{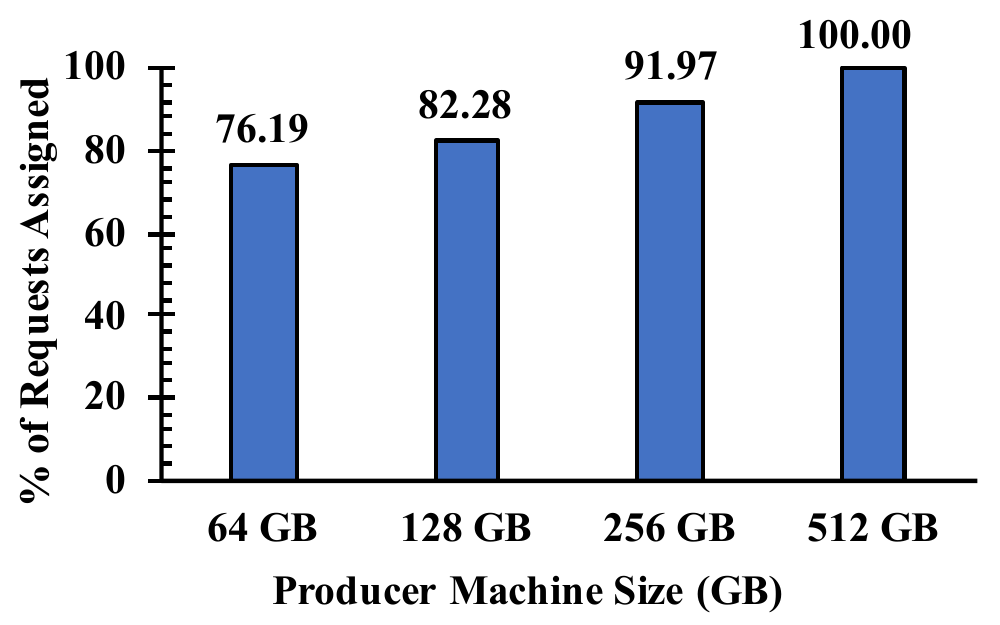}
	}
	\subfloat[][\textbf{Cluster-wide Utilization}]{
	\label{fig:simulation-utilization}
		\includegraphics[width=0.5\columnwidth]{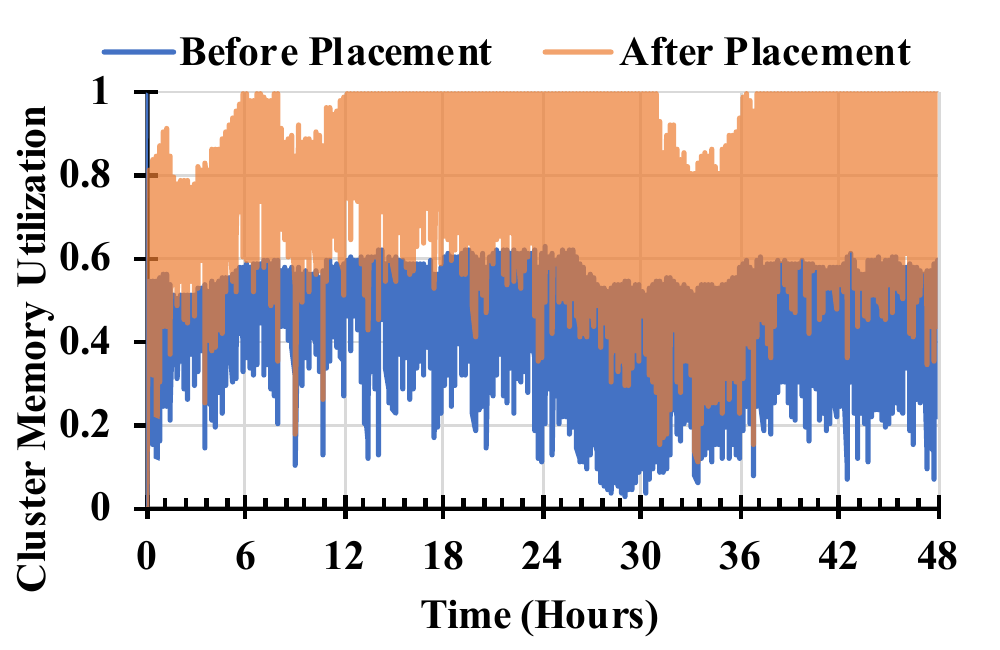}
	}
%	\subfloat[][\textbf{Memory Usage Prediction}]{
%	\label{fig:simulation-arima}
%		\includegraphics[width=0.5\columnwidth]{analysis/arima-prediction}
%	}
	\caption{Simulation of remote memory usage shows the {\broker} allocates most requests and improves cluster-wide utilization.}
\end{figure}

\begin{figure}[!t]
	\centering
	\subfloat[][\textbf{Average Latency}]{
	\label{fig:dalmatian-benefit-avg}
		\includegraphics[scale=0.46]{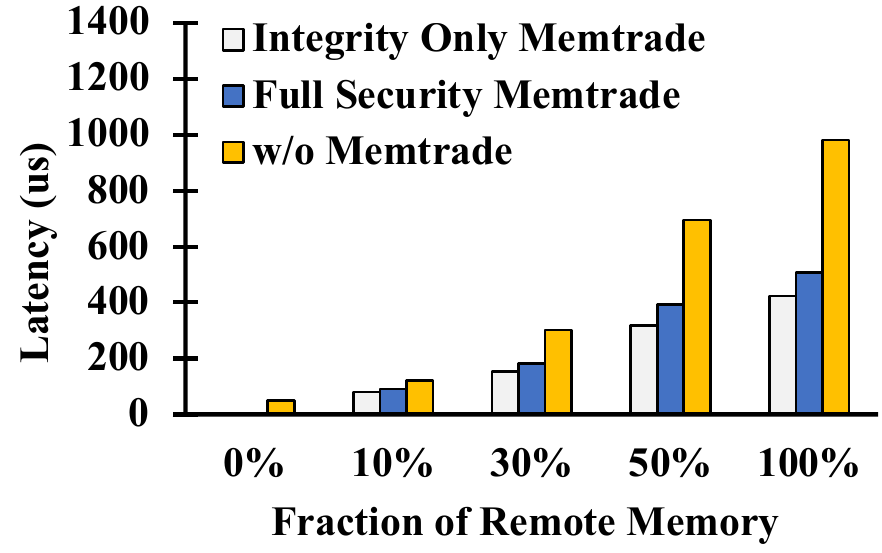}
	}
	\subfloat[][\textbf{p99 Latency}]{
	\label{fig:dalmatian-benefit-99}
		\includegraphics[scale=0.46]{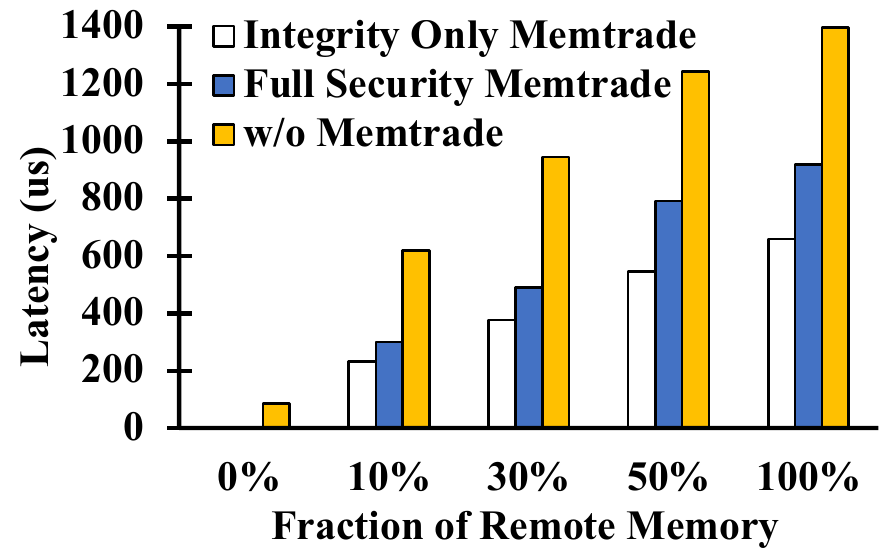}
	}
	\caption{Benefit of \spotframework with various configurations. Without \spotframework, remote requests are served from SSD.}
	\label{fig:dalmatian-benefit}
\end{figure}

\begin{table}[]
\resizebox{\columnwidth}{!}{%
\begin{tabular}{|l|r|r||l|r|r|}
\hline
\multirow{2}{*}{\begin{tabular}[c]{@{}l@{}}Producer\\ Application\end{tabular}} & \multicolumn{2}{c||}{Avg. Latency (ms)} & \multicolumn{1}{c|}{\multirow{2}{*}{\begin{tabular}[c]{@{}c@{}}Consumer\\ Application\end{tabular}}} & \multicolumn{2}{c|}{Avg. Latency (ms)} \\ \cline{2-3} \cline{5-6} 
 & \multicolumn{1}{c|}{\begin{tabular}[c]{@{}c@{}}w/o \\ Harvester\end{tabular}} & \multicolumn{1}{c||}{\begin{tabular}[c|]{@{}c@{}}w/ \\ Harvester\end{tabular}} & \multicolumn{1}{c|}{} & \multicolumn{1}{c|}{\begin{tabular}[c]{@{}c@{}}w/o \\ \spotframework \end{tabular}} & \multicolumn{1}{c|}{\begin{tabular}[c]{@{}c@{}}w/ \\ \spotframework \end{tabular}} \\ \hline
Redis & 0.08 & 0.08 & Redis 0\% & 0.62 & \multicolumn{1}{c|}{--} \\ \hline
Memcached & 0.82 & 0.83 & Redis 10\% & 1.10 & 0.71 \\ \hline
MySQL & 1.57 & 1.60 & Redis 30\% & 1.54 & 0.88 \\ \hline
Storm & 5.33 & 5.47 & Redis 50\% & 2.49 & 0.89 \\ \hline
\end{tabular}%
}
\caption{{\spotframework} benefits consumers at a small cost to producers.}
\label{table:eval-cluster}
\end{table}

\begin{figure*}[!t]
	\centering
	\subfloat[][\textbf{Representative MRCs}]{
		\label{fig:pricing-mrc-subset}
		\includegraphics[width=0.19\textwidth]{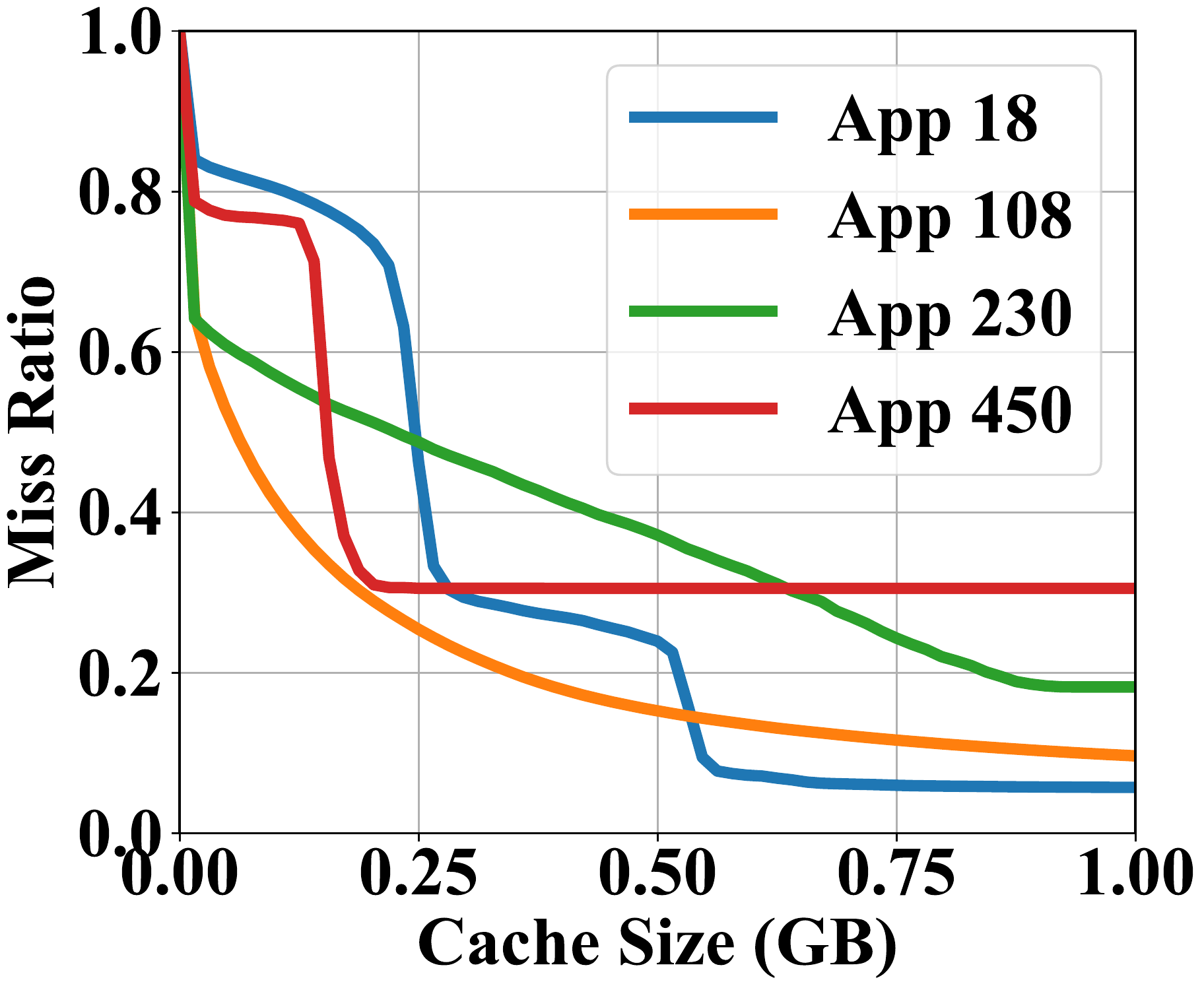}
	}
	\subfloat[][\textbf{Market Price}]{
		\label{fig:pricing-supply-price}
		\includegraphics[width=0.19\textwidth]{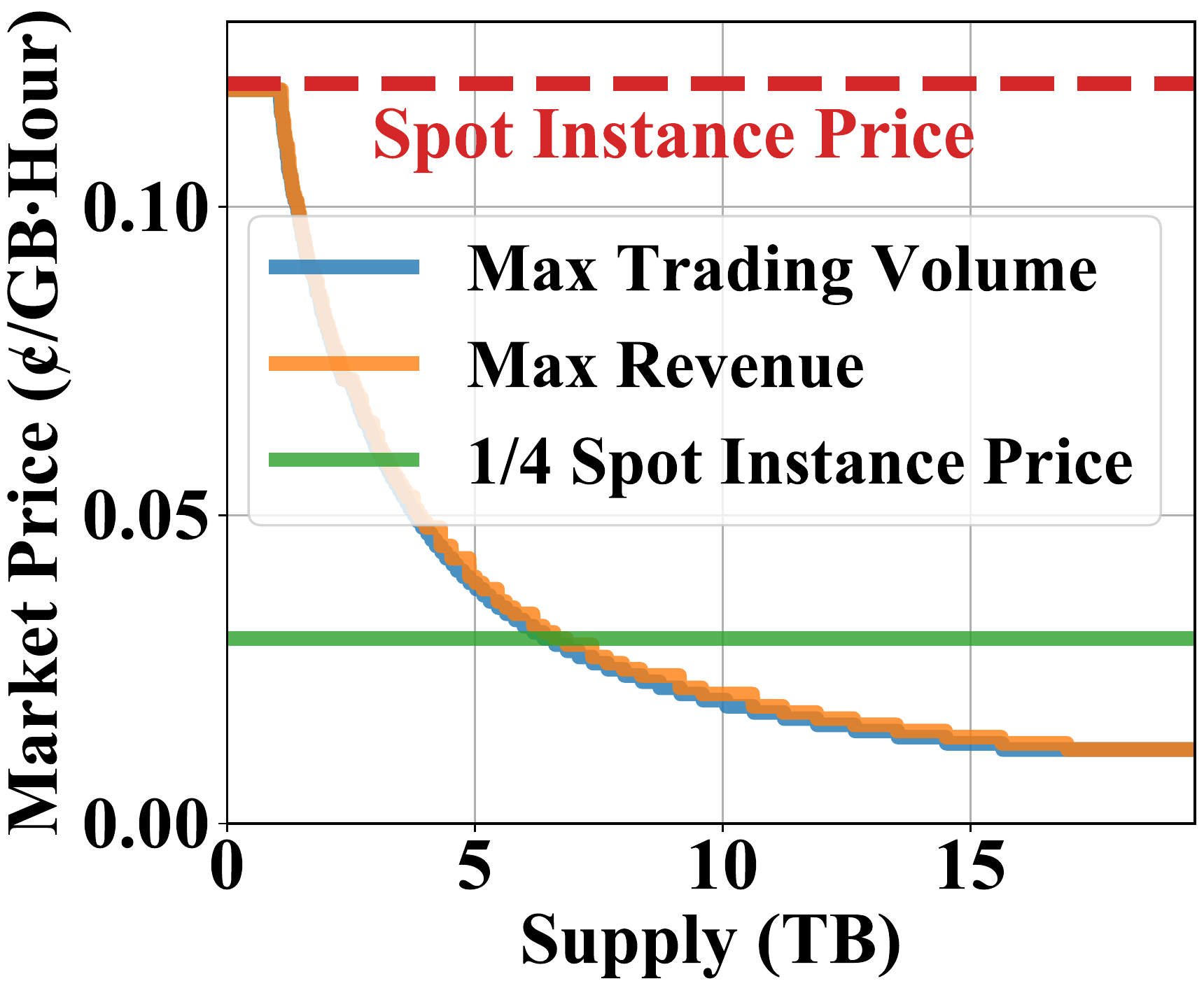}%
	}
	\subfloat[][\textbf{Trading Volume}]{
		\label{fig:pricing-supply-volume}
		\includegraphics[width=0.19\textwidth]{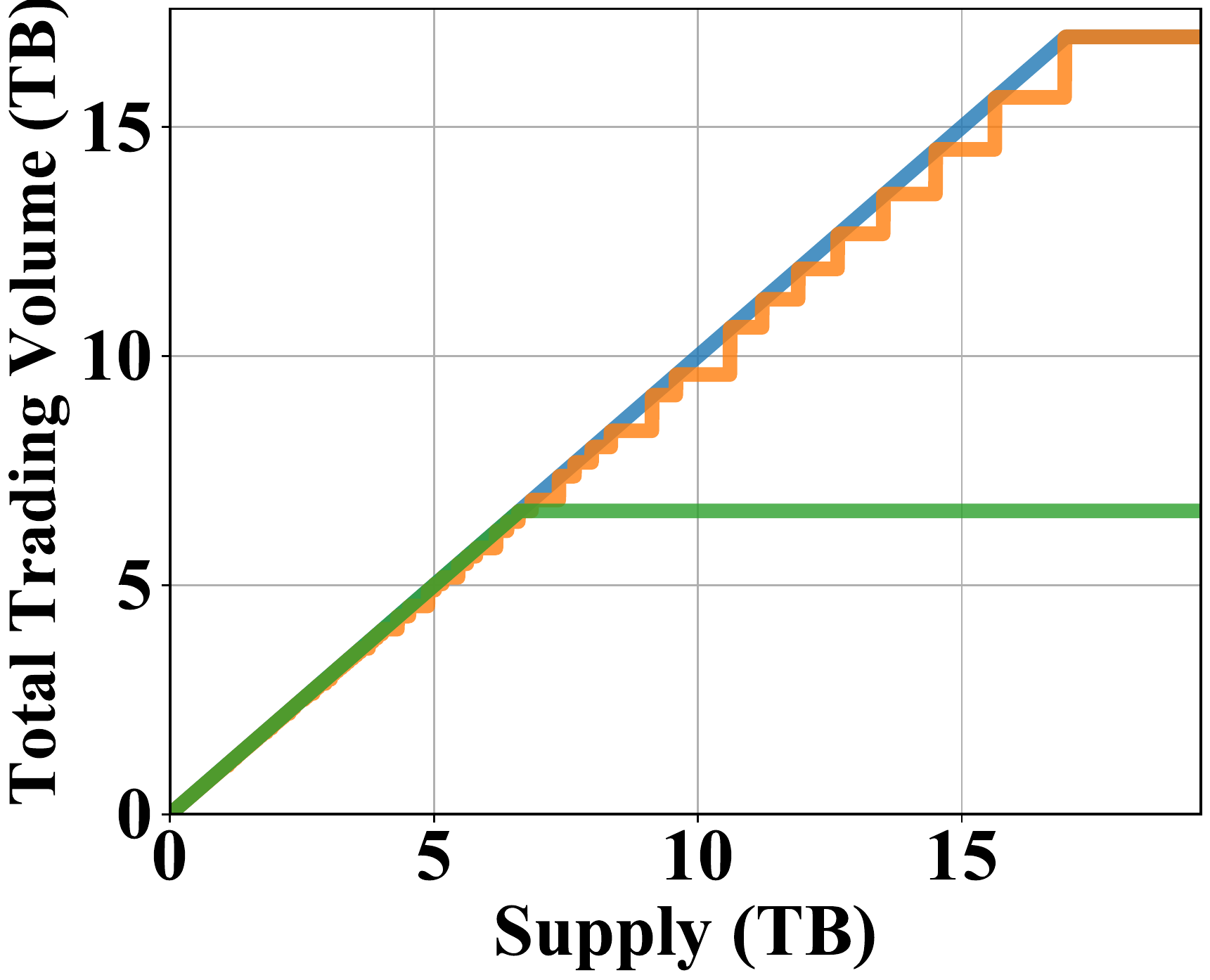}%
	}
	\subfloat[][\textbf{\Producer Revenue}]{
		\label{fig:pricing-supply-revenue}
		\includegraphics[width=0.19\textwidth]{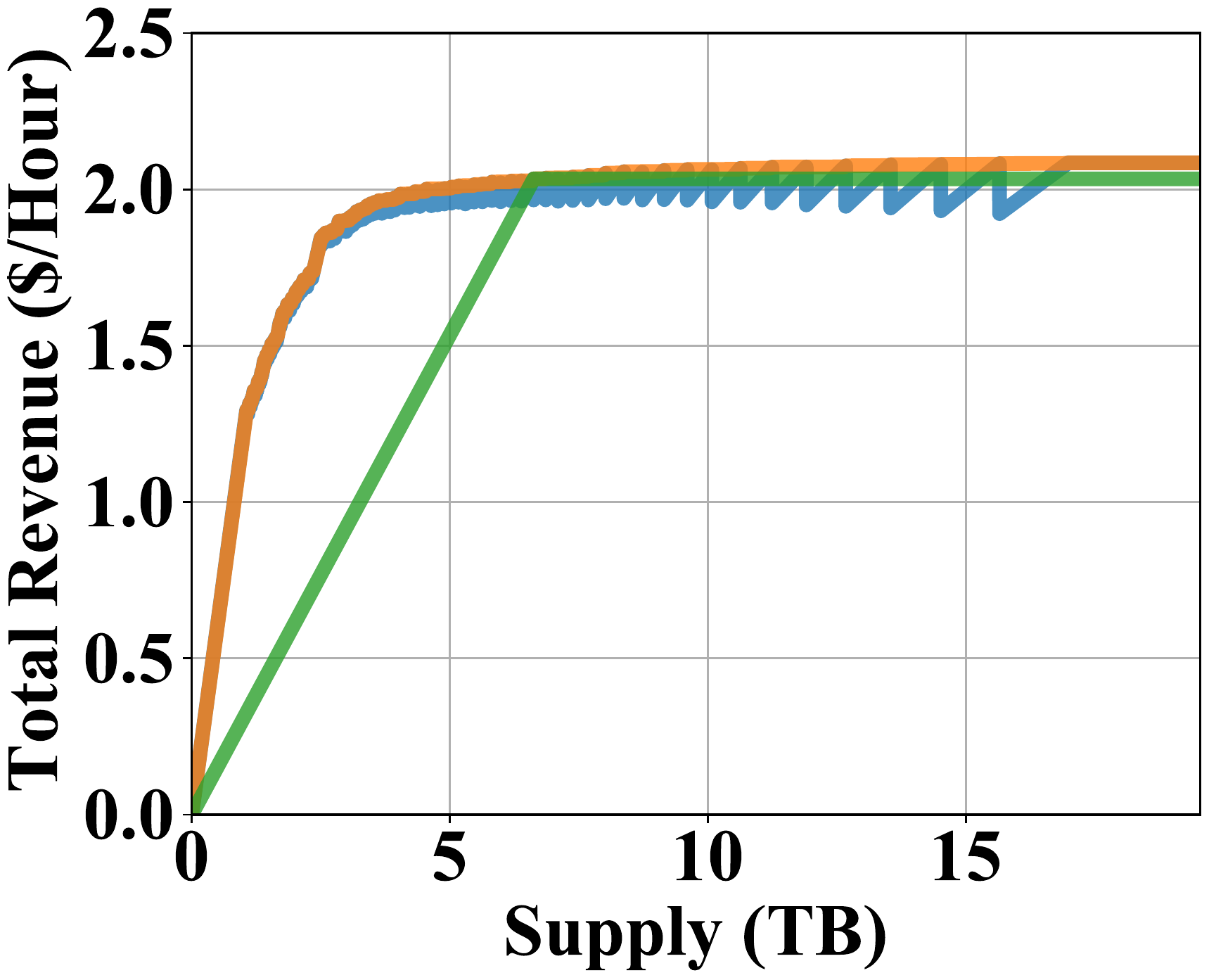}%
	}
	\subfloat[][\textbf{\Client Performance}]{
		\label{fig:pricing-supply-perf}
		\includegraphics[width=0.19\textwidth]{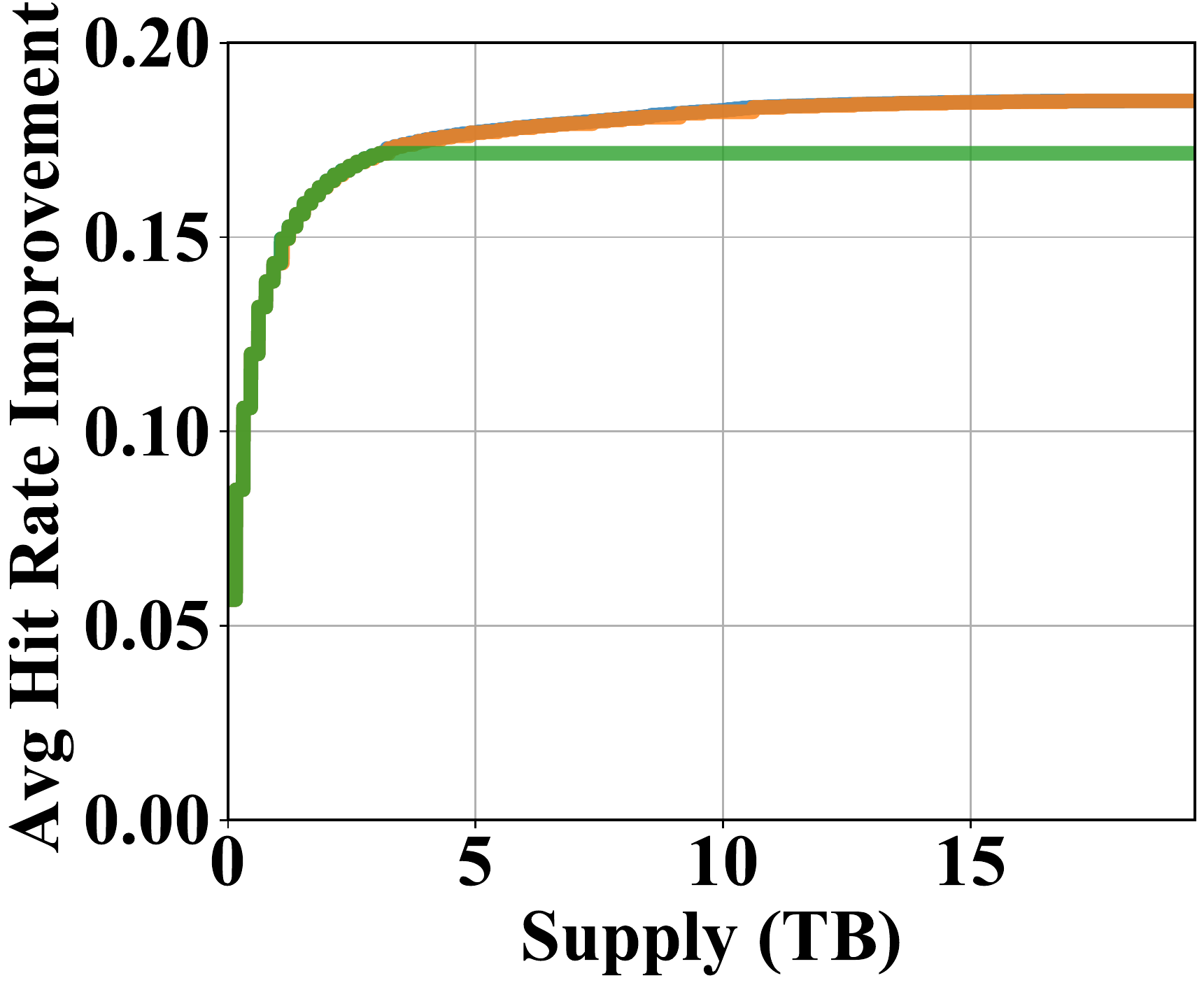}%
	}
	\caption{Effects of different pricing strategies. Revenue and trading-volume maximization both perform similarly.}
  \label{fig:pricing-supply}
\end{figure*}

\subsection{Broker Effectiveness}
\label{subsec:eval-broker}
To evaluate the effectiveness of the {\broker}, we simulate a disaggregated memory consumption scenario by replaying two days worth of traces from Google's production cluster~\cite{google-cluster-trace}. %, which include machines of different sizes (64--512~GB).
Machines with high memory demand -- often exceeding the machine's capacity -- are treated as consumers.
Machines with medium-level memory pressure (at least 40\% memory consumption throughout the trace period) are marked as producers.
When a consumer's demand exceeds its memory capacity, we generate a remote memory request to the \broker. 
We set the consumer memory capacity to 512~GB, the minimum remote memory slab size to 1~GB, and a minimum lease time of 10 minutes.
In our simulation, 1400 consumers generate a total of 10.7~TB of remote memory requests within 48 hours. 
On the producer side, we simulate 100 machines.

%Figure~\ref{fig:simulation-distribution} shows the aggregate remote memory demand for the cluster during the simulation.
On average, the \broker needs to assign 18~GB of remote memory to the producers per minute.
Our greedy placement algorithm can effectively place most requests.
Even for a simulation where producers have only 64~GB DRAM, it can satisfy 76\% of the requests (Figure ~\ref{fig:simulation-placement}). %~\carl{Not clear to me if (b) represents {\em different} simulation runs, or diverse producer host sizes within a single simulation?  Please clarify.}
With larger producers, the total number of allocations also increases.
As expected, \spotframework increases cluster-wide utilization, by 38\% (Figure~\ref{fig:simulation-utilization}).

% \paragraph{Availability Predictor Effectiveness.}
\paragraph{Availability Predictor.}
To estimate the availability of producer memory, we consider its average memory usage over the past five minutes
to predict the next five minutes.  
%Figure~\ref{fig:simulation-arima} demonstrates that 
ARIMA predictions are accurate when a producer has steady usage or follows some pattern; only 9\% of the predicted usage exceeds the actual usage by 4\%.
On average, 4.59\% of the allocated producer slabs get revoked before their lease expires.

%~\carl{We should report stats about premature revocations (revoked before lease expires) in the experiments, e.g. number of slabs and percentage of total slabs.}

%\paragraph{Load Balancing}
%Analysis of load balancing
%
%\paragraph{Benefit of Fair Bandwidth Share} 
%Application behavior during multiple different types of consumer application access the same producer

\begin{figure*}[!t]
	\centering
	\subfloat[][\textbf{Market Price}]{
		\label{fig:pricing-temporal-price}
		\includegraphics[width=0.20\textwidth]{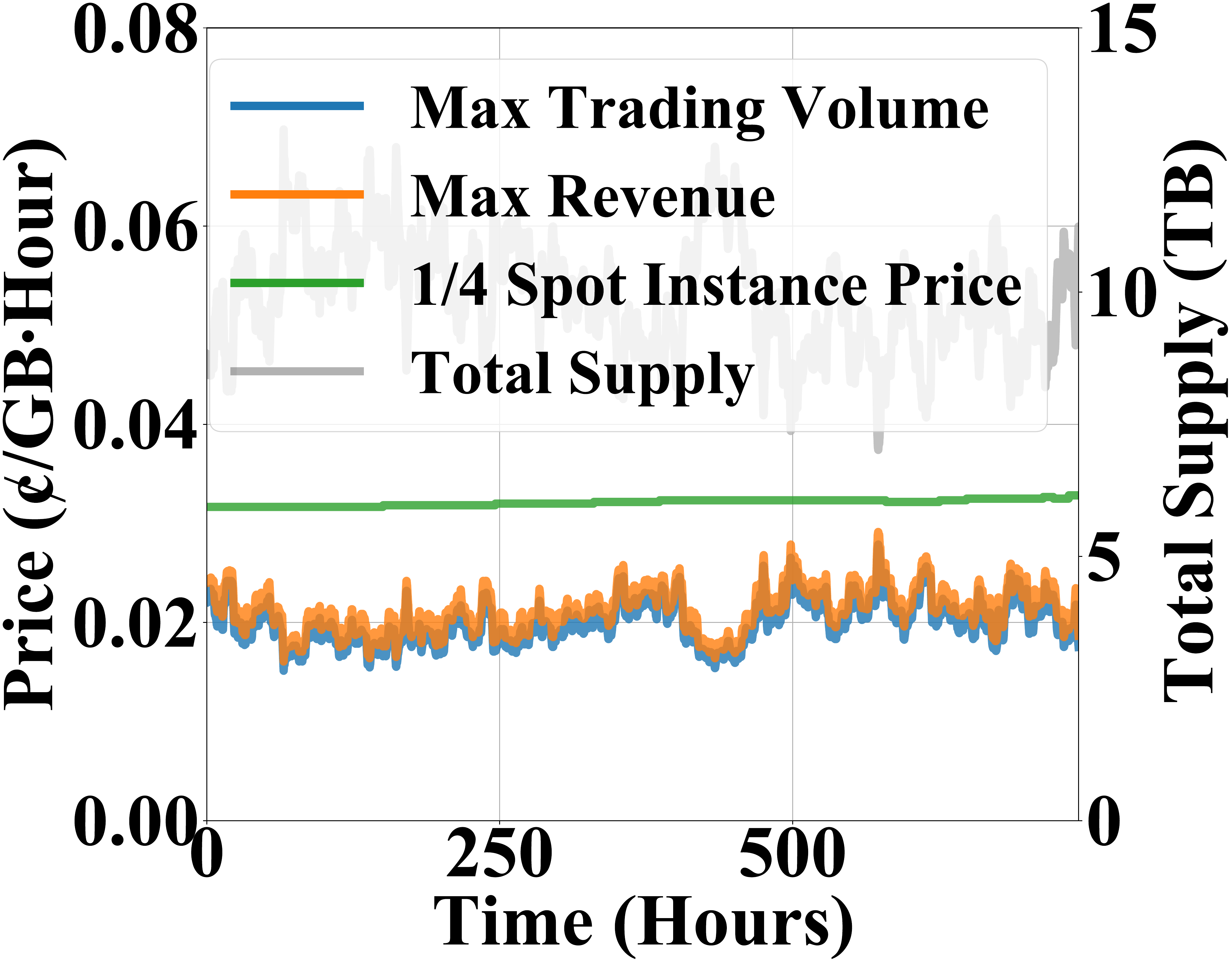}%
	}
	\subfloat[][\textbf{Trading Volume}]{
		\label{fig:pricing-temporal-volume}
		\includegraphics[width=0.192\textwidth]{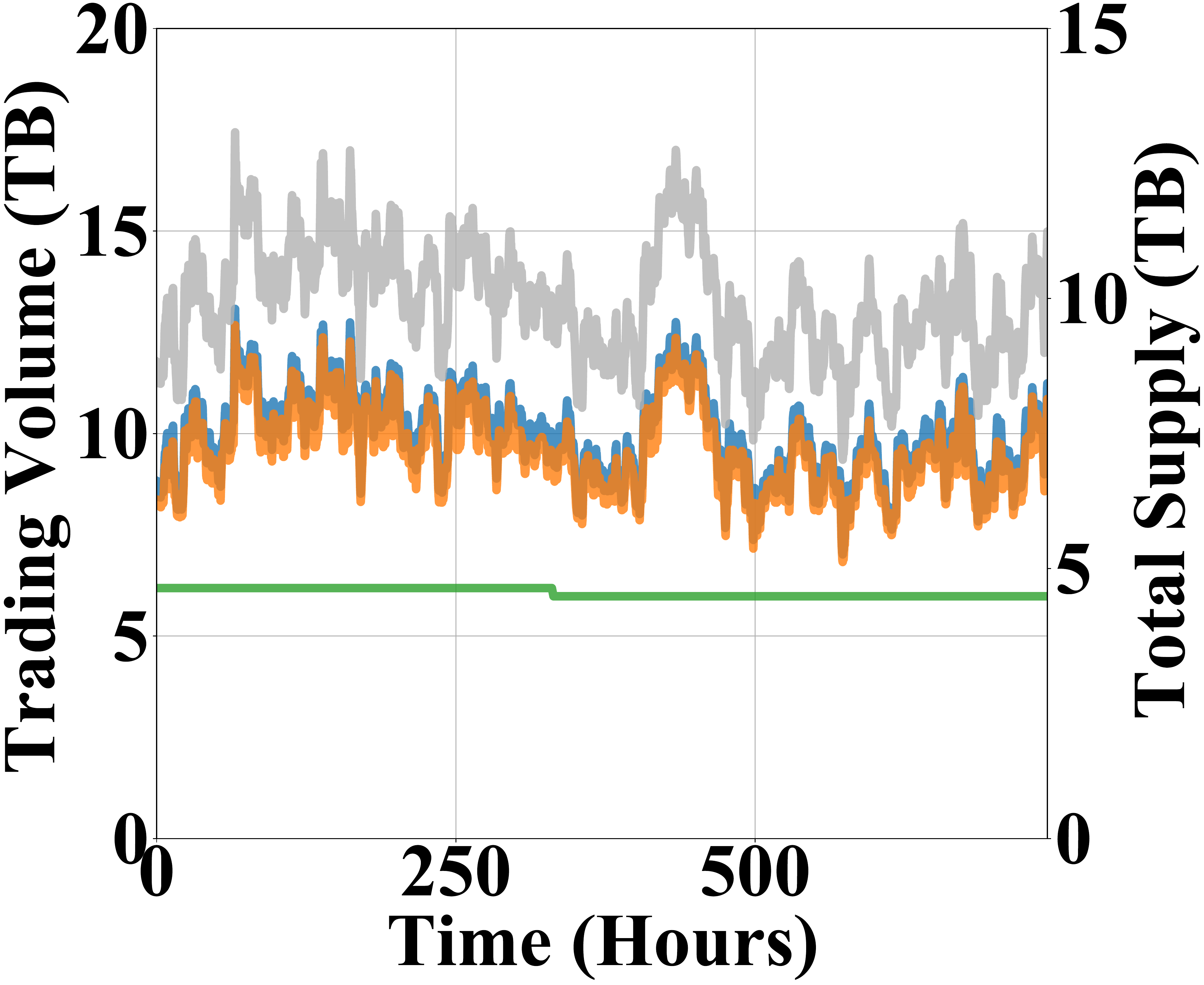}%
	}
	\subfloat[][\textbf{\Producer Revenue}]{
		\label{fig:pricing-temporal-revenue}
		\includegraphics[width=0.195\textwidth]{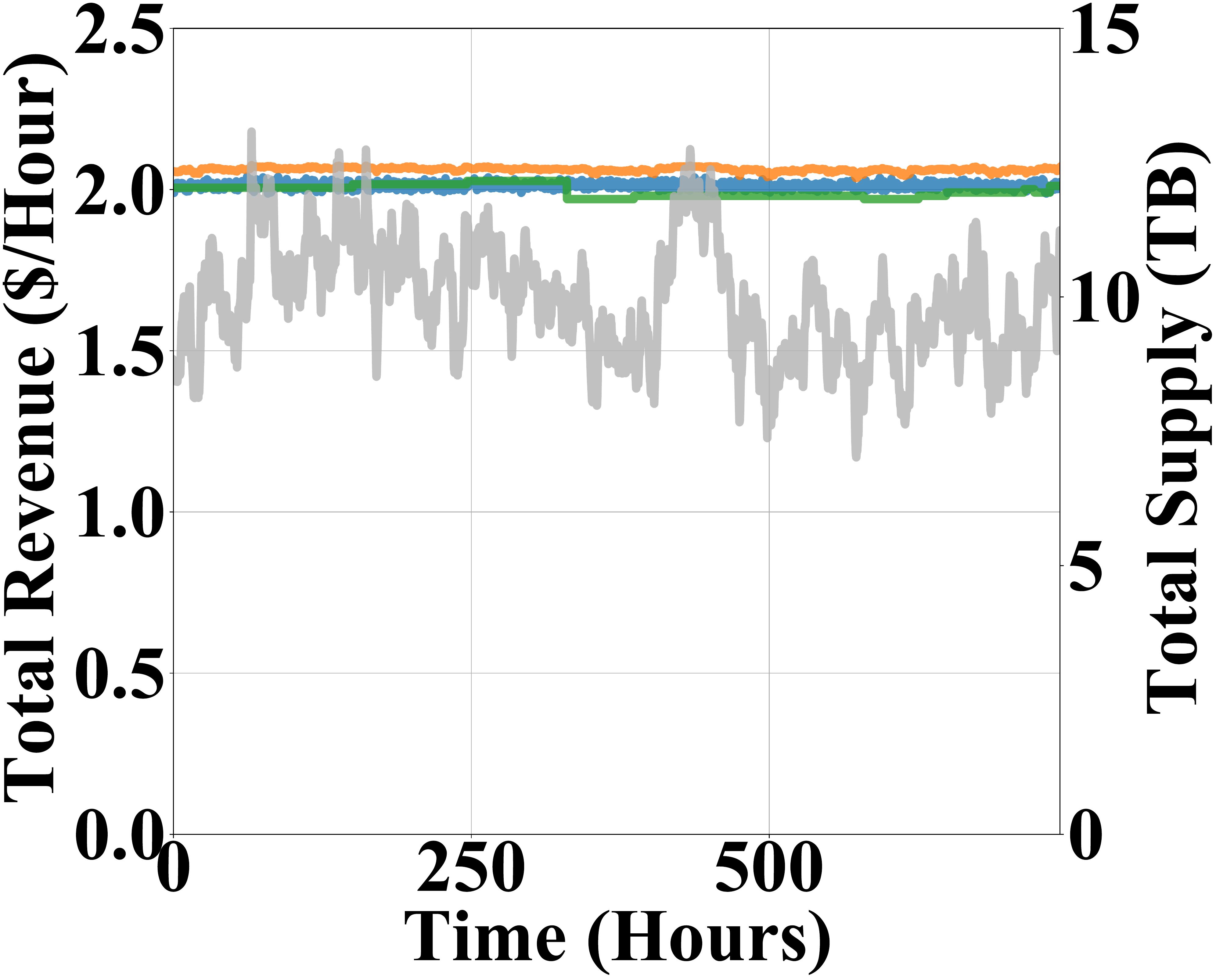}%
	}%\\[-1em]
	% \subfloat[][Perf Benefit of \Client{}s]{
	% 	\label{fig:pricing-temporal-perf}
	% 	\includegraphics[width=0.20\textwidth]{figure/pricing-temporal-perf.pdf}%
	% }
	\subfloat[][\textbf{Cluster Utilization}]{
		\label{fig:pricing-temporal-utilization}
		\includegraphics[width=0.18\textwidth]{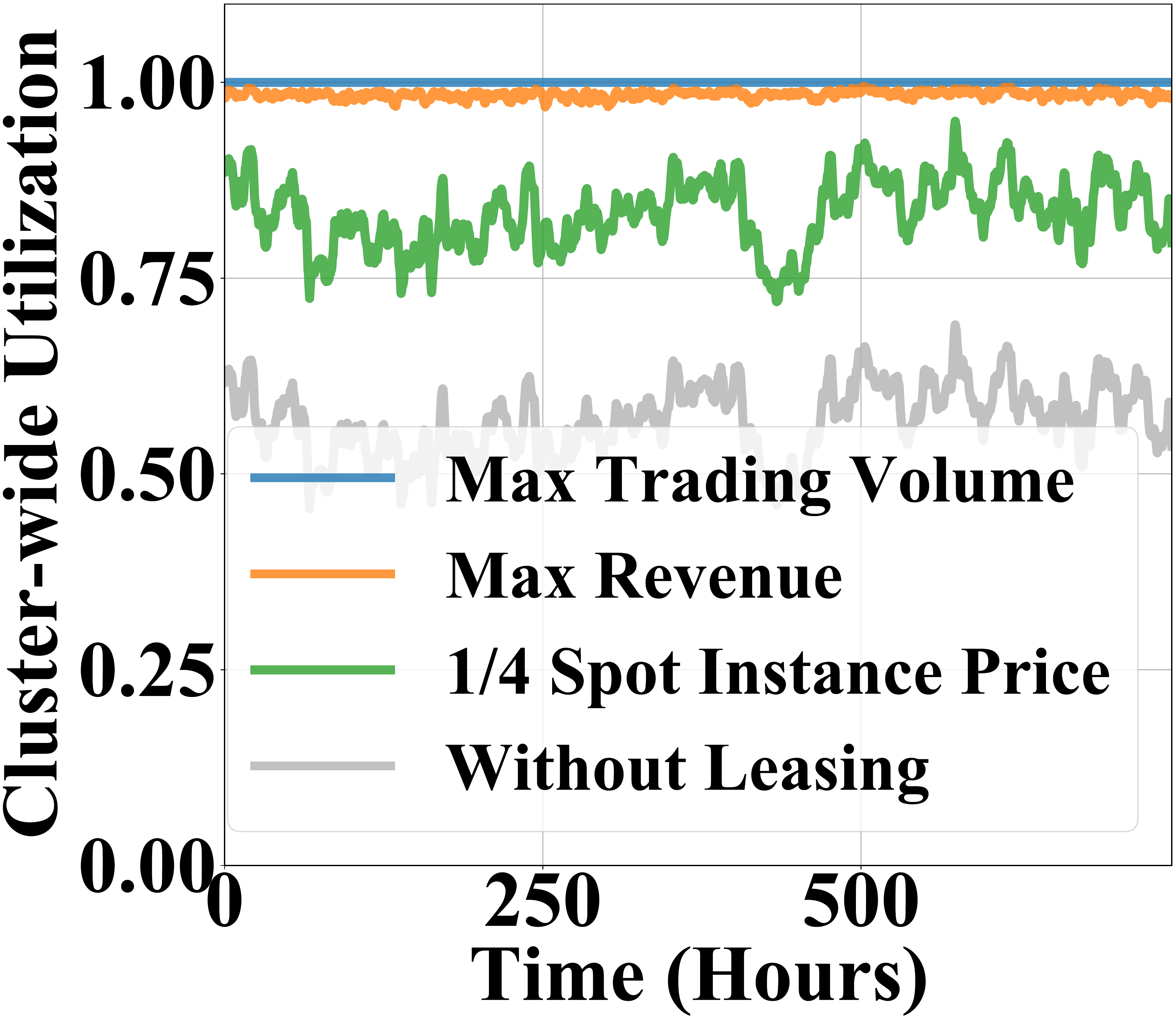}%
	}
	\subfloat[][\textbf{Price Adjustment}]{
		\label{fig:pricing-reactive}
		\includegraphics[width=0.18\textwidth]{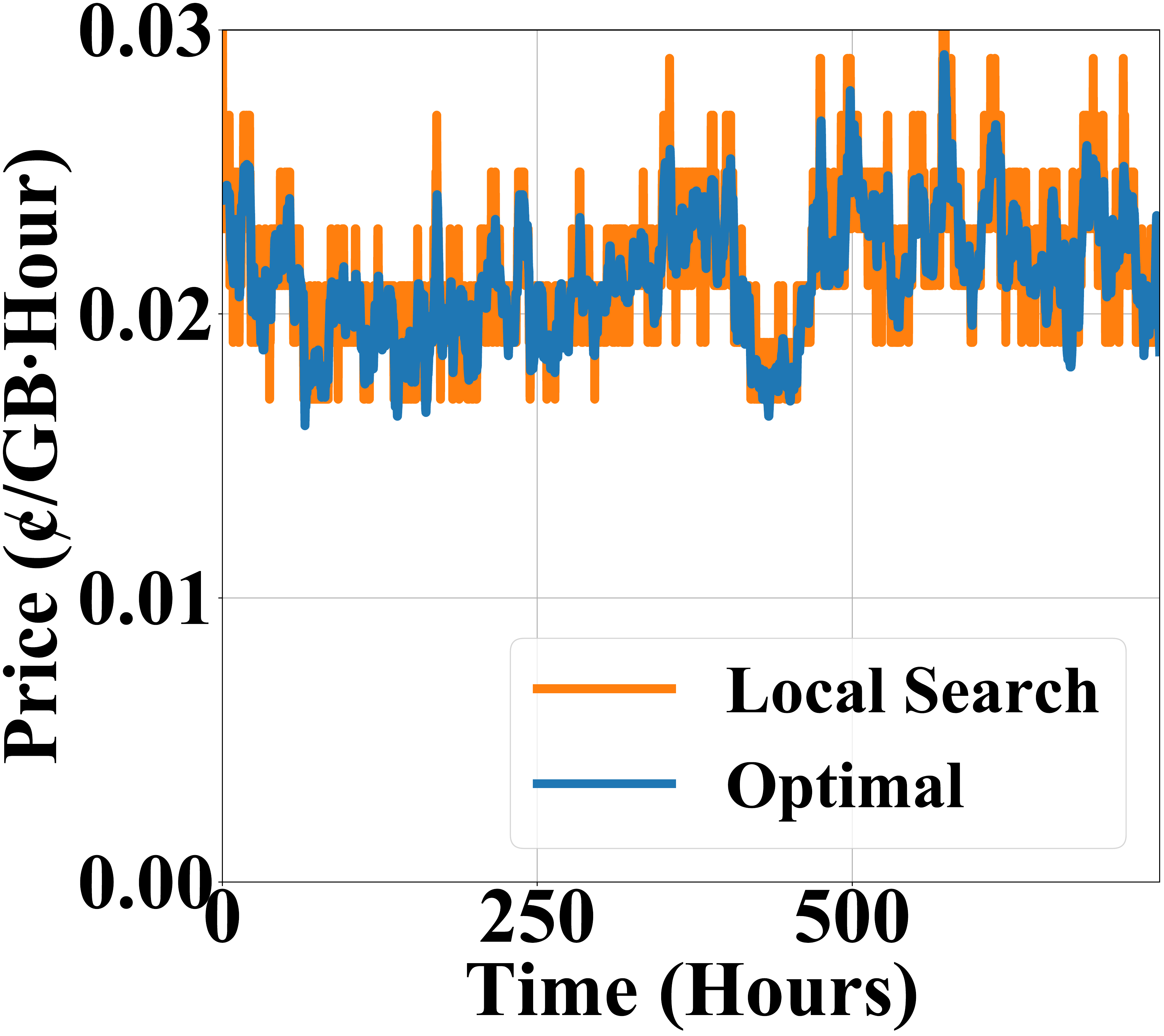}%
	}
	\caption{Temporal market dynamics with simulated supply time-series from idle memory sizes in Google Cluster Trace 2019.}
  \label{fig:pricing-temporal}
\end{figure*}

\subsection{{\spotframework}'s Overall Impact}
\label{subsec:eval-end2end}
\paragraph{Encryption and Integrity Overheads.}
To measure the overhead of {\spotframework}, we run the YCSB workload over Redis with 50\% remote memory. 
Integrity hashing increases remote memory access latency by 24.3\% (22.9\%) at the median (p99).
Hashing and key replacement cause 0.9\% memory overhead for the {\client}.
Due to fragmentation in the \producer VM,  the {\client} needs to consume 16.7\% extra memory over the total size of actual KV pairs. 
Note that this fragmentation overhead would be same if the consumer had to store the KVs in its local memory.

Encryption and key substitution increase latency by another  19.8\% (14.7\%) at the median (p99).
Due to padding, encryption increases the memory overhead by another 25.2\%. 
Fragmentation in the  \producer VM causes 6.1\% memory overhead.
Note that for trusted producers, or for non-sensitive consumer data, encryption can be disabled.

\paragraph{Application-Level Performance.}
We run YCSB over Redis with different consumer VM configurations and security modes. 
The consumer memory size is configured such that Redis needs at least x\% ($x \in \{0, 10, 30, 50, 100\}$) of its working set to be in remote \spotframework memory.
If remote memory is not available, the I/O operation is performed using SSD.

For the 0\% configuration, the entire workload fits in the consumer VM's memory, and is fully served by its local Redis.
Figure~\ref{fig:dalmatian-benefit} shows that an application benefits from using remote memory as additional cache instead of missing to disk.
% during memory constraints, instead of disk, if the application accesses spot memory using  \spotframework
For the fully-secure KV cache interface, \spotframework improves average latency by 1.3--1.9$\times$, and p99 latency by 1.5--2.1$\times$.
In the case of non-sensitive data with only integrity checks, \spotframework provides 1.45--2.3$\times$ and 2.1--2.7$\times$ better average and p99 latencies, respectively.

We also implemented a transparent swap-based consumer interface, built on top of Infiniswap~\cite{infiniswap}. When consuming remote memory via fully-secure remote paging, \spotframework{}'s performance drops due to hypervisor swapping overhead -- average and p99 latency drop by 0.95--2.1$\times$ and 1.1--3.9$\times$, respectively.
However, given a faster swapping mechanism~\cite{leap,AIFM} or faster network (\eg, RDMA), {\spotframework} swap is likely to provide a performance benefit to consumers.

\paragraph{Cluster Deployment.}
To measure the end-to-end benefit of \spotframework, we run 110 virtualized applications on CloudLab --
%Each application is randomly distributed across the cluster. 
64 producer and 46 consumer VMs, randomly distributed across the cluster.
Producers run the six workloads described earlier, while consumers run YCSB on Redis, configured with 10\%, 30\%, and 50\% remote memory.
The total consumer and producer memory footprints are 0.7~TB and 1.3~TB, respectively; our cluster has a total of 2.6~TB.
Table~\ref{table:eval-cluster} shows that {\spotframework} benefits the consumers even in a cluster setting.
{\spotframework} improves the average latency of consumer applications by 1.6--2.8$\times$,
while degrading the average producer latency by 0.0--2.1\%.

\subsection{Pricing Strategy}
\label{subsec:eval-pricing}
To study the impact of pricing strategy on the market, we simulate several pricing strategies with different objectives, such as maximizing the total trading volume, and maximizing the total revenue of \producer{}s.
Our baseline simply sets the remote memory price to one quarter of the current spot-instance price~(Figure~\ref{fig:pricing-supply-price}). 
We simulate 10,000 \client{}s that use \spotframework as an inexpensive cache. 
To estimate the expected performance benefit for \client{}s,
we select 36 applications from the MemCachier trace, generate their MRCs (Figure~\ref{fig:apdx-pricing-mrc} in the appendix for the interested), and randomly assign one to each \client.
%Figure~\ref{fig:pricing-mrc-subset} presents a subset of those MRCs.
For each {\client} we allocate memory such that local memory serves at least 80\% of its optimal hit ratio.

% \begin{figure}[!t]
% 	\centering
% 	\includegraphics[width=0.7\columnwidth]{figure/pricing-reactive-price.pdf}
% 	\caption{The market price set by the reactive approach and the optimal market price of the max-trading-volume strategy.}
%   \label{fig:pricing-reactive}
% \end{figure}

%\begin{figure}[]
%	\centering
%	\includegraphics[width=0.67\columnwidth]{figure/subset-pricing-mrc.pdf}
%	\caption{A subset of MemCachier application MRCs}
%  \label{fig:pricing-mrc-subset}
%\end{figure}

%With a given total remote memory supply, each pricing strategy explores different prices and sets the one which maximizes its objective. 
%Figure~\ref{fig:pricing-supply} shows the effect of each pricing strategy on the market with different supply when its economic objective is achieved.

% The max-trading-volume strategy guarantees maximal total trading volume (Figure~\ref{fig:pricing-supply-volume}), while the max-revenue strategy maximizes the total revenue of \producer{}s (Figure~\ref{fig:pricing-supply-revenue}) --
The strategies that maximize total trading volume (Figure~\ref{fig:pricing-supply-volume}) and total \producer revenue (Figure~\ref{fig:pricing-supply-revenue})
both adjust the market price dynamically to optimize their objectives.
Significantly, when the remote-memory supply is sufficient,
all three pricing strategies can improve the relative hit ratio
for \client{}s by more than 16\% on average
(Figure~\ref{fig:pricing-supply-perf}).
%~\carl{Seems like a profit-maximizing broker would want to employ the {\em max-revenue} strategy to optimize its own commission?}
% For the {\broker}s which prefer high utilization and/or efficient markets, the max-trading-volume strategy is the best fit.

We also examine temporal market dynamics using a real-world trace to simulate a total supply of remote memory which varies over time. We use the idle memory statistics from the Google Cluster Trace 2019 -- Cell~C~\cite{google-cluster-trace-2019} to generate the total supply for each time slot, and assume one Google unit represents 5~GB of memory. 
For the spot instance price, we use the AWS historical price series of the spot instance type r3.large in the region us-east-2b~\cite{aws-spot-instance-price}.
Figure~\ref{fig:pricing-temporal} plots the results.
Consistent with the earlier pricing-strategy results in Figure~\ref{fig:pricing-supply}, the max-trading-volume and max-revenue strategies effectively adjust the market price based on both supply and demand (Figure~\ref{fig:pricing-temporal-price}).
The behavior of all the three pricing strategies with different economic objectives show similar levels of consistency~(Figure~\ref{fig:pricing-temporal-volume}--\ref{fig:pricing-temporal-utilization}).
%They have similar behavior in terms of the total trading volume (Figure~\ref{fig:pricing-temporal-volume}), total revenue of \producer{}s (Figure~\ref{fig:pricing-temporal-revenue}), performance benefits of \client{}s (Figure~\ref{fig:pricing-temporal-perf}), and the cluster-wide utilization (Figure~\ref{fig:pricing-temporal-utilization}).

We also examine a more realistic scenario where consumers consider the probability of being evicted when using MRCs to calculate their demand. If the eviction probability is 10\%, the total revenue will decrease by 7.6\% and 7.1\% with the max-trading-volume strategy and the max-revenue strategy, respectively. Also, the cluster-wide utilization reductions corresponding to the max-trading-volume strategy and the max-revenue strategy are 0.0\% and 5.8\%.

In practice, the \broker may have no prior knowledge regarding the impact of market price on \client demand.
In this case, we adjust the market price by searching for a better price locally with a predefined step size (0.002 cent/GB$\cdot$hour).
Figure~\ref{fig:pricing-reactive} demonstrates the effectiveness of this approach using a simulation with the Google trace;
%To measure how well this reactive approach approximates the optimal market price of the max-trading-volume strategy, we simulate the reactive approach reusing the total remote memory supply extracted from Google Cluster Trace. 
%Figure~\ref{fig:pricing-reactive} presents the market price set by the reactive approach and the optimal price of the max-trading-volume strategy, which shows that the reactive approach works well in this real-world scenario and is only different than the optimal one by 4.4\% on average. 
% With this approach,
the market price deviates from the optimal one by only 3.5\% on average.
Cluster-wide utilization increases from 56.8\% to 97.9\%, \client hit ratios improve by a relative 18.2\%, and the \client{}'s cost of renting extra memory reduces by an average of 82.1\% compared to using spot instances.

\section{Related Work}

%Our work is informed by four avenues of related research: memory disaggregation, public cloud spot marketplaces, single-server memory management, and autoscaling and rightsizing of cloud instances.

\paragraph{Memory Disaggregation.}
%A large body of recent work has explored the possibility of \emph{disaggregating} datacenter resources~\cite{LegoOS,flash-disaggre,Han:2013,networkrequirementsdisagg,googledisagg,zombieland,bladedisagg,redbox,remotememorysocc,farmemory-throughput, remote-regions,infiniswap,AIFM}.
% Disaggregation has been widely adopted for some resources, like flash storage~\cite{Klimovic:2016,ebs,storagedisagg}.
% However, memory has proven to be a more challenging resource to disaggregate for several reasons.
% First, memory has stringent performance requirements: it has very low latency (10s-100s of nanoseconds) and high bandwidth (100s of GB/s)~\cite{memoryspecs}.
% Second, operating systems tend to consume the entirety of the physical memory of the server over time~\cite{ESX}. 
% Therefore, unlike storage, it is not always clear how much memory is actually needed for the application~\cite{ESX,memshare,infiniswap,googledisagg,farmemory-throughput}.
% Third, unlike storage, memory is volatile. 
% This presents a unique challenge in terms of managing consistency and handling unexpected errors~\cite{infiniswap,remoteregions}.
Existing work on \emph{disaggregating} datacenter resources~\cite{LegoOS,flash-disaggre,Han:2013,app-performance-disagg-dc,googledisagg,zombieland,bladedisagg,redbox,remotememorysocc,farmemory-throughput, remote-regions,infiniswap,AIFM}
%However, prior work assumes 
assume that \emph{memory is contained within the same organization and shared among multiple cooperative applications}. %~\cite{infiniswap,googledisagg,remoteregions,farmemory-throughput,AIFM}.
Given the large amount of idle memory  and diverse consumer applications and workloads, public clouds serve as a promising environment to exploit remote memory.
%However, implementing disaggregated memory on a public cloud raises the challenges outlined in Section~\ref{subsec:challenge}. 
% Public cloud environments are typically virtualized.
% This complicates both performance, since network requests need to traverse the hypervisor, as well as the memory estimation, since the semantic gap between the hypervisor and the guest limits visibility into application behavior.
% In addition, we need to avoid crashing the application if it unexpectedly needs to reclaim its harvested memory. 

\paragraph{Public Cloud Spot Marketplaces.}
Amazon AWS, Microsoft Azure, and Google Cloud offer spot instances~\cite{spotinstance} -- a marketplace for unutilized public cloud VMs that have not been reserved, but have been provisioned by the cloud provider.
AWS allows customers to bid on spot instances while Azure and Google Cloud \cite{azurespot,googlespot} sell them at globally fixed prices.
Prior research has explored the economic incentives for spot instances~\cite{agmon2013deconstructing,spot-cache,zheng2015bid,xu2013dynamic,song2012optimal,
zhang2011dynamic,raas,ginseng1,ginseng2,transient, cloudex}.
%Wang \etal's use of spot instances for improving the cost-efficiency of in-memory caches~\cite{spot-cache} is most closely related to our work from a consumer perspective.
However, prior work used full spot instances to produce memory; \spotframework is more generic, enabling
fine-grained consumption of excess memory from any instance type.

%Amazon AWS was the first major cloud provider to introduce a marketplace for unutilized public cloud VMs, called spot instances~\cite{spotinstance}.
%With spot instances, AWS allows customers to bid on VMs that have not been reserved, but have been provisioned by AWS.
%Azure and Google Cloud have introduced similar services. 
%Instead of offering VMs via an auction~\cite{azurespot,googlespot}, they sell them at a globally fixed price.
%In addition, there is a large body of academic work exploring the economic incentives for spot instances~\cite{agmon2013deconstructing,zheng2015bid,xu2013dynamic,song2012optimal,zhang2011dynamic,raas,ginseng1,ginseng2,transient}.
%The public cloud system most closely related to our work is Wang \etal's use of spot instances for improving the cost-efficiency of in-memory caches~\cite{spot-cache}.
%From a consumer perspective, their approach is similar to ours. 
%However, they use full spot instances to produce memory; \spotframework consumes excess memory from
%other existing instances.
% instances running other workloads.

\paragraph{Single-Server Memory Management.}
Reclaiming idle memory from applications %for various uses 
has been explored in many different contexts, \eg, physical memory allocation across processes by an OS~\cite{workingset,ostheory,thermostat}, ballooning in hypervisors~\cite{ESX,hotplug,sharma2016per}, transcendent memory for caching disk swap~\cite{transmemory1,transmemory2,transmemory3}, \etc
Our harvesting approach is related to application-level ballooning~\cite{application-level-ballooning}. %, in which a Java runtime reclaims application memory.  
However, in most prior work, applications belong to the same organization while \spotframework harvests and grants memory across tenants.
% to applications from different tenants.
% Work on
Multi-tenant memory sharing~\cite{ginseng1,ginseng2,memshare} has considered only single-server settings, limiting datacenter-wide adoption.

\paragraph{Resource Autoscaling and Instance Rightsizing.}
Cloud orchestration frameworks and schedulers~\cite{borg,omega,kubernetes,CRED,Yarn} can automatically increase or decrease the number of instances based on the task arrival rate.
However, users still need to statically determine the instance size before launching a task, which may lead to resource overprovisioning.

Instance rightsizing~\cite{autopilot,kubernetes-vpa,cherrypick, harvestvm, smartharvest} automatically determines the instance size based on the resource consumption of a task.
%While this is an effective way to reclaim underutilized resources (including memory), it may be difficult to deploy in a public cloud.
In existing solutions, the cloud provider is fully responsible for the resource reclamation and performance variability in the producer VMs.
\spotframework is by design more conservative: 
%producers generate cache capacity from their idle memory, can reclaim it at any time, and the memory is inherently consumed as a cache that can disappear.
{\producer}s can generate and reclaim the cache capacity at  any time.
Even with rightsizing, servers may have idle memory that can be harvested and offered to  remote applications.

\section{Concluding Remarks}

We presented \spotframework,  a readily deployable system for the realization of memory disaggregation in public clouds.
With the rising popularity of serverless and disaggregated storage, we believe there will be increased demand for offering disaggregated computing computing resources in public clouds, and our work attempts to apply a similar approach to memory.
This opens several interesting research directions for future work, including %exploring market and pricing mechanisms for disaggregated memory, as well as 
exploring whether other resources, such as CPU and persistent memory, can be offered in a similar manner via a disaggregated-resource market.
\label{lastpage}

\bibliographystyle{abbrv}
\bibliography{spot-memory}
\appendix
\clearpage
\onecolumn 
\section{APPENDIX}

\begin{figure}[!ht]
	\centering
	 \subfloat[][\textbf{Zipfian on Redis}]{
	 	\label{fig:apdx-memcachier-rocksdb}
	 	\includegraphics[width=0.33\textwidth]{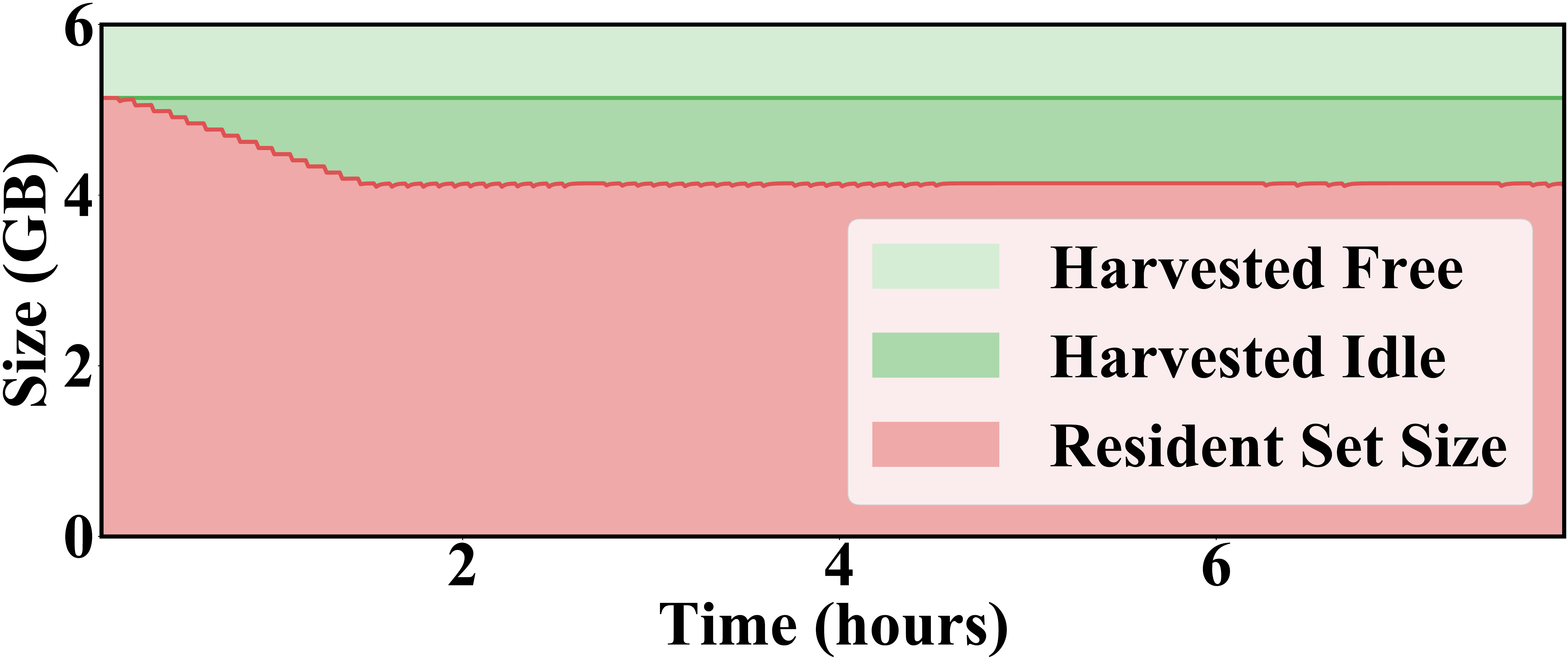}%
	 }
	\subfloat[][\textbf{MemCachier on memcached}]{
		\label{fig:apdx-memcachier}
		\includegraphics[width=0.33\textwidth]{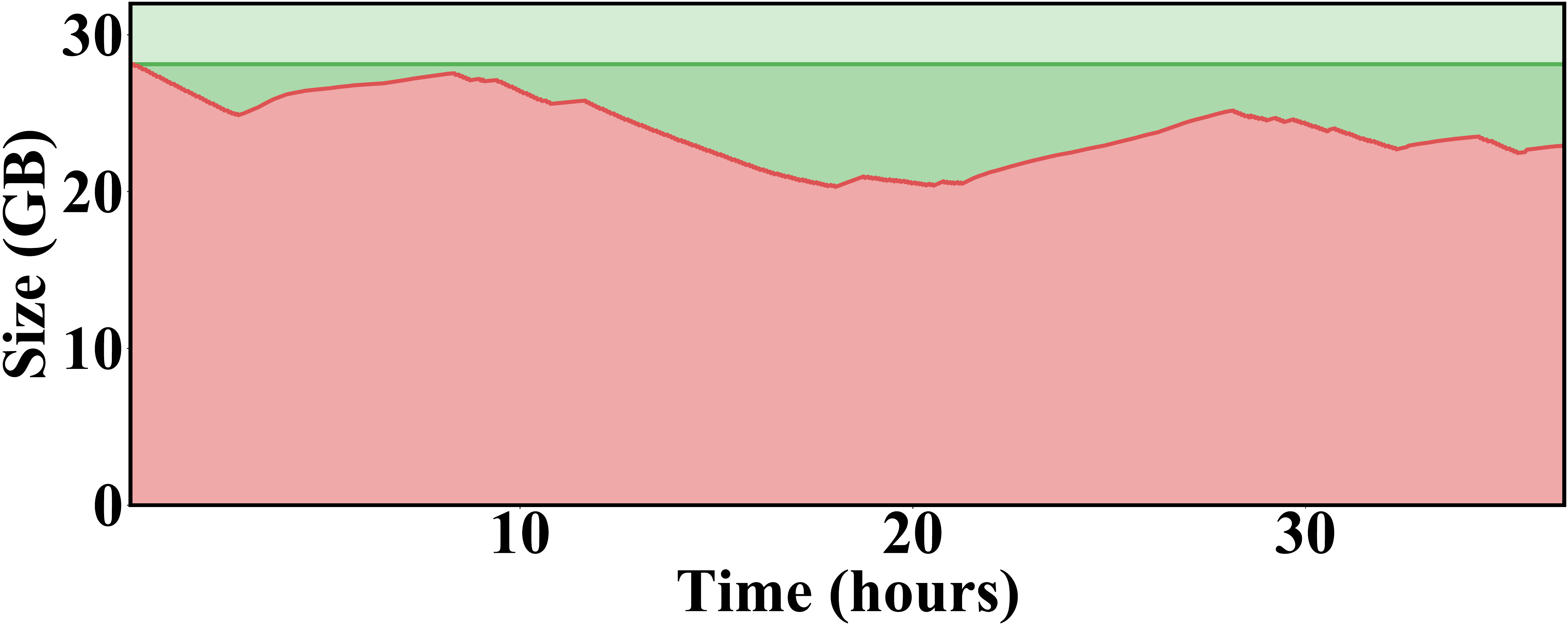}%
	}
	 \subfloat[][\textbf{MemCachier on MySQL}]{
	 	\label{fig:eapdx-snowset}
	 	\includegraphics[width=0.33\textwidth]{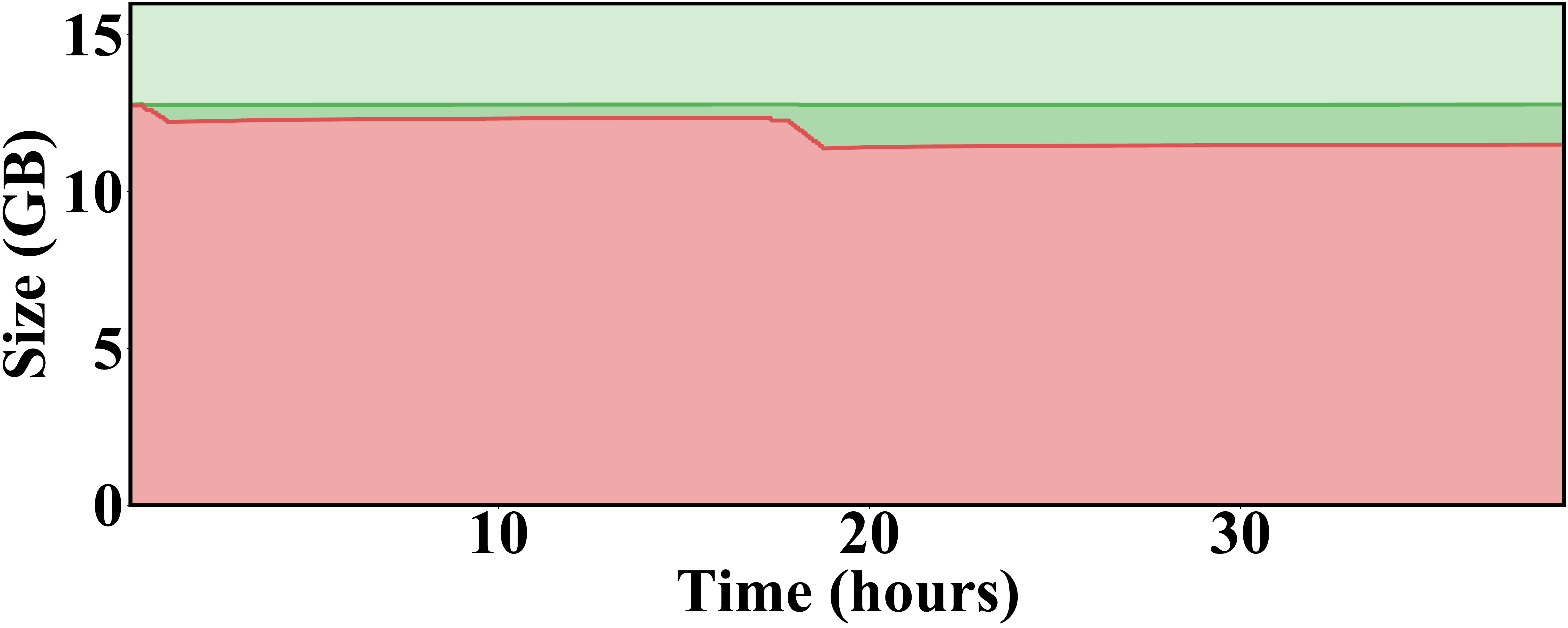}%
	 }\\[-1em]
	\subfloat[][\textbf{Image classification on XGBoost}]{
		\label{fig:apdx-tf}
		\includegraphics[width=0.33\textwidth]{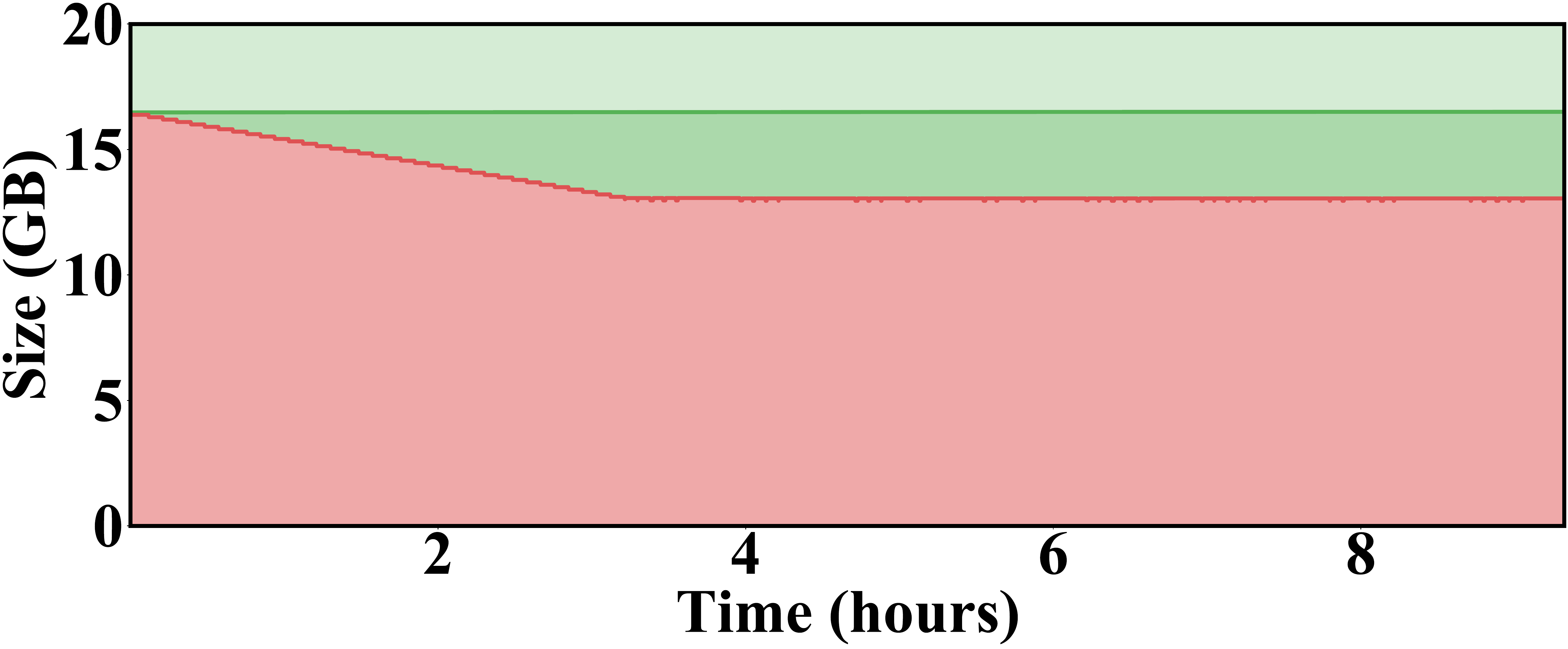}%
	}
	 \subfloat[][\textbf{Yahoo Streaming on Apache Storm}]{
	 	\label{fig:apdx-storm}
	 	\includegraphics[width=0.33\textwidth]{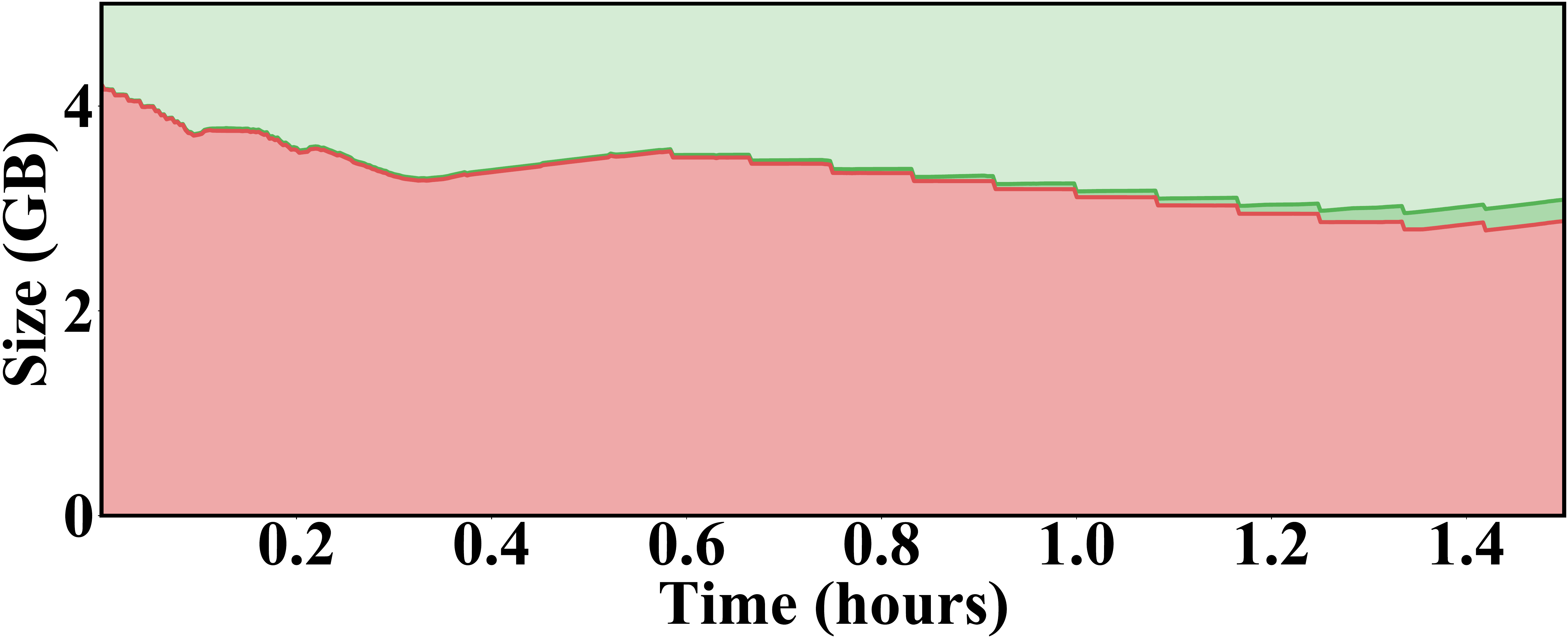}%
	 }
	 \subfloat[][\textbf{CloudSuite Web Serving}]{
	 	\label{fig:apdx-cloudsuite}
	 	\includegraphics[width=0.33\textwidth]{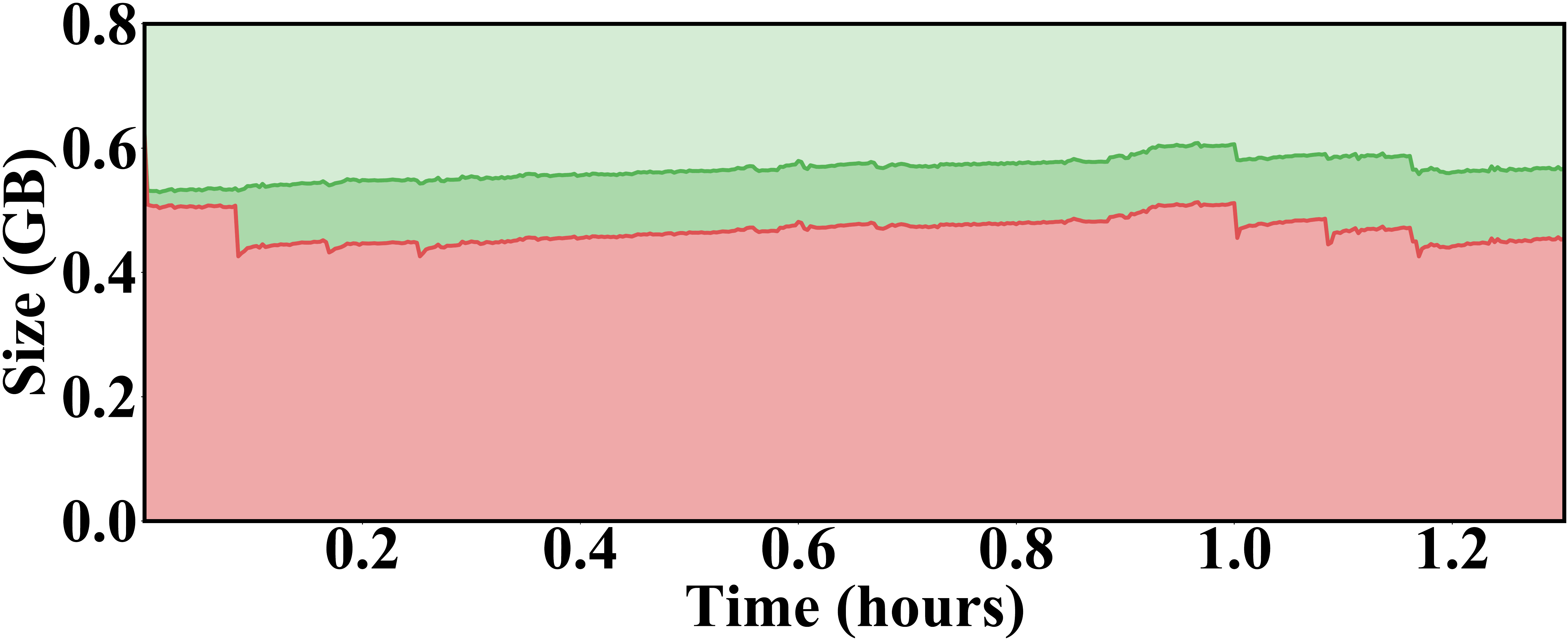}%
	 }
	\caption{Unallocated represents the part of memory not allocated to the application; harvested means the portion of application's memory which has been swapped to disk; \tswap denotes the part of memory used by \tswap to buffer reclaimed pages; RSS consists of application's anonymous pages, mapped files, and page cache that are collected from the cgroup's stats file. }
	\label{fig:apdx-harvester-over-time}
\end{figure}

\begin{figure}[!ht]
	\centering
	\includegraphics[width=\textwidth]{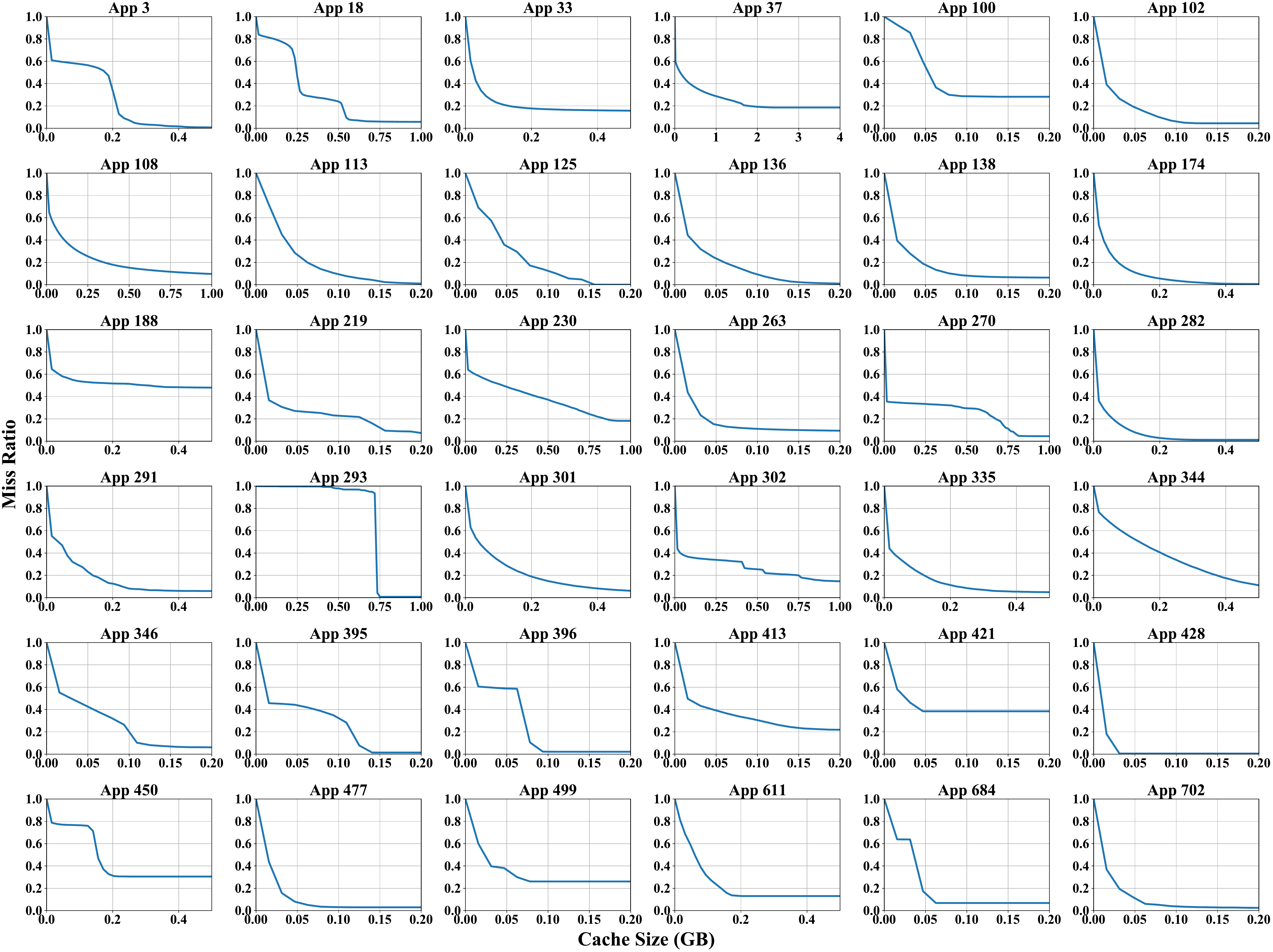}
	\caption{Miss Ratio Curves of 36 MemCachier Applications}
  \label{fig:apdx-pricing-mrc}
\end{figure}

\end{document}